%% file: ms.tex
\documentclass[twocolumn]{aastex6}

\usepackage{aas_macros}
\usepackage{amsfonts}
\usepackage{amsmath}
\usepackage{graphicx}
\usepackage{subfigure}
\usepackage{grffile}
\newcommand\ut{1}
\newcommand\sagan{2}
\newcommand\ww{3}

\shorttitle{ZEIT V}
\shortauthors{Rizzuto et al.}

\begin{document}
\title{Zodiacal Exoplanets in Time (ZEIT) V: A Uniform Search for Transiting Planets in Young Clusters Observed by K2}

\author{
Aaron C. Rizzuto\altaffilmark{\ut},
Andrew W. Mann\altaffilmark{\ut},
Andrew  Vanderburg\altaffilmark{\sagan},
Adam L. Kraus\altaffilmark{\ut},
Kevin R. Covey\altaffilmark{\ww}
}

\altaffiltext{\ut}{Department of Astronomy, The University of Texas at Austin, Austin, TX 78712, USA}
\altaffiltext{\sagan}{Sagan Fellow, Department of Astronomy, The University of Texas at Austin, Austin, TX 78712, USA}
\altaffiltext{\ww}{Department of Physics \& Astronomy, Western Washington University, Bellingham, WA 98225, USA}

\begin{abstract}

Detection of transiting exoplanets around young stars is more difficult than for older systems due to increased stellar variability. Nine young open cluster planets have been found in the \emph{K2} data, but no single analysis pipeline identified all planets. We have developed a transit search pipeline for young stars which uses a transit-shaped notch and quadratic continuum in a 12 or 24\,hour window to fit both the stellar variability and the presence of a transit. In addition, for the most rapid rotators (P$_\mathrm{rot}<2$\,days) we model the variability using a linear combination of  observed rotations of each star. To maximally exploit our new pipeline, we update the membership for four stellar populations observed by \emph{K2} (Upper Scorpius, Pleiades, Hyades, Praesepe), and conduct a uniform search of the members.  We identify all known transiting exoplanets in the clusters, 17 eclipsing binaries, one transiting planet candidate orbiting a potential Pleiades member, and three orbiting unlikely members of the young clusters. Limited injection-recovery testing on the known planet hosts indicates that for the older Praesepe systems, we are sensitive to additional exoplanets as small as 1-2\,R$_\oplus$, and for the larger Upper Scorpius planet host (K2-33) our pipeline is sensitive to $\sim$4\,R$_\oplus$ transiting planets.  The lack of detected multiple systems in the young clusters is consistent with the expected frequency from the original Kepler sample, within our detection limits. With a robust pipeline that detects all known planets in the young clusters, occurrence rate testing at young ages is now possible.

\end{abstract}

\maketitle

\section{Introduction}

Young exoplanets in all regimes of size and separation ranging from short-period super-Earths and Neptunes (e.g., K2-33\,b; \citealt{zeit3}) to ultra-wide orbit gas-giants (e.g., GSC 6214-210\,b; \citealt{mikegsc6214}, $\beta$-Pic\,b; \citealt{lagrange09}) are key benchmark systems that inform our understanding of the evolution of  planetary systems. Because 
planetary systems undergo the majority of their evolution in the first few hundreds of millions of years after formation (e.g., \citealt{adams06,ida10}), young systems  ($\lesssim$1\,Gyr) are essential to probe the primary drivers of planetary evolution. Comparison of the atmospheres and physical properties (such as radius and mass) of young planets with their older and more numerous counterparts can be used to constrain planet formation and evolution (e.g., HR~8799; \citealt{marois08}, 51~Eri\,b; \citealt{macintosh15}, $\beta$~Pic\,b; \citealt{lagrange09}). For the case of transiting young exoplanets, mapping a change in the radius distribution of planets of differing age can inform our understanding of atmospheric mass-loss. Similarly, the occurrence of exoplanets as a function of age may directly constrain the processes that cause migration of exoplanets from their formation zones to other regions within their systems (e.g., \citealt{lubow_migration,chatterjee08,naoz12}).  

The original Kepler mission \citep{keplermission} has discovered thousands of transiting exoplanets \citep{mullally15}, however the vast majority of these are associated with older ($>$1\,Gyr) host stars, or have poorly constrained ages (e.g., \citealt{walkowicz13,silva_aguirre15}). The repurposed Kepler mission (\emph{K2}) is beginning to bridge this age gap by observing several young open clusters, moving groups, and star-forming regions. These regions include the Upper Scorpius subgroup of Sco-Cen (C2/C15, $\tau$=10\,Myr, d$\sim$145\,pc; \citealt{myfirstpaper,pecaut12}), the Hyades open cluster (C4/C13, $\tau$=800\,Myr, d$\sim$50\,pc; \citealt{leeuwen09,brandt15}),  Praesepe open cluster (C5/C16, $\tau$=800\,Myr, d$\sim$\,180pc; \citealt{leeuwen09,brandt15}) , the Pleiades open cluster (C4, $\tau$=110\,Myr, d$\sim$135\,pc;  \citealt{gaiadr1_clusters,dahm15}), and the Taurus-Auriga star-forming region (C13, $\tau$=1-5\,Myr, d$\sim$145\,pc; \citealt{torres09,taueco1}). 

The utility of young stellar clusters and populations to probe exoplanet evolution has motivated a plethora of searches in these regions using multiple techniques (e.g., \citealt{yeti,palms1,mikegsc6214,krilandlkca15,quinn14,john-krull16}). The relatively well-determined cluster ages, distances and metallicities yield improvements in host star property estimations (\citealt{zeit1}) which are accompanied by a similar improvement of the exoplanet properties. Furthermore any detected planets inherit the bulk cluster properties of their host star, including metallically and age, allowing studies to control for the role of these factors on exoplanet properties and occurrence.

To make full use of the \emph{K2} dataset, we have launched the Zodiacal Exoplanets in Time (ZEIT) survey, with the aim of identifying, characterizing, and exploring the statistical properties in nearby young clusters, star-forming regions, and OB associations. Nine planets transiting members of young populations of known ages have been discovered in the first five \emph{K2} campaigns, all of them as part of the ZEIT survey. Six confirmed and one candidate transiting exoplanets, with periods ranging from 1.5-22\,days have been identified in the Praesepe observations from C5 \citep{zeit4,obermeier16,pepper17}, one Neptune-sized planet orbiting a M4.5 star in the Hyades cluster from C4 (K2-25\,b, R$=$3.43\,R$_\oplus$, P$=$3.43\,day; \citealt{zeit1,david16_hyades}), and the current youngest known transiting exoplanet, K2-33\,b, a member of Upper Scorpius ($\tau$=11\,Myr,P=5.425\,days, R=5.04\,R$_\oplus$; \citep{zeit3,trevor_k233}).  The number and size of these planets indicates a radius distribution  that is significantly larger than exoplanets hosted by similar stars in older populations \citep{zeit4}, suggesting that some close-in planets lose significant portions of their atmosphere over many hundreds of millions of years. Additionally, K2-33\,b orbits near the corotation radius of its host star system in the Upper Scorpius OB association, suggesting that short timescale migration mechanisms may be responsible for at least some close-in planets.

Previous searches for young exoplanets have utilized a mix of methods to produce \emph{K2} lightcurves from pixel data, and search for planets within these lightcurves (e.g., \citealt{Petigura_TERRA,vanderburgk2,zeit1,zeit3,zeit4,trevor_k233,crossfield16,everestk2,aigrain_k2sc}). However, the majority of these pipelines are designed for older, less-active stars and as such their utility can be limited for use with the members of the young clusters observed by \emph{K2}. Indeed, these different pipelines are all sensitive to different subsets of the full sample of known transiting exoplanets in the young clusters, with the exception of the \citet{zeit1} pipeline that required hand tuning of parameters to detect the full sample. 

A number of factors specific to the case of young stars contribute to the need for a more robust and specialized detrending and planet search pipeline. Unlike old field stars, young stellar rotation signals are often $1-2$ orders of magnitudes greater than a typical planetary transit (1-10\% peak-to-peak, \citealt{rebull16_plei}), with rotational periods comparable to the orbital periods of most detectable transiting planets in the \emph{K2} dataset (1-20\,days). High-pass filtering techniques are thus difficult to implement without adversely affecting real transits. Furthermore, some of the existing transit-search pipelines are only sensitive to the largest planet-like signal for any given system \citep{Petigura_TERRA,zeit1}. This is problematic for young, active stars where poorly detrended or complicated rotational variability can produce significant, spurious periodic signals which prevent detection of real transits if only the strongest signal is considered. More sophisticated techniques such as Gaussian Process-based detrending has shown some promise \citep{aigrain_k2sc}, however the failure rate for these methods is higher than that of simpler pipelines, and the stellar variability signals can be difficult to model without also removing transits even with these sophisticated probabilistic models. Data from future planet search missions such as the Transiting Exoplanet Survey Satellite ($\emph{TESS}$), which will observe thousands of nearby young stars, will suffer from the same issues.

Utilizing many reduction and search algorithms makes measurement of overall detection completeness difficult. Injection-recovery or similar completeness correction tests (e.g., \citealt{petigura13,christiansen15,christiansen16,dressing15}) would need to be done over the multitude of \emph{K2} reduction pipelines and transit-search algorithms and then combined afterwards, which is computationally impractical. A more self-contained method, capable of detecting the full sample of known young transiting planets observed by K2, would open the possibility of a uniform measurement of detection efficiency, and hence the calculation of occurrence rates in the clusters. 

Here we present two new methods for detrending significant rotational variability on $\sim$day timescales, one for stars with rotational periods of $>$1\,day, and one for more rapid rotators. We describe the detrending methods in Sections~\ref{sec:notchfilter} and \ref{sec:lcr} and the transit search in Section~\ref{sec:bls}. In Section~\ref{sec:membership} we identify members of the young clusters observed by K2, and in Section~\ref{sec:detections} we demonstrate that our method detects all known young planets from \emph{K2}, as well as a number of other transit-like signals that we conclude are not due to planets. We highlight the power of a single search algorithm with a limited injection-recovery test which we describe in Section~\ref{sec:limits}. We use this information to set strong limits on the size and period of any additional planets in these systems.

\section{Notch Filtering of \emph{K2} Lightcurves}\label{sec:notchfilter}

The lightcurves of young stars feature periodic  variability of typically 1-5\% phased at the stellar rotation period, in addition to significant aperiodic variability. This rotational variability is often temporally complex, with evolution over the course of a $\sim$70\,day \emph{K2} observing campaign. We have developed a detrending model that uses a windowed region of data and a transit-shaped box to remove the rotational signal without adversely affecting transits.

\begin{figure}
\includegraphics[trim={0 0 12mm, 10mm},clip,width=0.48\textwidth]{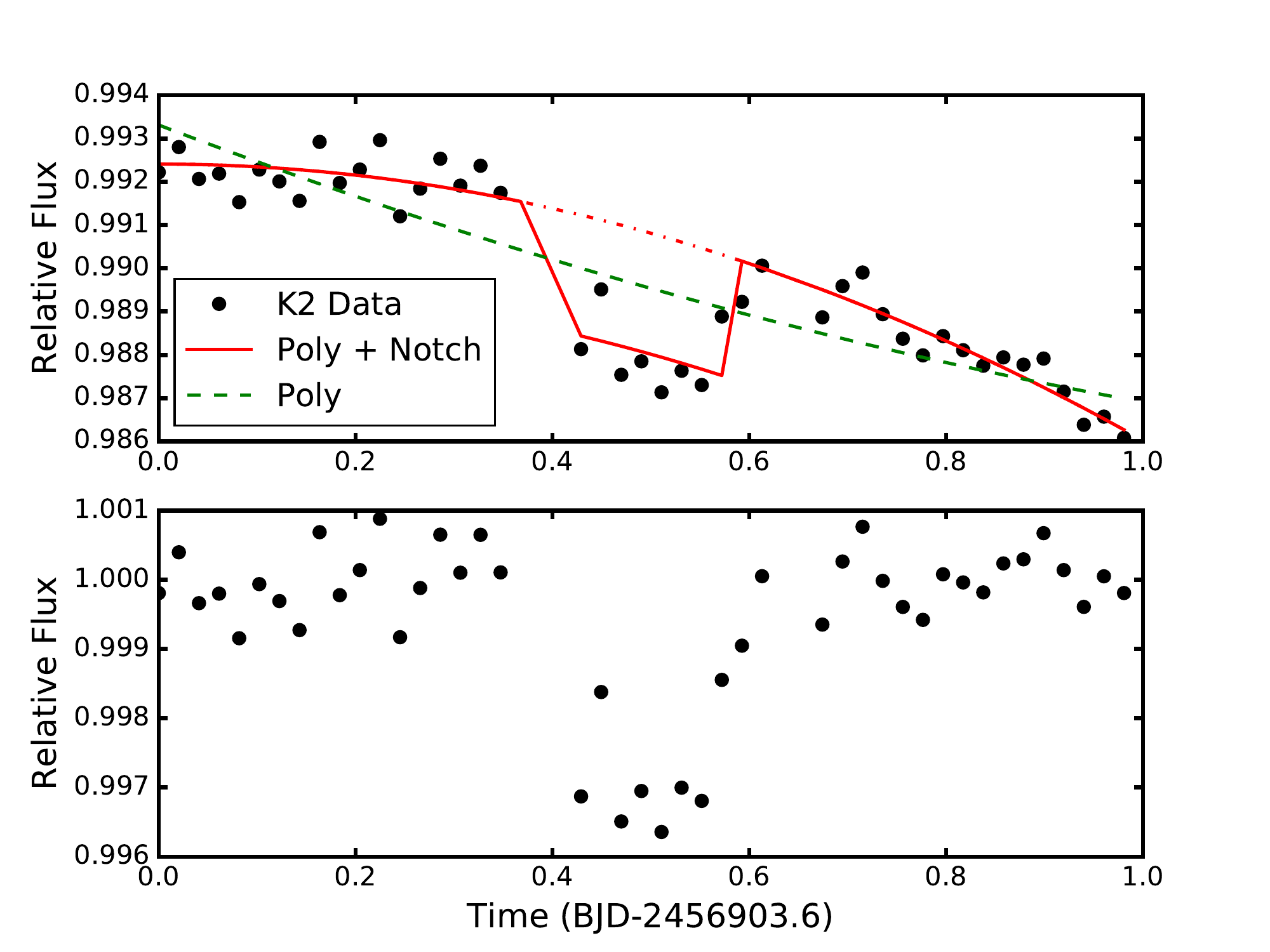}
\caption{Example lightcurve detrended using the notch filter pipeline we have developed, showing a 1-day window centered on a single transit of K2-33\,b ($\tau \sim10$\,Myr planet in Upper Scorpius; \citealt{zeit3}). The upper panel is the raw light curve, corrected for pointing effects using \citet{vanderburgk2}. The full model comprising a polynomial and a transit notch (red solid line) leaves the transit untouched, and completely corrects the rotational variability after division (lower panel). Applying a polynomial without the notch (green dashed line) results in a poor fit and a shallower transit after division, and may completely obscure smaller transits. Running median pipelines (e.g., \citealt{zeit1}) suffer from similar issues. The 1-day window is moved along the entire light curve for the full correction.}
\label{k233b_examp}
\end{figure}

\subsection{Notch Filter Model}

\begin{figure*}
\centering
\includegraphics[width=0.7\textwidth]{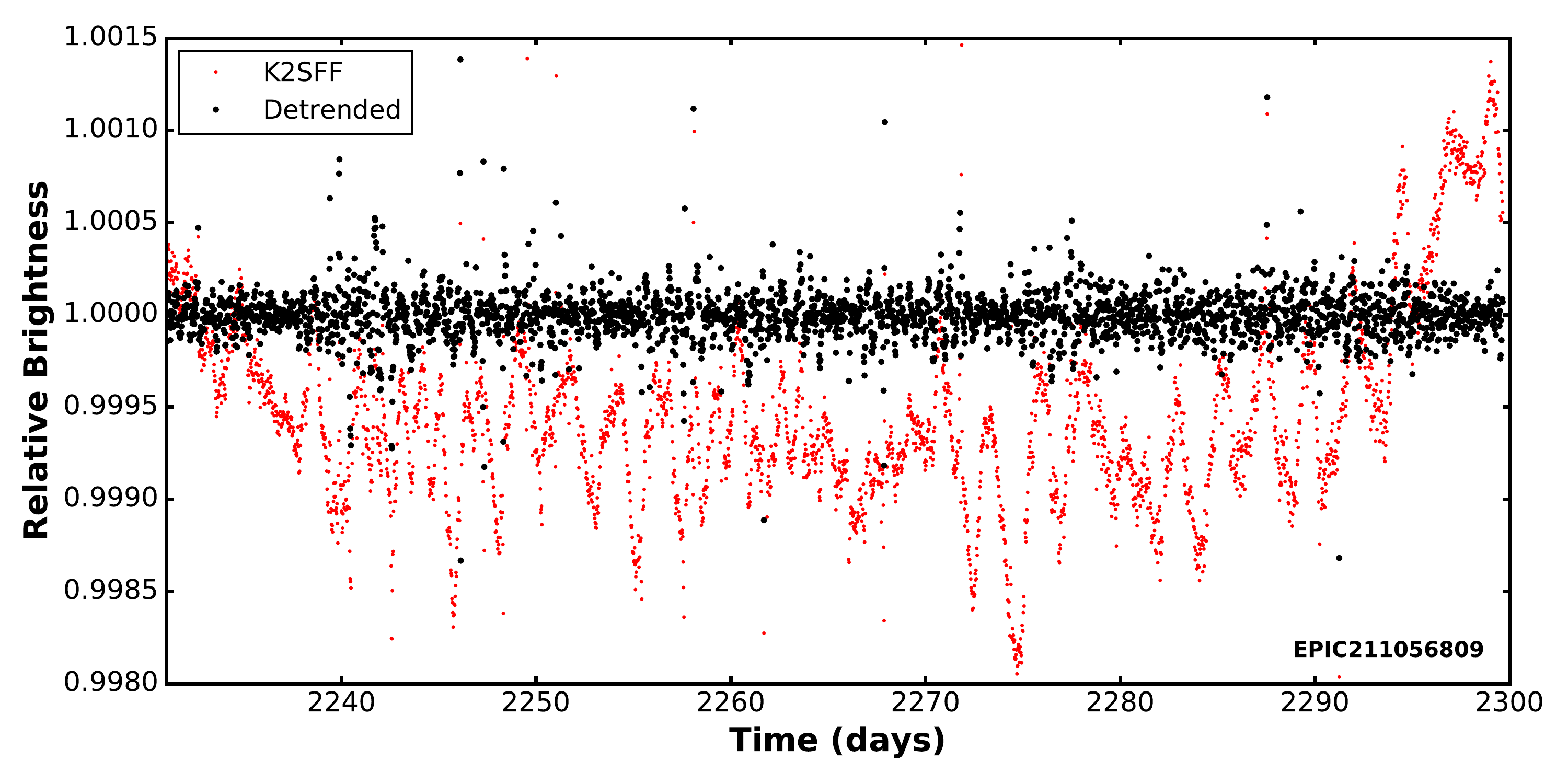}\\
\includegraphics[width=0.7\textwidth]{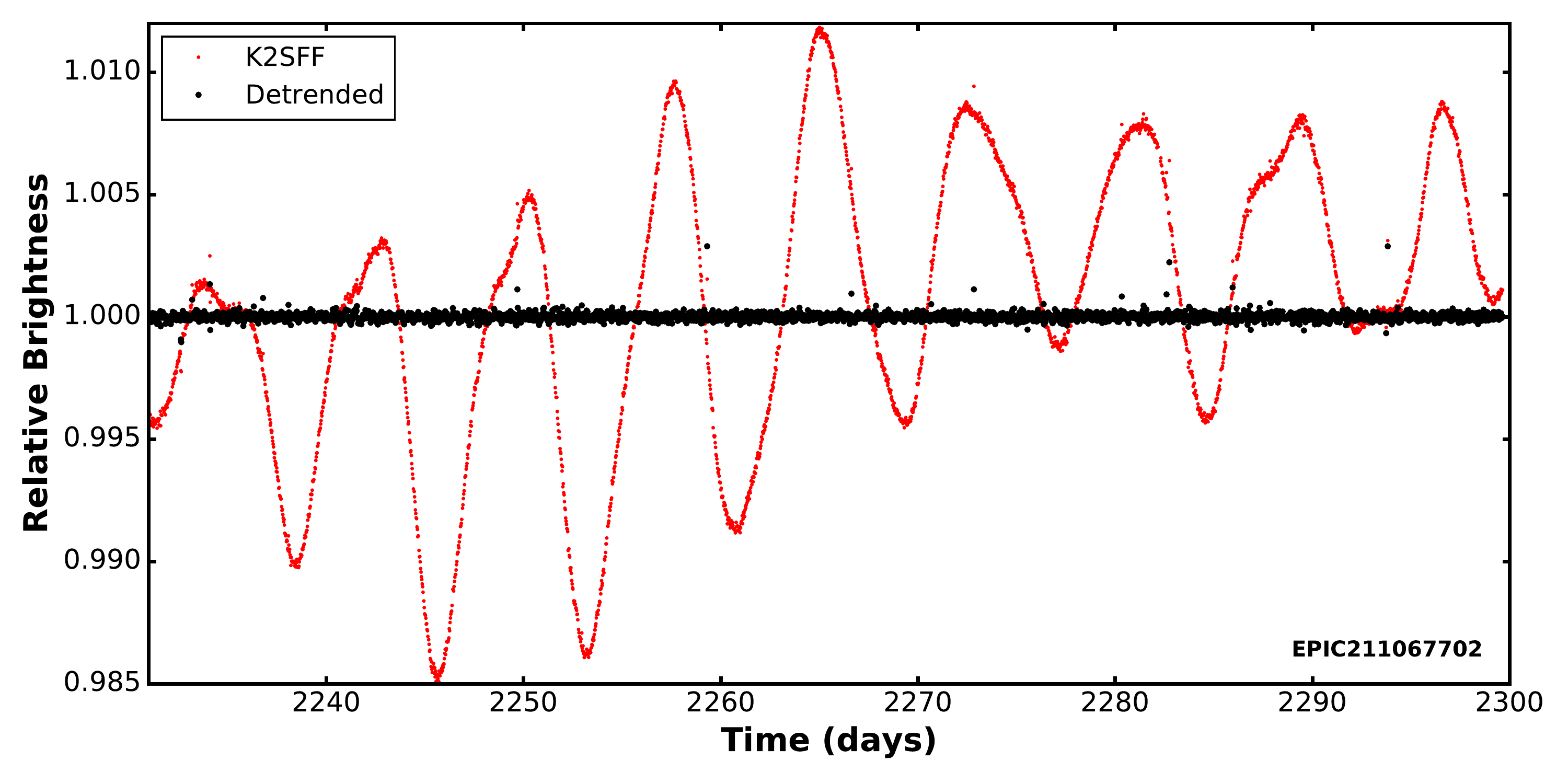}\\
\includegraphics[width=0.7\textwidth]{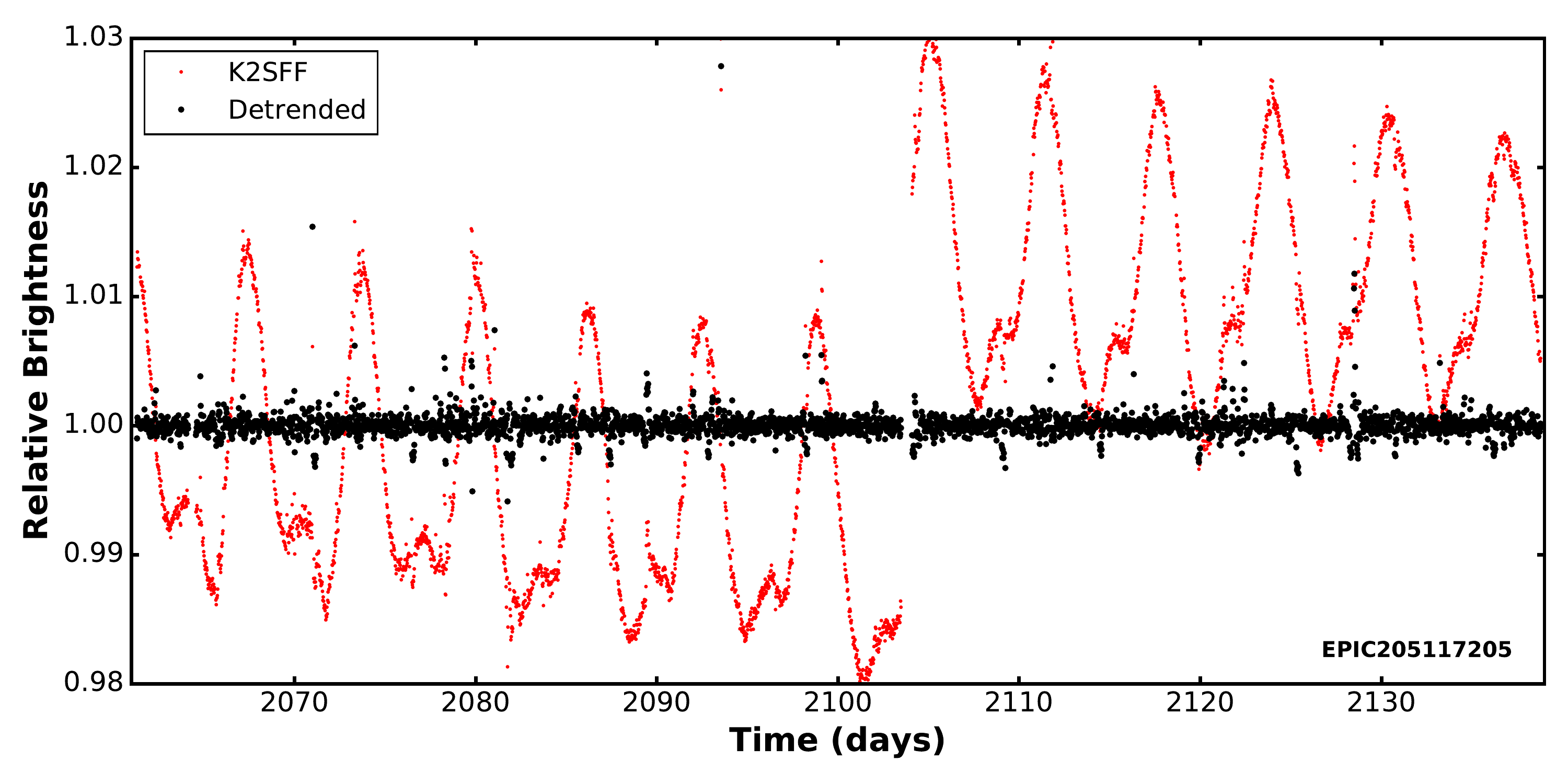}
\caption{K2SFF (red, pointing systematic removed; \citealt{vanderburgk2}), and detrended lightcurves produced with the notch filter method described above (black) for two Pleiades members (upper and middle), and Upper Scorpius member K2-33.  The x-axis in each plot is in Kepler time (BJD-2454833). The notch-filter detrending process removes the rotational variability, while maintaining short timescale dips and increases in flux that could potentially be signatures of transiting planets. EPIC 211056809 (\emph{Upper}) shows significant complex variability, with frequent dips, and amplitude changes on a $\sim$2.4\,day period. The detrending process functions by considering only small stretches of time and thus reducing the overall complexity of the problem. EPIC 211067702 (\emph{middle}) features almost uniform variability on a 7.8\,day period \citep{rebull16_plei}, which is completely removed using the 1-day window. For EPIC 205117205 (\emph{lower}), the planetary transits of K2-33\,b are clearly preserved after detrending of the stellar variability.}
\label{detrend_examp}
\end{figure*}

We start by producing \emph{K2} light curves using the method of \citet{vanderburgk2} (K2SFF), which are already corrected for the \emph{K2} pointing systematic, but not stellar rotation or variability. By default, the K2SFF pipeline models stellar variability using a basis spline with knots spaced every 1.5\,days, but this is too slow to adequately model photometric features on rapidly rotating stars in young clusters. In this work, we instead model stellar variability in K2SFF with a spline with knots every 0.5\,days for the younger two clusters, and 1.0\,days for the older two clusters. See \citealt{vanderburgk2} for more detail. 

After removal of the \emph{K2} pointing systematic using the K2SFF pipeline, the next step was then to remove astrophysical signals such as the stellar rotational variability. To remove the rotational variability, we fit a model consisting of a quadratic polynomial in time with a box-shaped transit notch to a short (0.5-1\,day) window of data centered on the transit notch. This can be thought of as applying a matched-filter to the lightcurve data and is similar to the model used by the \emph{MEarth} project \citep{missmarple} and also \citet{foreman-mackey_search15}. Our model consists of five parameters: Three polynomial coefficients, and both the notch depth and duration. Detrending using only a small portion of the full \emph{K2} lightcurve significantly reduces the complexity required to closely model the rotational variability. 

The first step in the detrending process is to perform an initial outlier rejection on the data in the fitting window by fitting a linear function to any overall slope in flux across the window. We then flag any points more than 5 times the RMS offset from the line in the positive direction as outliers. Due to the large rotational variability, this outlier condition is highly conservative and rarely met, and tends to remove only measurements associated with strong flares that might affect the fitting statistics in later steps.

Following the initial removal of strong positive outliers, we test a grid of four different transit durations (0.75, 1, 2 and 4\,hours) allowing the polynomial parameters and notch depth to vary. We then use the best solution (chosen on the basis of $\chi^2$) from the grid as a starting-point for a fit over all five parameters. We iterate this fit five times, clipping and flagging outliers of more than five times the RMS scatter about the model at each iteration. Typically, convergence is achieved after one or two iterations.  Points removed as outliers in this step are only removed from the model fit, and not the final detrended lightcurve. Because the notch filter is designed to remove in-transit points from the out-of-transit variability fit, the method is robust to falsely identifying in-transit points as outliers.

To test if the notch was needed to adequately fit the data, we also fit a model consisting of just the quadratic polynomial to the data in the fit window and calculate the Bayesian information criterion (BIC; \citealt{schwartz_BIC}) for both this null model and the best fit transit-notch model. We adopt the model with the smallest BIC. This ensures that the notch is only applied when there is statistical evidence that it significantly improves the fit. The quadratic polynomial from the selected model is then used to detrend the observation at the centre of the 1-day window. The process is then repeated for the entire \emph{K2} lightcurve, with points previously determined to be in transits being excluded from fits if they fall in the first third of future fitting windows. All fitting is done with the python implementation of the Levenberg-Marquardt least-squares minimization \emph{mpfit}. Figure \ref{k233b_examp} illustrates the detrending process for a single transit of the K2-33\,b system \citep{zeit3}, and Figure \ref{detrend_examp} shows example notch-filter responses for rotating cluster stars and the full lightcurve of K2-33.

\subsubsection{The Choice of Window Size}

\begin{figure}
\includegraphics[width=0.5\textwidth]{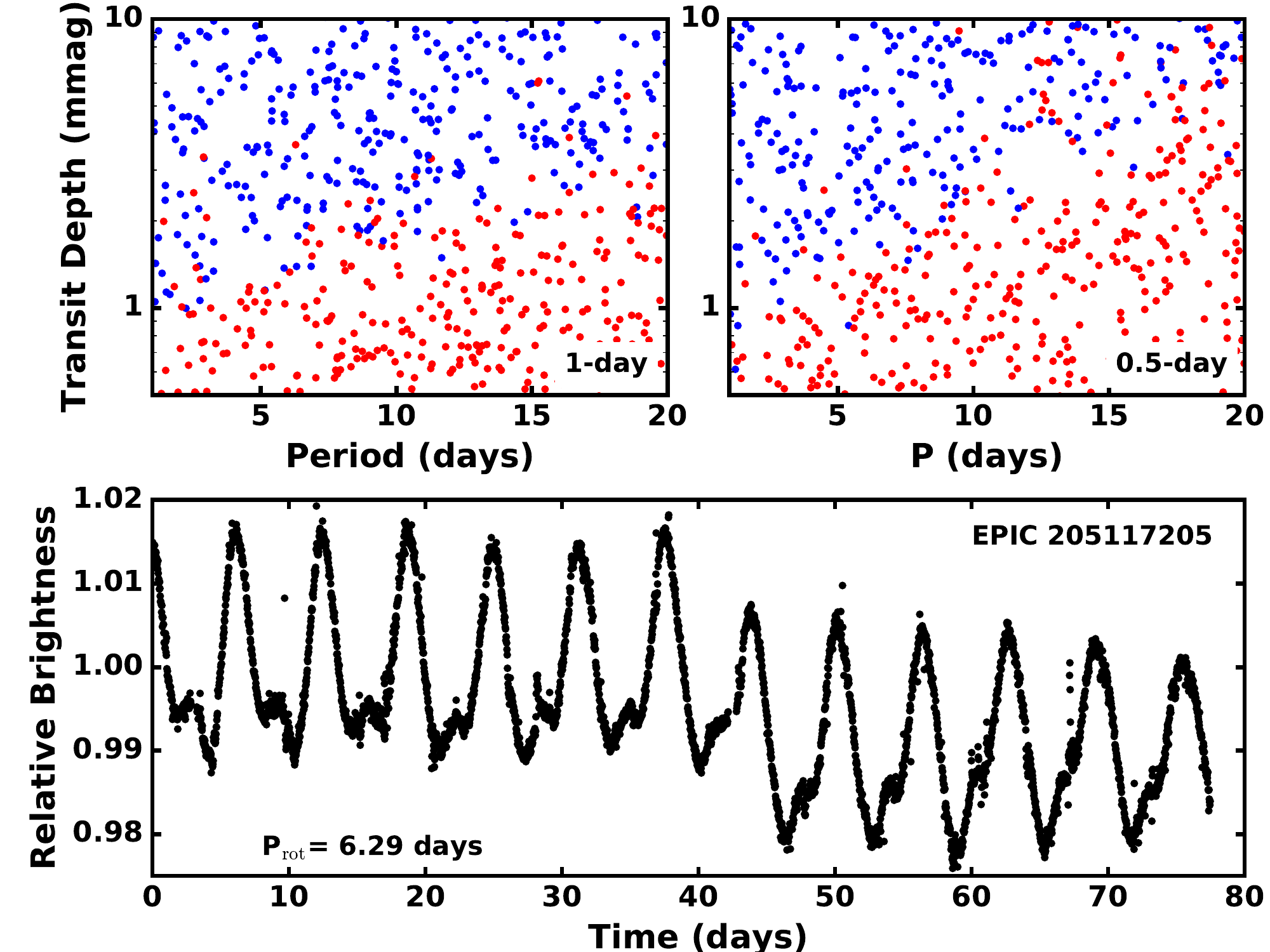}
\includegraphics[width=0.5\textwidth]{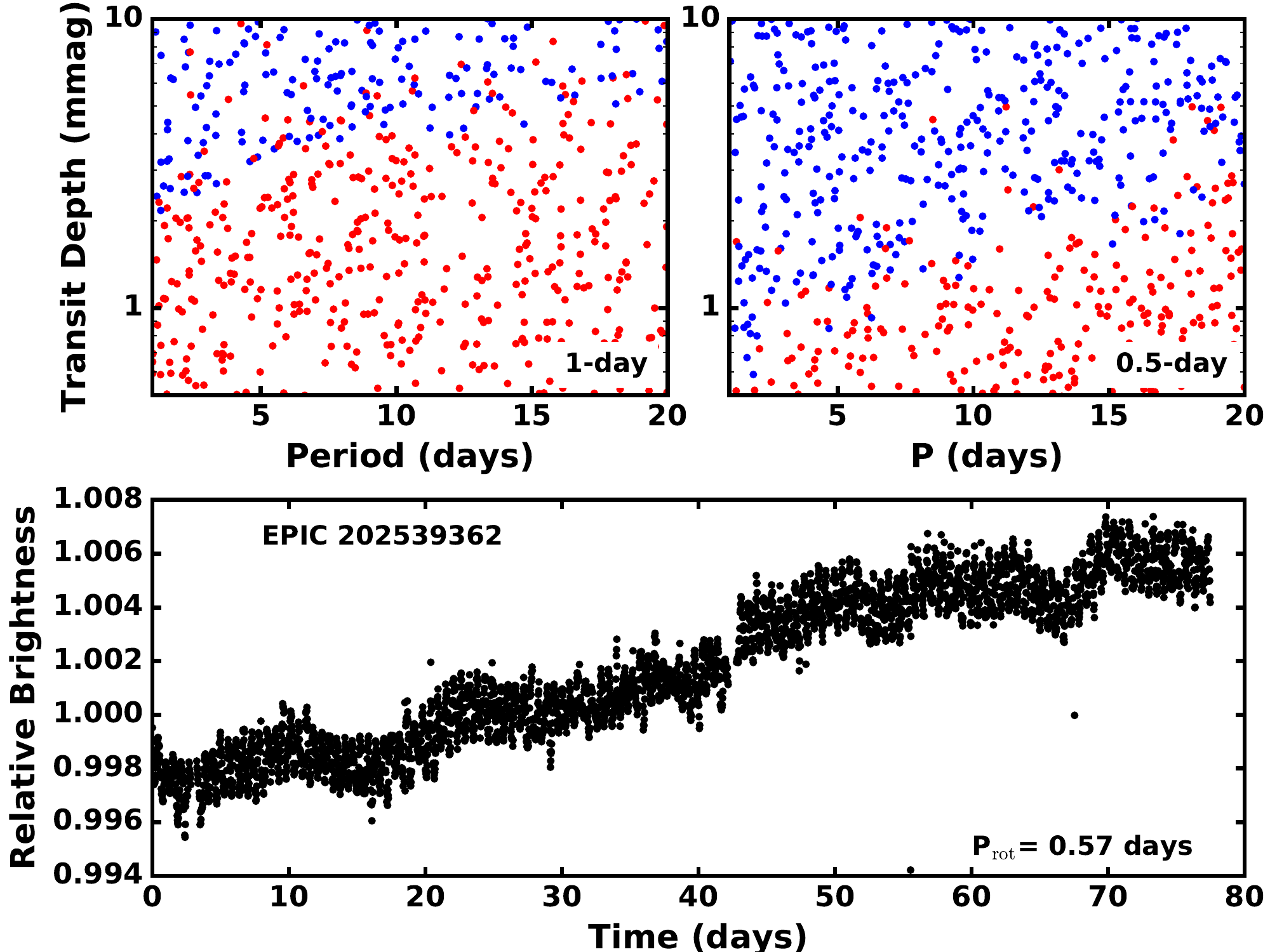}
\caption{Injection recovery tests on two members of Upper Scorpius with different rotation periods, EPIC 205117205 (top, P$_\mathrm{rot}$=6.29\,days)  and EPIC 202539362 (bottom, P$_\mathrm{rot}$=0.57\,days). In each panel, the left and right scatter plots show the recovery of injected transit signals (blue) with 1 and 0.5\,day fitting windows in the notch filter detrending pipeline. For the faster rotator, limiting the window size provides significant improvement of the sensitivity of the pipeline to smaller radius planets. However, for the slower rotator, the smaller window size results in more missed planets at the longer periods, where the larger transit duration relative to the window length starts to become degenerate with out-of-transit rotational variability.}
\label{windowsizeplots}
\end{figure}

The fitting window for the detrending model is chosen to reduce the complexity of the stellar rotation/activity signal to a level that allows it to be well-modeled by a simple polynomial. Typically the window should be less than approximately half the rotation period of the star. The window must also be sufficiently large such that a planetary transit does not span a significant fraction of the data in the window, or degeneracy between the transit notch parameters and the polynomial will be introduced into the model, leading to the removal of real transits or the persistence of stellar variability. This is particularly important for longer orbital periods where the transit duration can become a more significant fraction of the fitting window.

To determine the most appropriate window size, we have performed an injection-recovery test for stars of varying rotation periods in Upper Scorpius, using the methodology described in Sections \ref{sec:notchfilter}. We tested 1000 random orbits for 10 different USco stars with both a 12 and 24\,hour window size. Figure \ref{windowsizeplots} displays the results for two representative objects EPIC 205117205 (K2-33, P$_\mathrm{rot}$=6.29\,days) and EPIC 202539362 (P$_\mathrm{rot}$=0.57\,days). For the faster rotator, decreasing the window size to less than the rotation period increased the overall recovery fraction by a factor of more than 2 is some cases. However, for K2-33, the shorter window reduced the sensitivity, in particular to planets with orbital period longer than $\sim$10\,days. This is because when a shorter window length is employed and the number of measurements informing the fit is reduced, the change in observed flux due to transits with longer durations starts to become degenerate with rotational variability. For the search for transit signals in the K2 cluster observations, we use a 1\,day window for objects with rotation period longer than two days, and a 0.5\,day window for faster rotators. Using a window larger than 1\,day for stars with longer rotation periods does not offer any improvement over the 1\,day window at the orbital periods of interest for the \emph{K2} dataset (1-30\,days), and below 0.5\,days, there are insufficient data for a robust fit to the rotation and possible transit. A sliding window inside of this range would offer only marginal improvement, and so we opt for the simpler 0.5 and 1\,day windows for our search. Rotation periods for the cluster members are sourced from \citet{douglas16,douglas17} for Hyades and Praesepe, \citet{rebull16_plei} for Pleiades, and Covey et al., (private communication) for Upper Scorpius. For the cases where cluster members (Section \ref{sec:membership}) do not have an available rotation period measurement in the literature, we inspect the lightcurves and flag them for processing with the shorter detrending window if they show clear rotational variability with very short periods.

\subsection{Limitations of the Notch-Filter Method}
While our notch-filter detrending method offers improvement over other existing detrending methods for the rotationally active young stars of interest in this study, there are still limitations for the most rapidly rotating stars in the samples. For objects with rotation periods P$_\mathrm{rot}\lesssim1-1.5$\,days, even the use of the 0.5\,day detrending window does not completely remove the rotational systematics. Further reducing the window size to better sample the rotation during the detrending process leads to significant degeneracy between transits and rotational variability, resulting in no significant improvement to detectability of transits. Our search is thus only sensitive to transits with depths equal to or greater than the amplitude of the rotational variability for the most rapidly rotating stars. These objects tend to be the late M-dwarfs in the young clusters, and as such, we suggest that for future missions (e.g., \emph{TESS}) such young stars are observed with a faster cadence. The additional information contained in resolved ingress and egress points for a transit will facilitate the use of a smaller fitting window and a resolved transit model in place of the notch-box for identifying transits over the null rotation model for lower-mass rapid rotators.

\section{Locally Optimized Combinations of Rotations}\label{sec:lcr}
\begin{figure}
\includegraphics[width=0.49\textwidth]{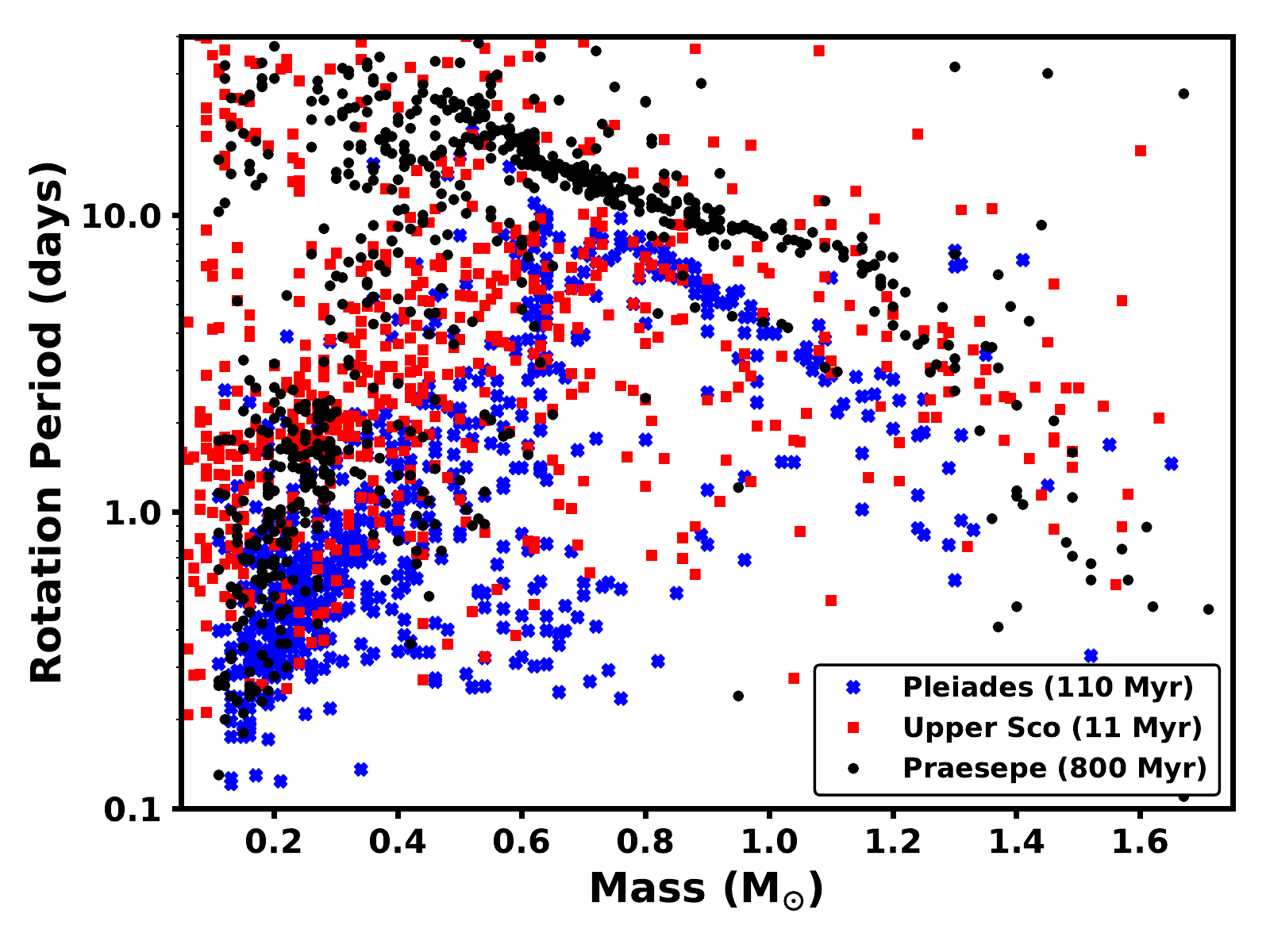}
\caption{Rotation periods measured from \emph{K2} lightcurves for Upper Scorpius (10\,myr), Pleiades (110\,Myr, \citealt{rebull16_plei}) and Praesepe (800\,Myr, \citealt{douglas17}) as a function of model stellar mass from the isochrones of \citet{feiden16_usco}. A significant fraction of the members of these young populations (20-40\%) have rotation periods shorter than 1\,day.}
\label{cluster_mass_rot}
\end{figure}

The notch sliding window detrending method has proven highly effective at removing complex and varying rotational signatures for young stars with rotation period longer than $1-1.5$\,days. Due to an insufficient number of observation points to support robust fitting to windows smaller the $\sim$0.5\,days in the long-cadence \emph{K2} data, it is unable to remove rotational signals at shorter periods. Given the slower spin-down of lower mass stars in the clusters of interest (e.g., \citealt{rebull16_plei,covey16,douglas17}), rotational variability for lower-mass M-dwarfs in the younger Pleiades and Upper Scorpius groups is often poorly corrected and hence sensitivity to smaller planets transiting these stars is limited. It is not uncommon for M$=$0.3\,M$_\odot$ dwarfs in the Pleiades to have rotation periods of P$_\mathrm{r}=0.2- 0.7$\,days \citep{rebull16_plei}.  Figure \ref{cluster_mass_rot} shows the rotation periods for the cluster members observed by \emph{K2}. The lower mass members of these young populations have not yet had time to lose angular momentum, and hence the number of stars with very rapid rotation period (P$_\mathrm{rot}$$\lesssim1$\,day) varies as a function of both stellar mass and cluster age. For the four clusters of interest to this work, Pleiades hosts the largest fraction of such rapid rotators. To recover precision and sensitivity to transits at the expected level for \emph{K2}, we introduce a secondary detrending method designed specifically for the most rapidly rotating cases. This method is based on the Locally Optimized Combination of Images (LOCI) algorithm of \citet{lafreniere07_loci}. We hereafter call this method Locally Optimized Combination of Rotations (LOCoR).

\begin{figure}
\includegraphics[width=0.47\textwidth]{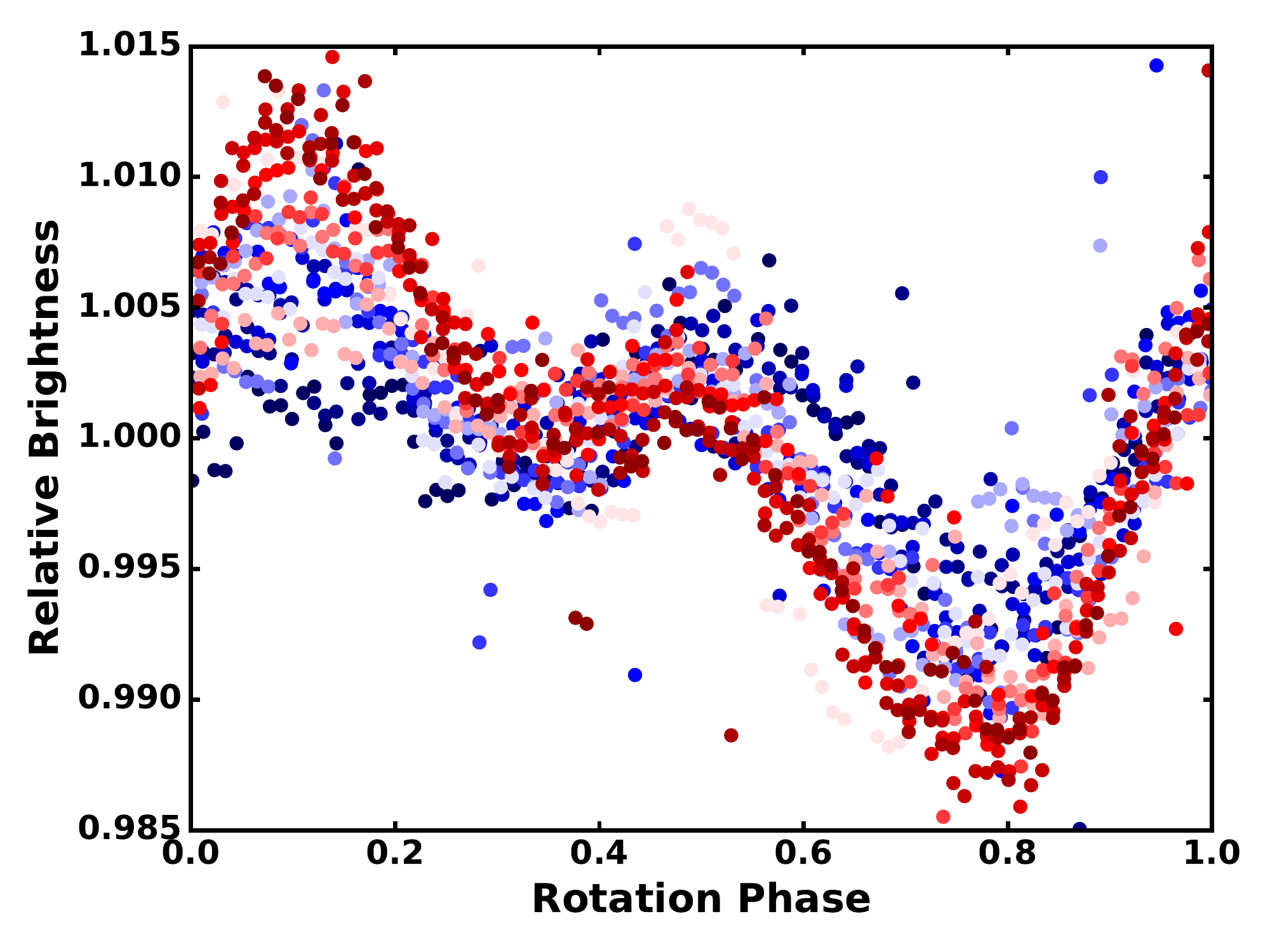}
\caption{Every second rotation period for K2-25 (EPIC 210490365), a $\sim$0.3\,M$_\odot$ Hyades member with rotation period P$_\mathrm{R}=1.88$\,days hosting a $\sim$3.5\,M$_\oplus$ transiting planet in a 3.48\,day orbit \citep{zeit1}.  Each rotational period is color coded by time and median corrected. Evolution of the rotational profile as well as significant flaring can be clearly seen. Division by an average rotational profile thus does not completely address this complexity.}
\label{hyades_spotevol}
\end{figure}

It is intuitively the case that past or future rotation periods of the same object are useful models for the rotational variability in any given period. This is partially true, as spot evolution, flaring activity, and potential transits can make division by an average rotation profile problematic. Figure \ref{hyades_spotevol} displays example rotation periods for EPIC 210490365 (K2-25), a $\sim$0.3\,M$_\odot$ Hyades member with rotation period P$_\mathrm{R}=1.88$\,days \citep{zeit1}. There is clear evolution of the rotational profile over the course of the \emph{K2} campaign.

An improvement on a simple mean-rotation division, which is non-ideal in the presence of spot evolution, is to use all available rotations of a star as references to model all other rotations. For a given cluster member observed by \emph{K2} with known rotation period, we model each individual rotation as a linear combination of other rotations of the same star in the dataset. First we label each rotation in the \emph{K2} dataset of the star of interest. We then apply a median to each rotation individually to remove any long-term trend in the data. Following this bulk correction we take the entire median-corrected lightcurve and bin it into 20 phase bins, and identify and mask $3-\sigma$ outliers using the robust mean statistic on each bin. We then alias the rotation period by an integer multiple to the smallest possible alias greater than 2\,days. This step is significant because for stars with shorter rotation periods e.g., ($<$0.5\,days), there are very few measurements per rotation  at the \emph{K2} 30\,minute cadence, making a robust fit difficult. Aliasing to 2\,days results in approximately $\sim$90 measurements in each aliased ``pseudo-rotation'' to be fit (excepting outliers) ensuring a robust fit.

\begin{figure}
\vspace{1mm}
\includegraphics[trim={0mm 0mm 15mm 12mm},clip,width=0.47\textwidth]{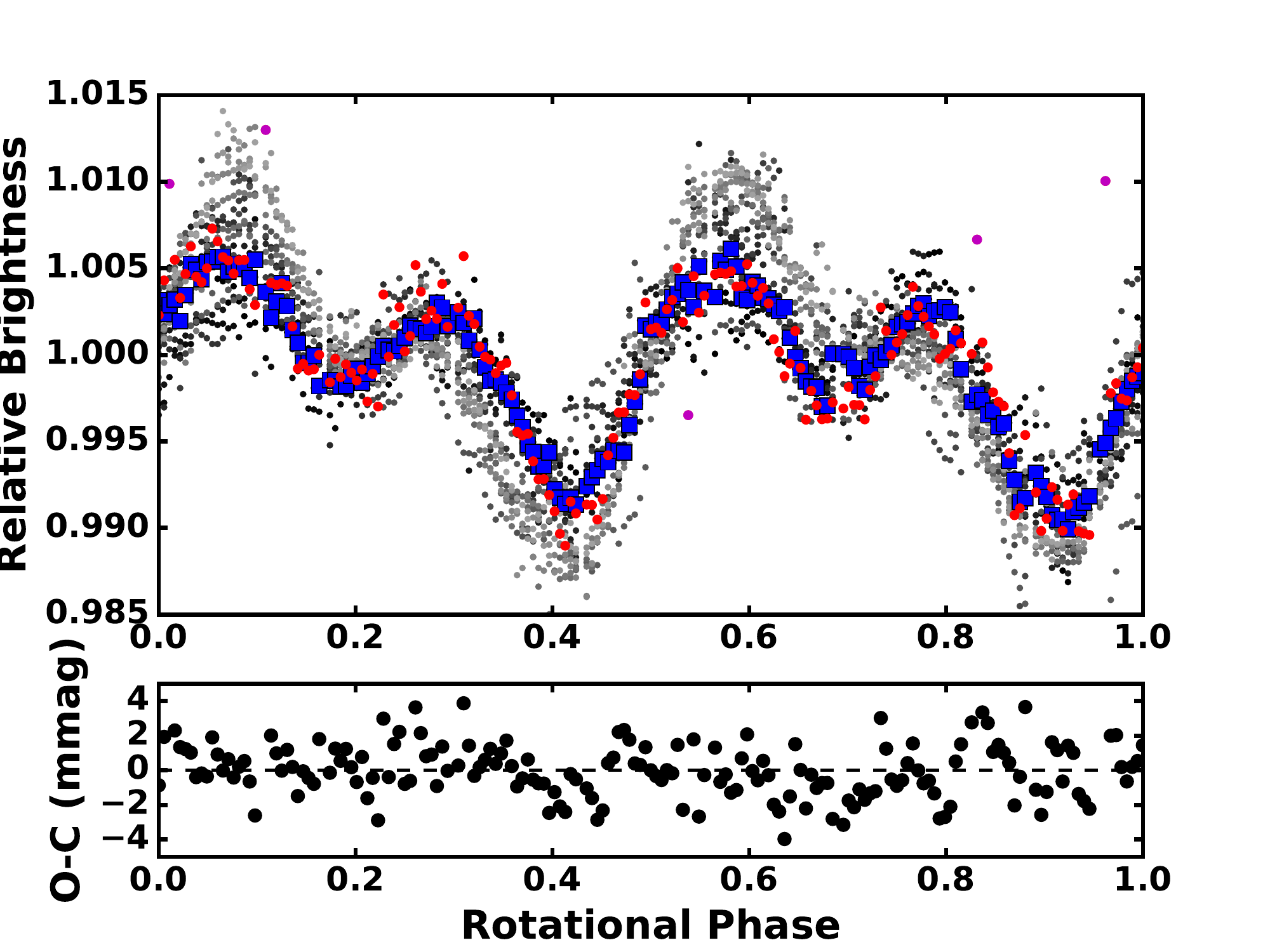}
\caption{Example linear combination fit for two rotation periods of K2-25 (EPIC 210490365, P$_\mathrm{Rot} = 1.88$\,days). Data being fit (red circles), outliers (magenta circles), the model (blue squares) and the reference pseudo-periods interpolated onto the phase scale of the fitted data (grey circles) are shown.}
\label{hyades_lcr_example}
\end{figure}

Following this initial cleaning, for each pseudo-rotation $P_{T}$, we interpolate all other pseudo-rotations $\{P_{j}\}$ onto the phase sampling of $P_{T}$ and model the light curve as a linear combination of $\{P_{j}\}$;

\begin{equation}
M_T = \sum_{j \in J}a_j P_j,
\label{modeldef}
\end{equation}

where $M_T$ is the model for the target rotation $P_T$, and $a_j$ are the coefficients of the linear combination of all other periods $P_j$. We compute these coefficient by minimizing the $\chi^2$ statistic over all data points in the target pseudo-rotation;

\begin{equation}
\chi^2 = \sum_{i}m_i(P_{T,i} - \sum_j a_j P_{j,i})^2,
\label{chi2eq}
\end{equation}

where $m_i$ is a binary mask that excludes outliers from influencing the fit. Since the model used here is a linear combination it can be expressed as a set of linear equations which can be solved as an inverse-matrix problem by taking the partial derivative of Equation \ref{chi2eq} with respect to the coefficients $a_j$;
\begin{equation}
\sum_j a_j(\sum_i m_i P_{j,i} P_{k,i}) = \sum_i m_i P_{T,i} P_{k,i}, \forall k \in J,
\label{matrixeq}
\end{equation}

which takes the form $\mathbf{Ax}=\mathbf{b}$. Multiplying by the inverse of the bracketed portion of the left hand side of Equation \ref{matrixeq} gives an analytical expression for the best-fit coefficients. We repeat this process up to three times per target pseudo-rotation, masking in each iteration points three times the median-absolute deviation offset from the model fit. For those outlier points that were masked, we interpolate the final model to the relevant phases before applying the model for detrending. This process is then applied for all other aliased pseudo-rotations for the entirety of the $\emph{K2}$ lightcurve. Figure \ref{hyades_lcr_example} gives an example of the linear combination fit for one pseudo-rotation (1.88\,days aliased to 3.76\,days) of  K2-25 (EPIC 210490365), which hosts a $\sim$3.5\,R$_\oplus$ planet with orbital period of 3.48\,days \citep{zeit1}. As expected, this method provides similar detrending ability as subtracting an average rotation period (after median correction) as in Figure \ref{hyades_spotevol} when the spot evolution amplitude is below the photometric precision of the lightcurve data, but provides significant improvement when spot evolution is resolved.

\begin{figure}
\includegraphics[width=0.23\textwidth]{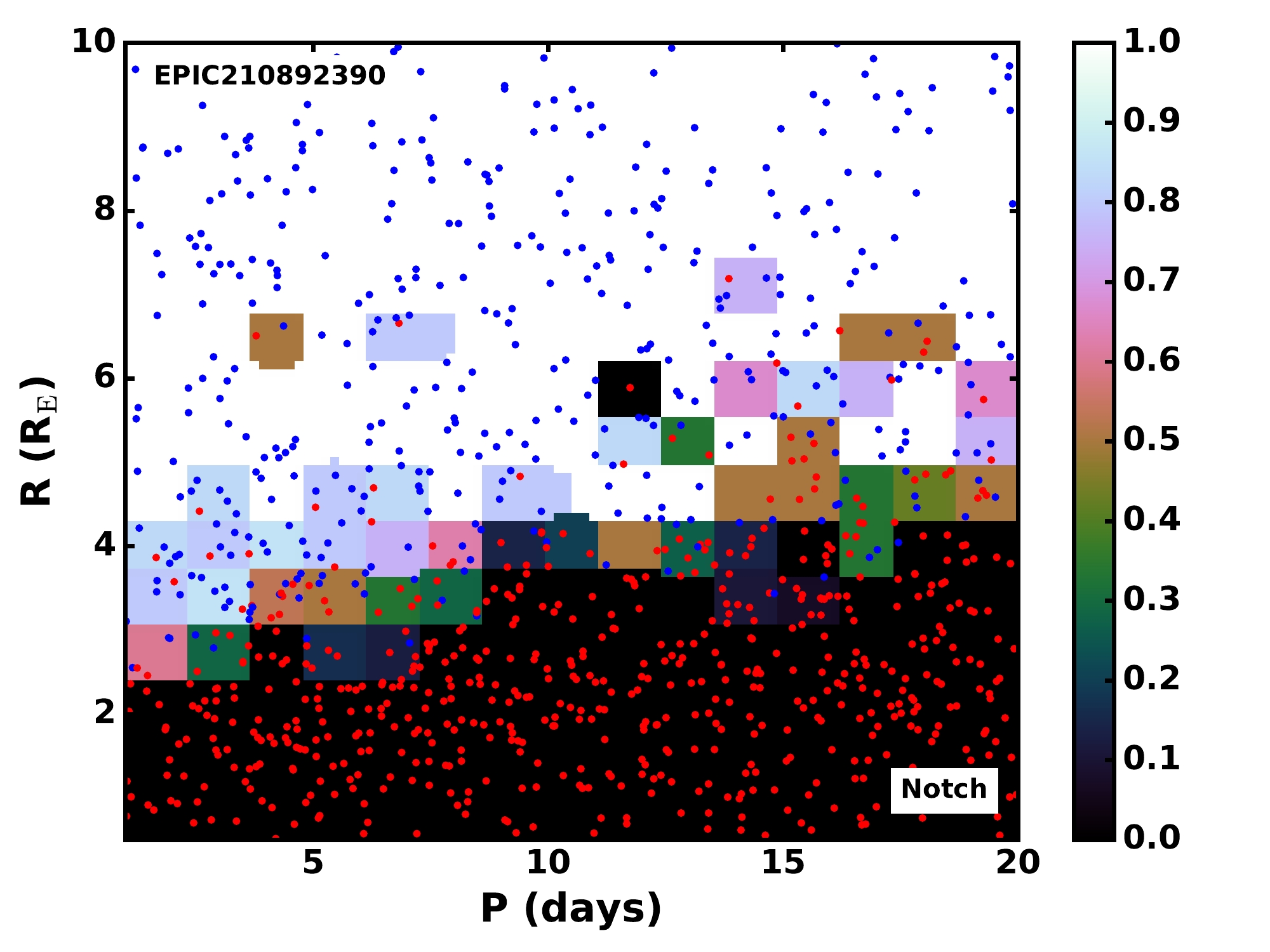}
\includegraphics[width=0.23\textwidth]{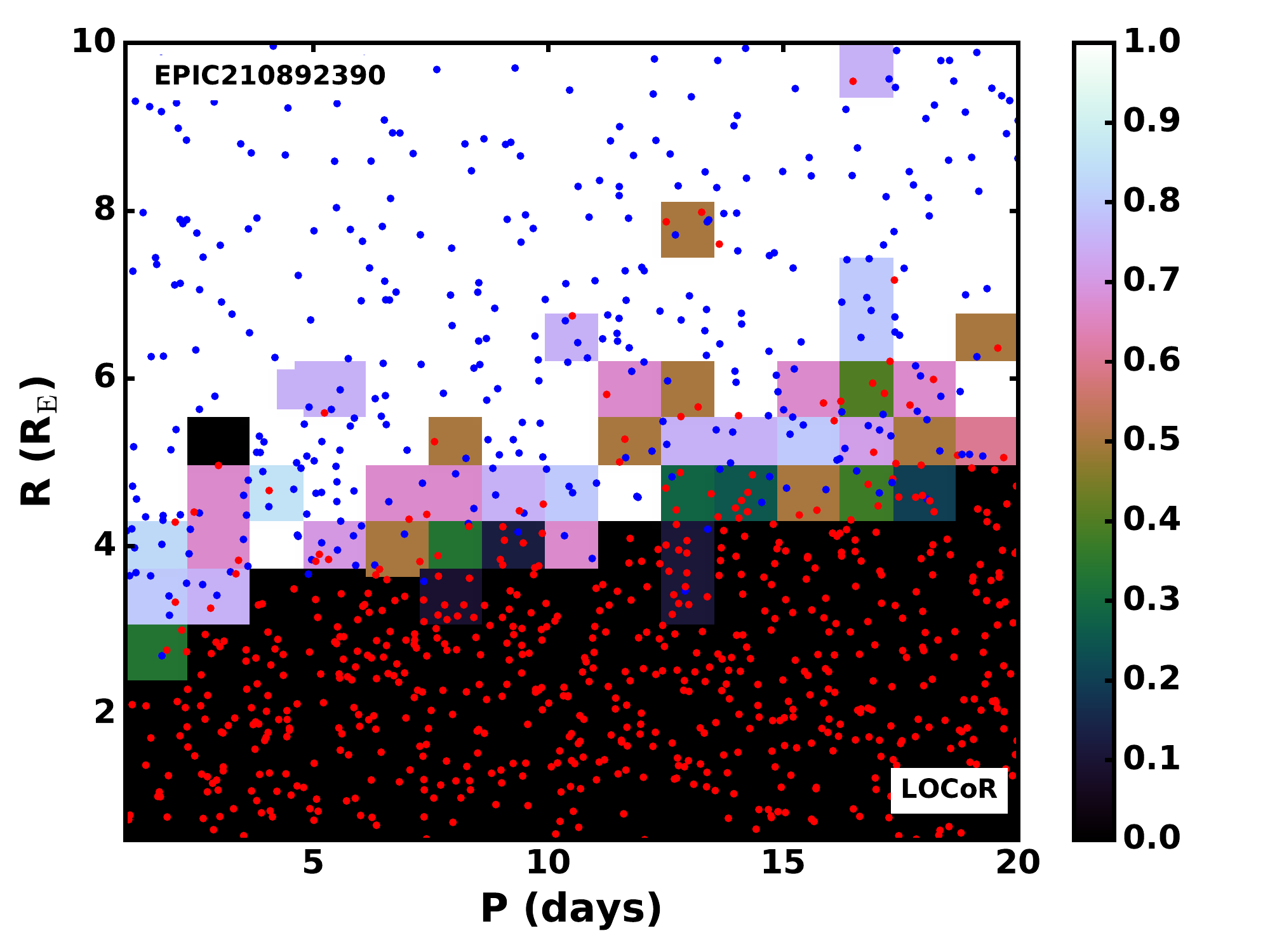}\\
\includegraphics[width=0.47\textwidth]{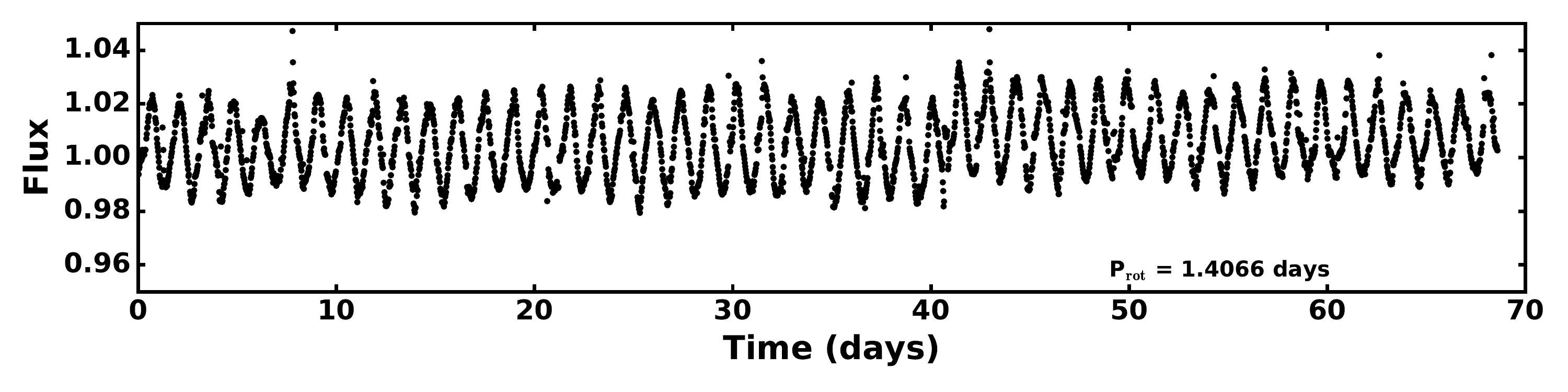}\\
\includegraphics[width=0.23\textwidth]{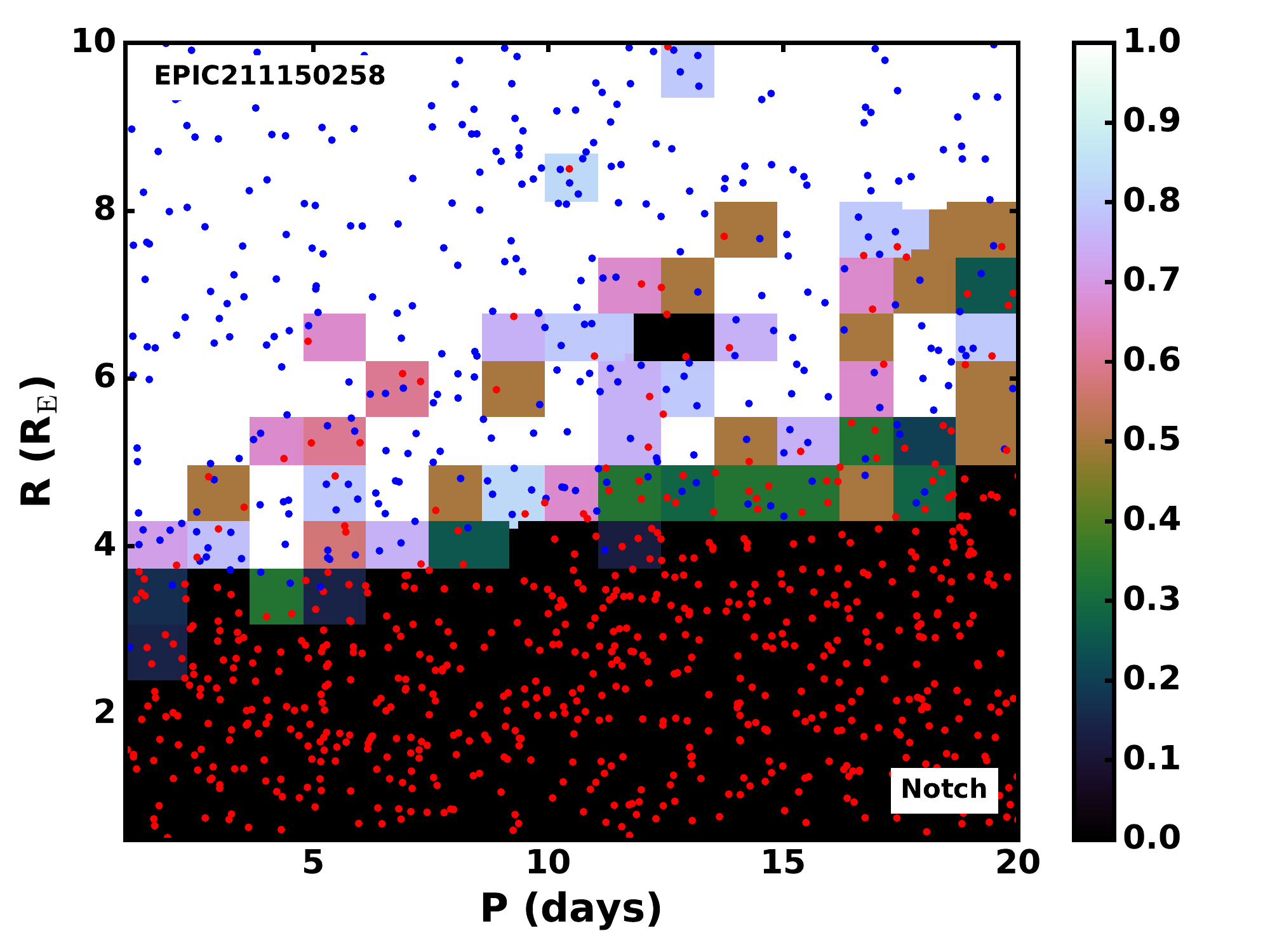}
\includegraphics[width=0.23\textwidth]{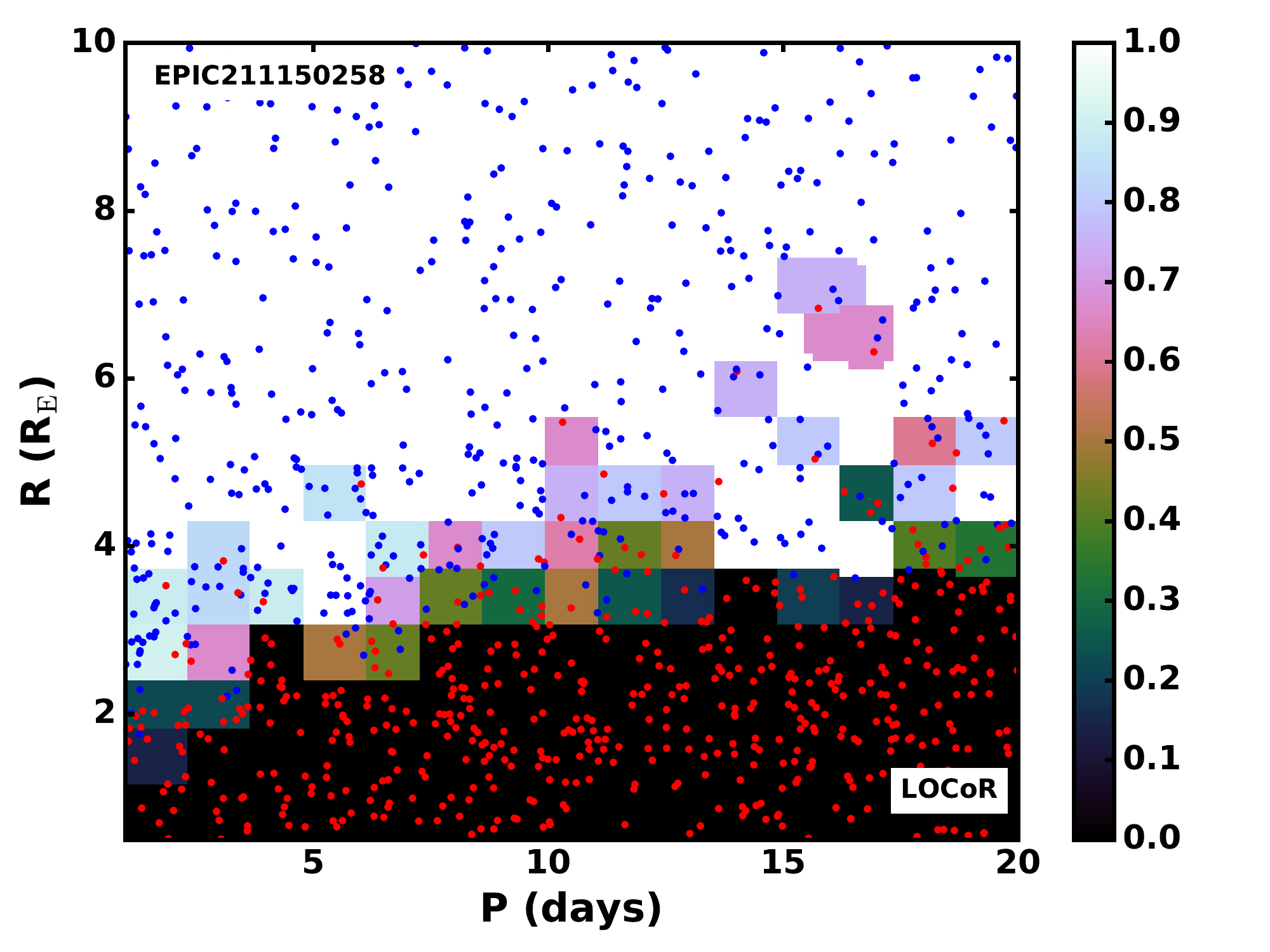}\\
\includegraphics[width=0.47\textwidth]{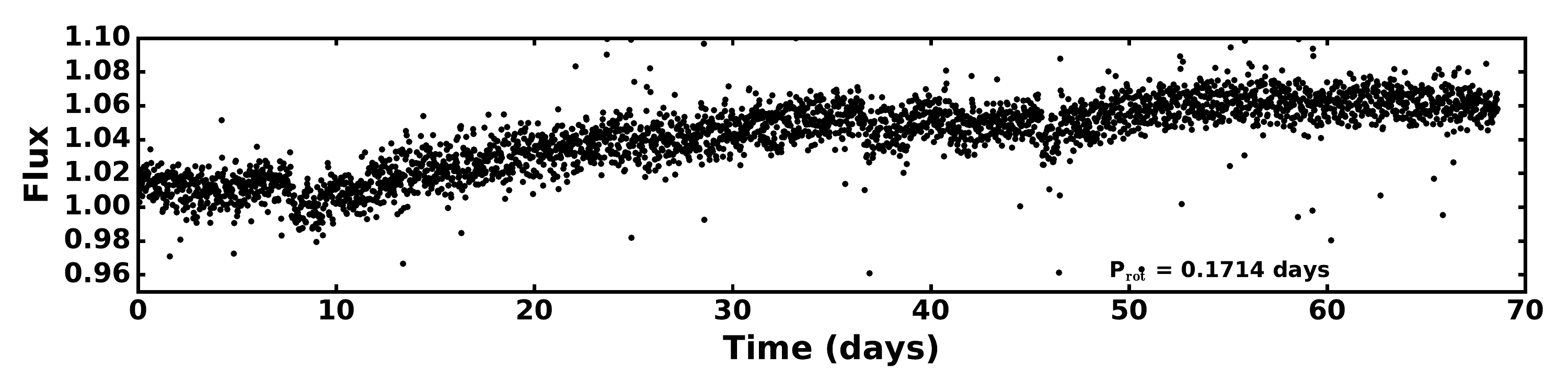}\\
\caption{Performance between the LOCoR and notch-filter pipelines for two rotators EPIC~211150258 (P$\mathrm{rot}$=0.1714\,days) and EPIC~210892390 (P$\mathrm{rot}$=1.4\,days) in the Pleiades cluster. For the slower rotator EPIC~210892390, the two detrending methods provide similar sensitivity to transiting planets, however for the rapid rotator EPIC~211150258, LOCoR provides a significant improvement in the smallest detectable planets.}
\label{locor_comparison}
\end{figure}

\subsection{Limitations of LOCoR}
LOCoR has two major limitations that directly relate to the rotational characteristics of a target of interest. The method performs poorly for objects where each rotation is a poor model for past or future rotations. This comprises objects where the spot evolution timescale as seen in the lightcurve data (i.e., the change in the rotational variability shape) is significantly faster than the rotation period. From inspection, this begins to occur for objects with rotation periods of $\gtrsim$3-5\,days in the \emph{K2}  data, and the notch-filter begins to match the sensitivity of LOCoR at rotation periods of $\gtrsim$1.5\,days. As such, we only apply this method to objects with rotation periods of $<$3\,days. Figure \ref{locor_comparison} shows the relative performance of the LOCoR and the notch-filter for two rotating Pleiades members.

Additionally, for transiting signals whose period is closely matched to the rotation period of the star, nearby alias, or the pseudo-rotation period using in LOCoR, any transit-induced features may be indistinguishable from the rotational variability in the application of the linear combination model unless they appear to be clear outlier points. This is similar to self-subtraction in the LOCI algorithm \citep{mawet12}. This is not as problematic as it may appear; while there are a small number of exoplanets whose host star rotates at an alias of the orbital period of the exoplanet, (e.g., HAT-P-11, 6:1; \citealt{beky14}), even if all short-period exoplanets migrate inward via the disk to the corotation radius \citep{lubow_migration}, the host star will continue to spin up as it approaches the zero-age main sequence ($\sim$100\,Myr for K-type stars), and then spin down as it ages. As such we do not expect to ever see a planet at exactly the corotation radius at the time of observation for the case of disk-migration. A clear example of this is K2-33\,b, a 10\,Myr old transiting exoplanet orbiting with period of $\sim$5.4\,days whose host star rotates with period 6.29\,days. Even a small difference between the orbital and rotation periods for a transiting exoplanet will allow detection via the LOCoR model. Limited injection-testing for the preferred pseudo-rotation period of 2\,days indicates that recovery of transits is possible for an overall shift in transit phase of $\gtrsim$3\,transit durations over the course of the campaign. This corresponds to a transit-to-transit shift of $\sim$20\% of the transit duration, or a loss of approximately 0.1-0.2\% of planets  with periods of 1-20\,days assuming a uniform period distribution.


\begin{deluxetable*}{lcccccccccccc}
\tabletypesize{\footnotesize}
\tablewidth{0pt}
\tablecaption{Cluster and Association Properties.}
\tablehead{
\colhead{} &\colhead{Age}  & \colhead{R.A.} & \colhead{Decl.}        & \colhead{D}& \colhead{$X$}& \colhead{$Y$}& \colhead{$Z$} & \colhead{$U$} & \colhead{$V$} & \colhead{$W$} & \colhead{$\sigma_\mathrm{int}$} & \colhead{Ref}\\
\colhead{} &\colhead{(Myr)} & \colhead{(J2000} & \colhead{(J2000)} & \colhead{(pc)} &  \colhead{(pc)} & \colhead{(pc)} & \colhead{(pc)} &\colhead{(km/s)} & \colhead{(km/s)} & \colhead{(km/s)}& \colhead{(km/s)} & \colhead{}\\
}
\startdata
Hyades     &800 & 04 26 54 & +15 52 &  46  & 43$\pm$6   & 2$\pm$7       & -17$\pm$8    & 41.1 & -19.2 & -1.4  & 2.0  & (1,a)    \\
Praesepe  &800 &08 40 24 & +19 40 &182 &138$\pm$11 &-67$\pm$13 & 96$\pm$12    & 41.5 & -19.8 & -9.7  & 1.0  & (1,a)  \\
Pleiades   &110 & 03 47 00 & +24 07 & 135 & 119$\pm$7 & 27$\pm$18  & -56$\pm$16   & 6.8 & -28.7 & -14.2  & 1.0  & (1,2,3,b)   \\
Upper Sco &11  &  16 12  & -23 24 &145 &                    &                     &                       & 5.8 & -17.6 & -6.6   & 3.0     & (4,5,c)  \\
\enddata
\tablecomments{Adopted kinematic properties for the four young clusters observed in K2 Campaigns 2-5. Space velocities are taken from (1) \citet{leeuwen09}, (2) this paper, (3) \citet{gaiadr1_clusters}, (4) \citet{chen11}, (5) \citet{myfirstpaper}. Cluster ages are taken from (a) \citet{dahm15}, (b) \citet{brandt15}, (c) \citet{pecaut12}. }
\label{clusteruvw}
\end{deluxetable*}

\section{Identifying Transits Following Detrending}\label{sec:bls}
Following detrending with either of the methods described above, we identify periodic transit signals using the Box-Least Squares (BLS) algorithm of  \citet{kovacs02} in an iterative process. In each iteration, we run the BLS search at 48335 periods spanning 1 to 32\,days spaced equally in frequency, and use 300 bins for the phase folding. We allow durations of 0.005-0.2 times the phased orbital period in the search. 

We take the output power spectrum from the BLS search, and subtract the low-frequency envelope profile by first binning the spectrum into 30 bins. This low-frequency envelope is reflective of the observational cadence, and adds spurious power at longer orbital periods. We then calculate the signal-to-noise ratio (SNR) of each point in the BLS power spectrum by treating the entire power spectrum as a half-normal distribution with standard deviation proportional to the median absolute deviation of the power spectrum from zero.

We then identify the strongest peak in the power spectrum with SNR$>$7 and record the period, phase, duration and transit time for later inspection. The in-transit points are then masked, and the transit search is repeated up to 10 times or until all peaks are below the 7-$\sigma$ threshold.  All candidate transiting planets are then inspected visually to ensure they are planet-like (i.e., flat-bottomed with consistent depth).

This entire process is repeated twice, once with all points included, and once with points of extreme $\emph{K2}$ pointing offsets removed. The K2SFF correction fit is applied in sections of $\sim$200 datapoints (marked in the K2SFF pointing correction \citealt{vanderburgk2}). We select only those points most well-fit by the pointing correction model by  removing the most extreme 8\% of points in terms of centroid offsets in each section and repeating the BLS search. This fraction of extreme points to remove was determined on the basis of inspection of the K2SFF correction. Sections of very poor pointing-systematic correction typically contain $\sim$10 significant outlier points, and so an 8\% cut ensures all poorly corrected points are removed. Repeating the search twice ensures that we identify transits at extreme pointing offsets, and smaller planets whose transit signals may be overwhelmed if observations with poorly corrected pointing are included. For the LOCoR detrended lightcurves, we search three times, with the additional step of removing all points flagged as significant outliers in the detrending process. Following the detection of a candidate periodic transit signal with the BLS algorithm, if the input lightcurve was detrended with the notch-filter method, we reapply the detrending, masking the in-transit points and forcing the null-model (no transit) to produce a detrended light-curve for visual inspection (as opposed to a filter-response, which is the initial output of the notch-filter pipeline).

\section{Young Cluster Membership in the K2 Fields}
\label{sec:membership}

There has been significant work in the literature on the membership of the four young clusters observed by \emph{K2}. For the three open clusters, Hyades, Praesepe and Pleiades, the stellar memberships based on proper motions and photometry are expected to be complete down to mid-M type stars \citep{roeser11,kraushillenbrand_comaber,dance1}, though some interlopers are expected to be included in the samples. While B, A and F-type stars in Upper Scorpius are also identified kinematically, its 10\,Myr age also allows for a more direct  assessments of youth and membership; the presence of spectroscopic youth indicators in the spectra of G-M type members (e.g., Li 6707\AA, \citealt{preibisch01}), the presence of circumstellar disks around some members (e.g., \citealt{luhman2012_disk}) and X-ray activity (e.g., \citealt{preibisch98}). There are currently $\sim$1500 Upper Scorpius members confirmed via these methods including $\sim$800 which were observed by \emph{K2}. However, the current membership is highly incomplete for stars of spectral type later than mid-G \citep{wifes1_2015}.

Since the most recent membership studies for these young populations were carried out, significant improvements in the availability and quality of photometry and astrometry in all-sky catalogs have become available. In particular the first Gaia data release  \citep{gaiadr1} provided optical G-band photometry for all cluster members down to brown-dwarfs, and parallaxes and proper motions for many stars in common with the TYCHO-2 catalog \citep{perrymantycho}. Furthermore, improved proper motions determined from the combination of PPMXL \citep{ppmxl}, 2MASS \citep{2mass} and Gaia astrometry data are now available \citep{hsoy}. As such, in order to produce a uniform membership with the latest available data for the young clusters, we re-examine the evidence for membership of \emph{K2} targets in the Hyades, Praesepe, Pleiades and Upper Scorpius clusters (C2, C4 and C5). We utilize some coarse cuts on photometry and kinematic data to thin out clear interlopers, and then apply the Bayesian membership selection methods previously used for the Sco-Cen association in \citet{myfirstpaper}. We note that upon the release of the next installment of the Gaia catalog in mid-2018, which will contain parallax measurements for all cluster candidate members, the assessment carried out here will need to be repeated.

\subsection{Input Data and Adopted Kinematic Models}

We begin with all objects with available data in the relevant observing campaigns in the K2 Ecliptic Plane Input Catalog of \citet{huber_epic}. This catalog includes positions, photometry from Tycho \citep{perrymantycho}, SDSS \citep{sloan_dr12}, and 2MASS \citep{2mass}, and proper motions and parallaxes from \emph{Hipparcos} \citep{leeuwen07} and UCAC4 \citep{ucac4}. We supplement this with improved proper motions and parallaxes from the recently available Tycho-Gaia Astrometric Solution (TGAS; \citealt{gaiadr1}) and the Hot Stuff for One Year catalog \citep{hsoy} where available, and PPMXL \citep{ppmxl} proper motions for object lacking entries in the catalog. We also adopt G-band magnitudes from Gaia DR1 where available \citep{gaiadr1}.

We take the kinematics of the older 650-800\,Myr clusters Hyades and Praesepe from the Hipparcos study of \citet{leeuwen09}. For the  $\sim$110\,Myr Pleiades cluster, we take the mean cluster proper motions and radial velocity from \citet{leeuwen09}, and calculate an updated velocity using the latest cluster parallax of 7.4\,mas \citep{melis14,gaiadr1_clusters}. For Upper Scorpius, we adopt the mean of the kinematic models of \citet{chen11} and \citet{myfirstpaper}, and the internal velocity dispersion of \citet{myfirstpaper}. For the open clusters which have relatively well-defined spatial extent, we derive Galactic position models on the basis of known member positions in the literature (e.g., \citealt{kraushillenbrand_comaber,roser11,rebull16_plei}), treating the spatial distributions as Gaussians with scale radii of 2$^\circ$ for Praesepe and Pleiades, and 10$^\circ$ for Hyades, with depths corresponding to the scale radii in physical units. For Upper Scorpius, we adopt the spatial model of \citet{myfirstpaper} that models spatial position and distance as a linear function of Galactic longitude. Table \ref{clusteruvw} lists the spatial and kinematic properties adopted for the four young clusters.

\subsection{Membership Selection}

The first step in our membership selection is to apply coarse cuts to remove clear non-members. We initially reject any objects that sit below the main sequence at the distance of each cluster in either the (G-J, G) or (J-K, K) color-magnitude diagrams with a conservative buffer of 1 magnitude. This rejects $<$5 members from the literature in each cluster, but removes the majority of the objects observed by \emph{K2} that are not literature members. We then calculate the expected proper motions for each star from the kinematic models of the clusters, using the parallax measurement of the object if available or the cluster distance otherwise. Any object with proper motion discrepancies of more than 5-$\sigma$ from this expected value were rejected as non-members (where $\sigma$ is both the velocity dispersion in angular units and the uncertainties in proper motion combined in quadrature). For Praesepe, Pleiades and Upper Scorpius candidates with measured parallaxes from either \emph{Hipparcos} or TGAS, we also reject objects with distances more than 50\,pc discrepant from the expected cluster distances.

\begin{figure}
\includegraphics[trim={8mm 0mm 15mm 14mm},clip,width=0.48\textwidth]{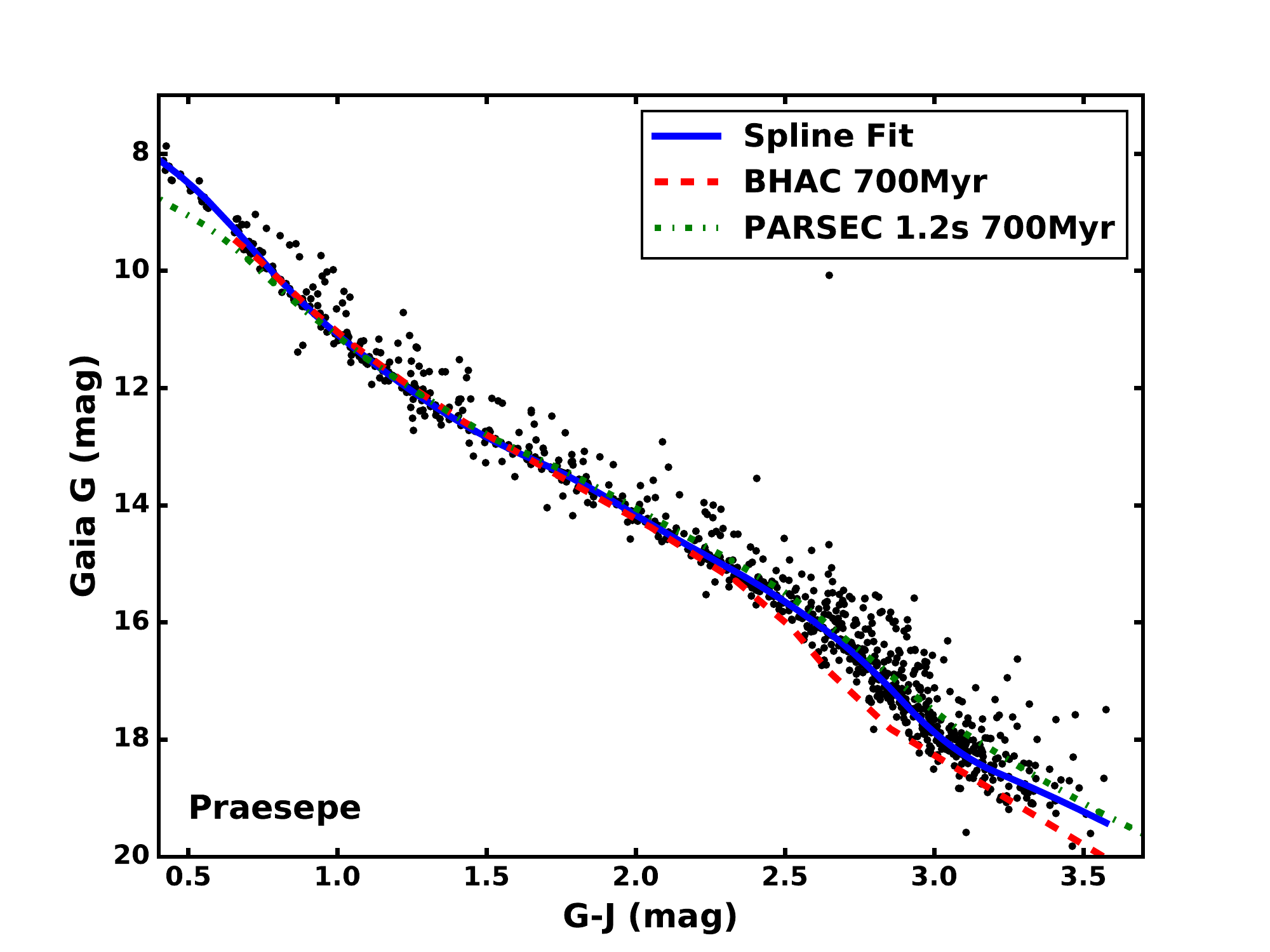}
\includegraphics[trim={8mm 0mm 15mm 14mm},clip,width=0.48\textwidth]{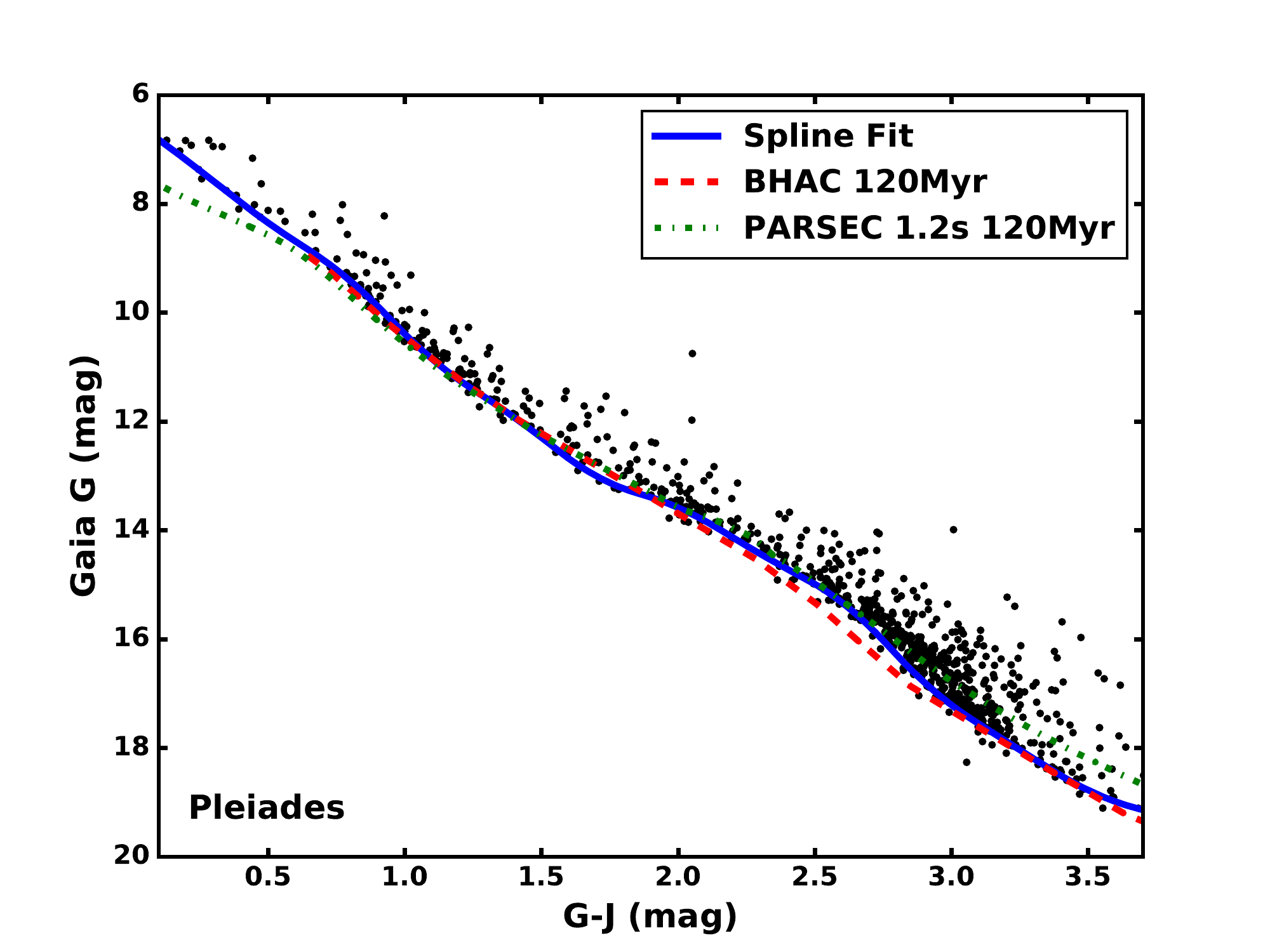}
\includegraphics[trim={8mm 0mm 15mm 14mm},clip,width=0.48\textwidth]{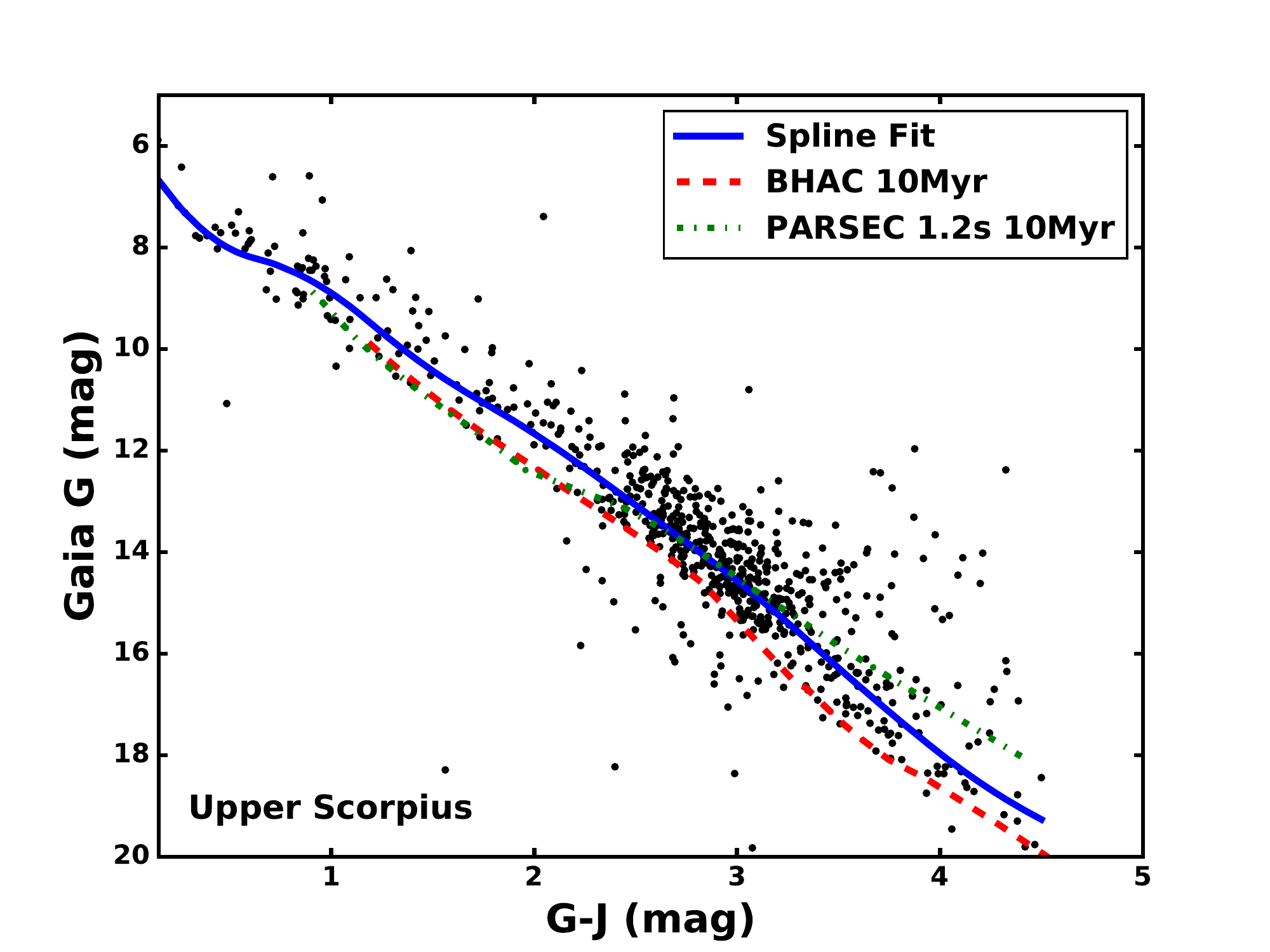}
\caption{G-J Color-magnitude diagrams for Praesepe (upper), Pleiades (middle) and Upper Scorpius (bottom) members with BHAC (dashed red; \citealt{bhac15}) and PARSEC 1.2s (dot-dashed green; \citealt{chen14_padova}) isochrones of the appropriate age. The model isochrones do not accurately reproduce the empirical cluster sequences, with offsets varying as a function of star color. We also show a BSpline fit to the cluster members (solid blue), with breakpoints every 0.4 magnitudes in color. }
\label{cmdspline}
\end{figure}

\begin{figure*}
\centering
\includegraphics[trim={8mm 0mm 15mm 13mm},clip,width=0.365\textwidth]{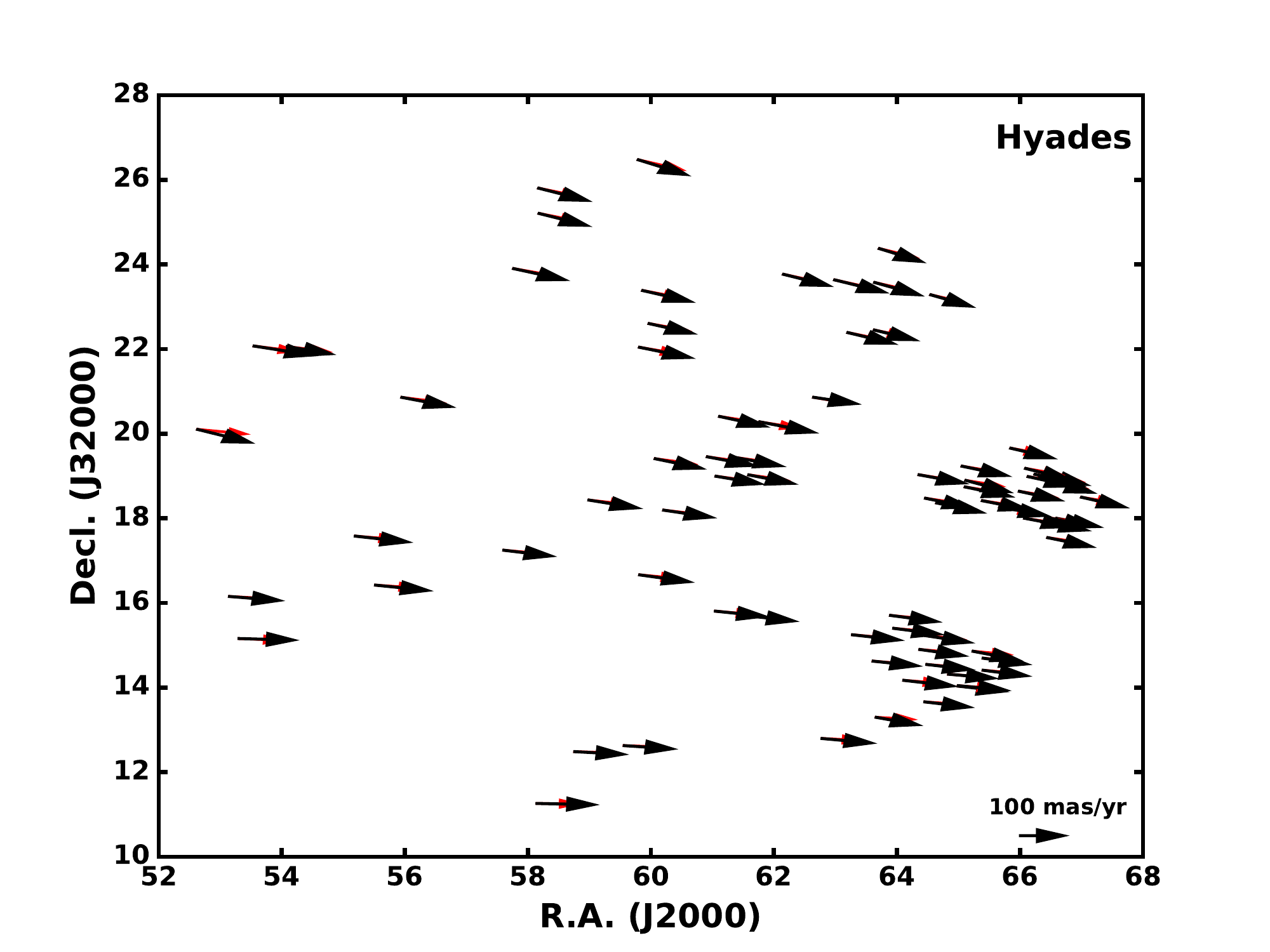}
\includegraphics[width=0.38\textwidth]{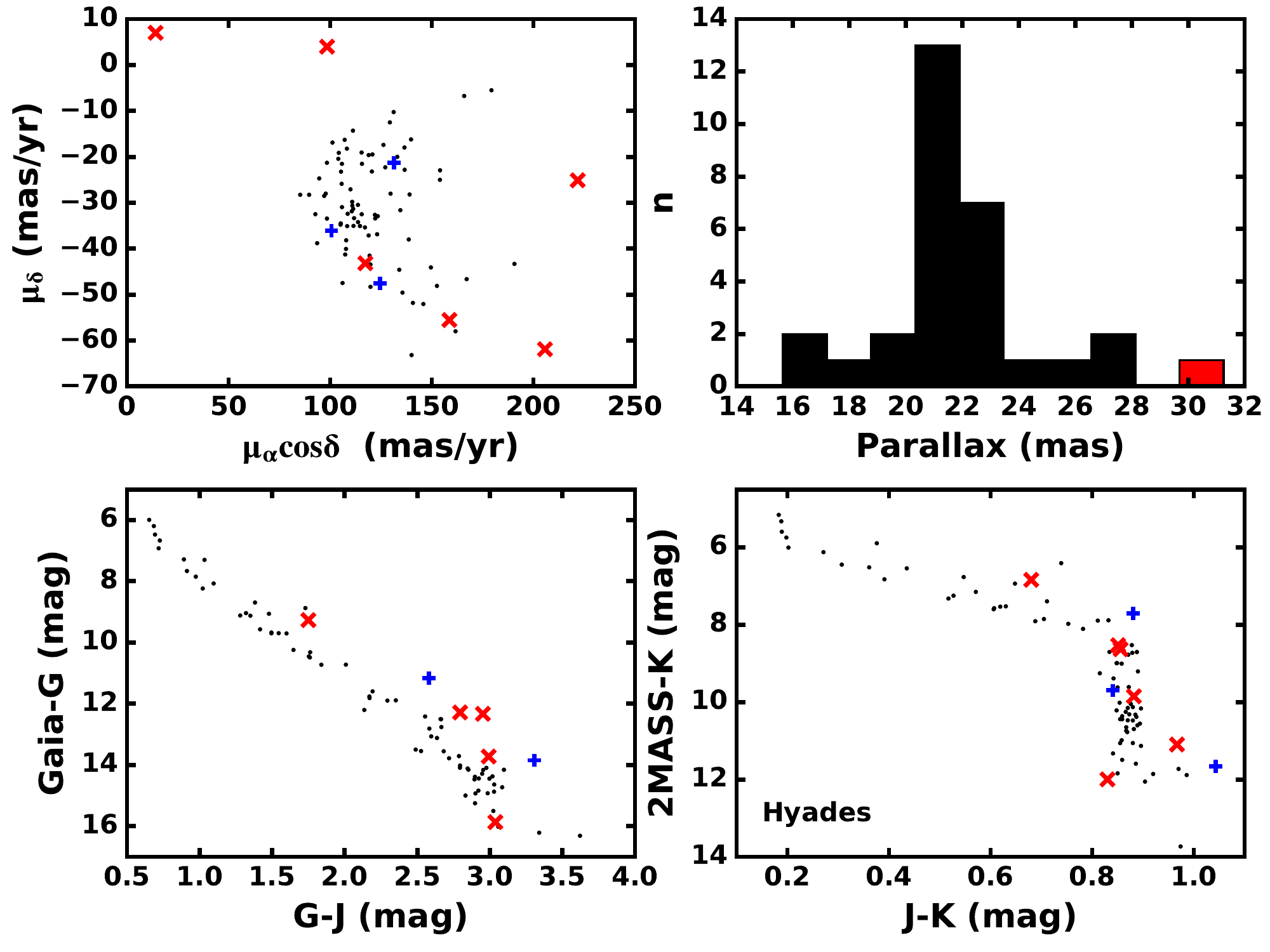}\\
\includegraphics[trim={8mm 0mm 15mm 13mm},clip,width=0.365\textwidth]{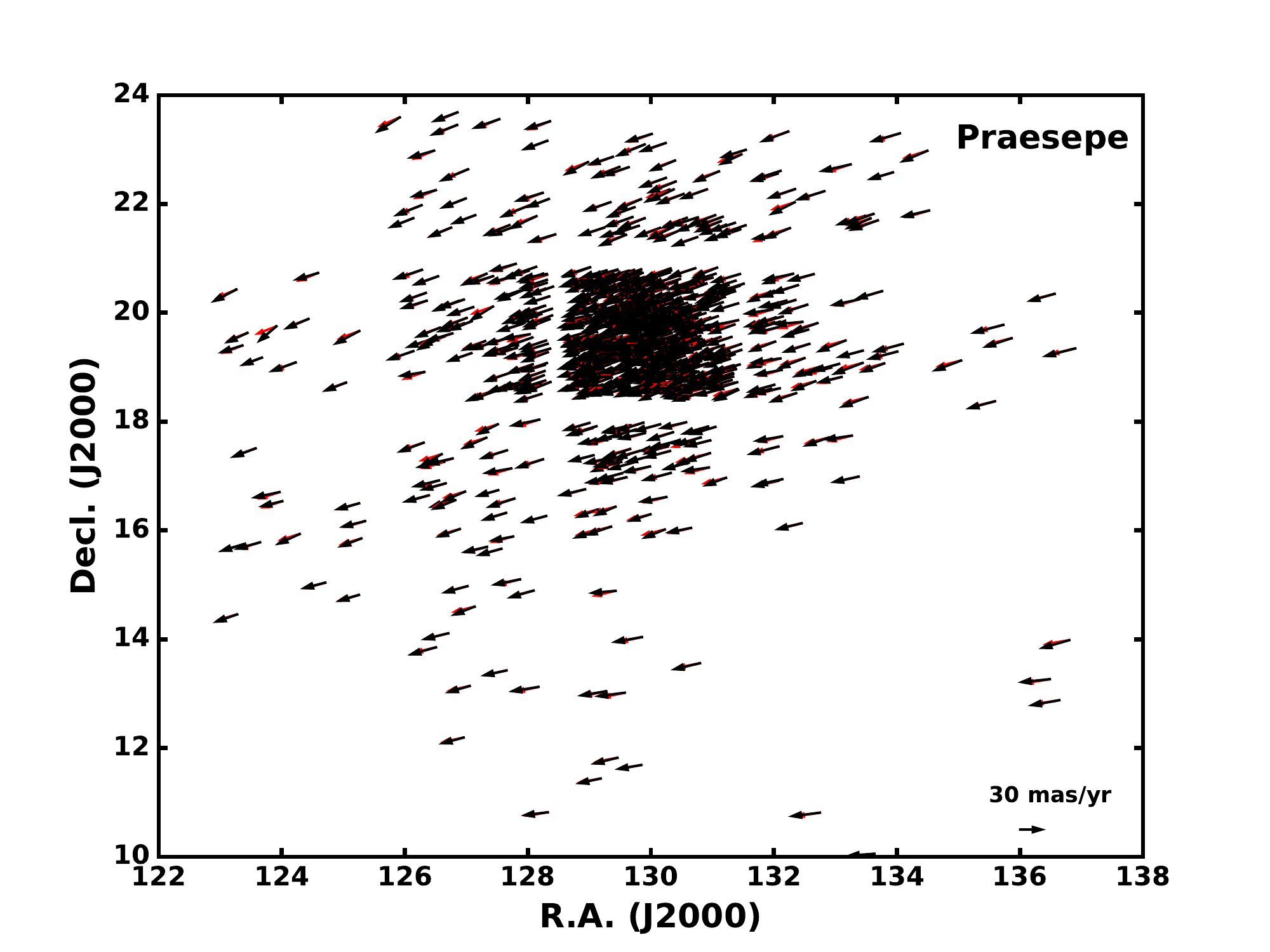}
\includegraphics[width=0.38\textwidth]{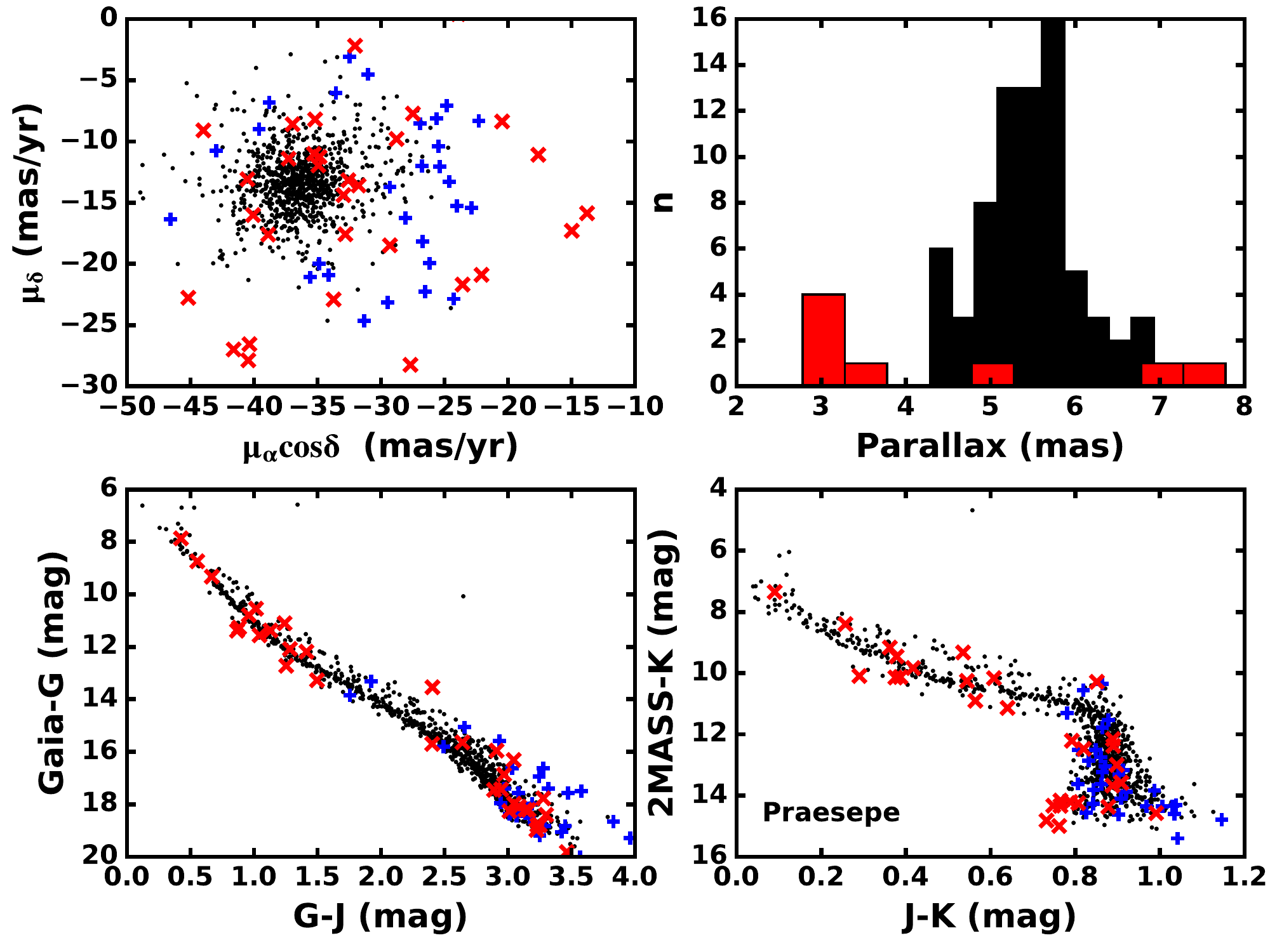}\\
\includegraphics[trim={8mm 0mm 15mm 13mm},clip,width=0.365\textwidth]{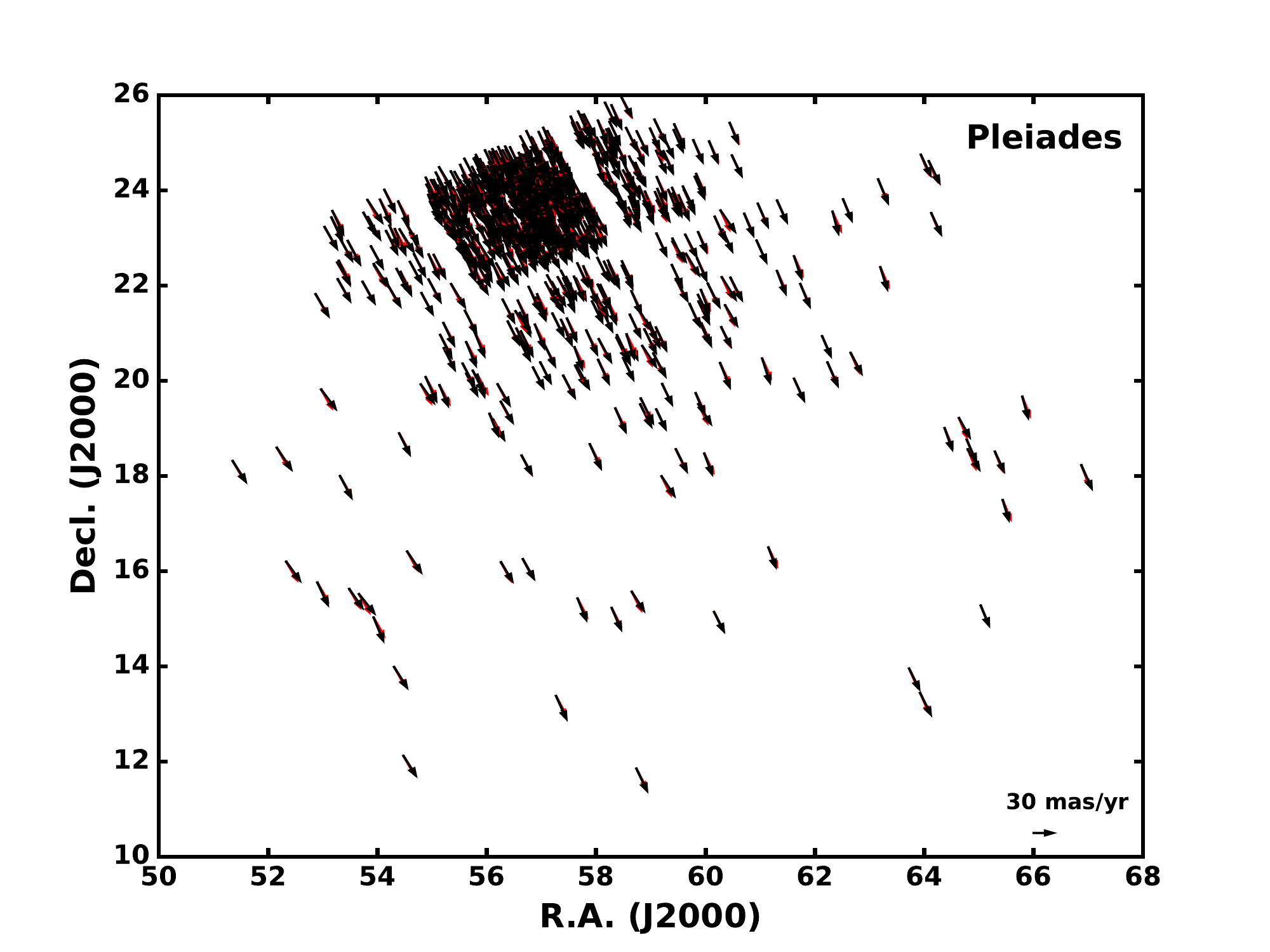}
\includegraphics[width=0.38\textwidth]{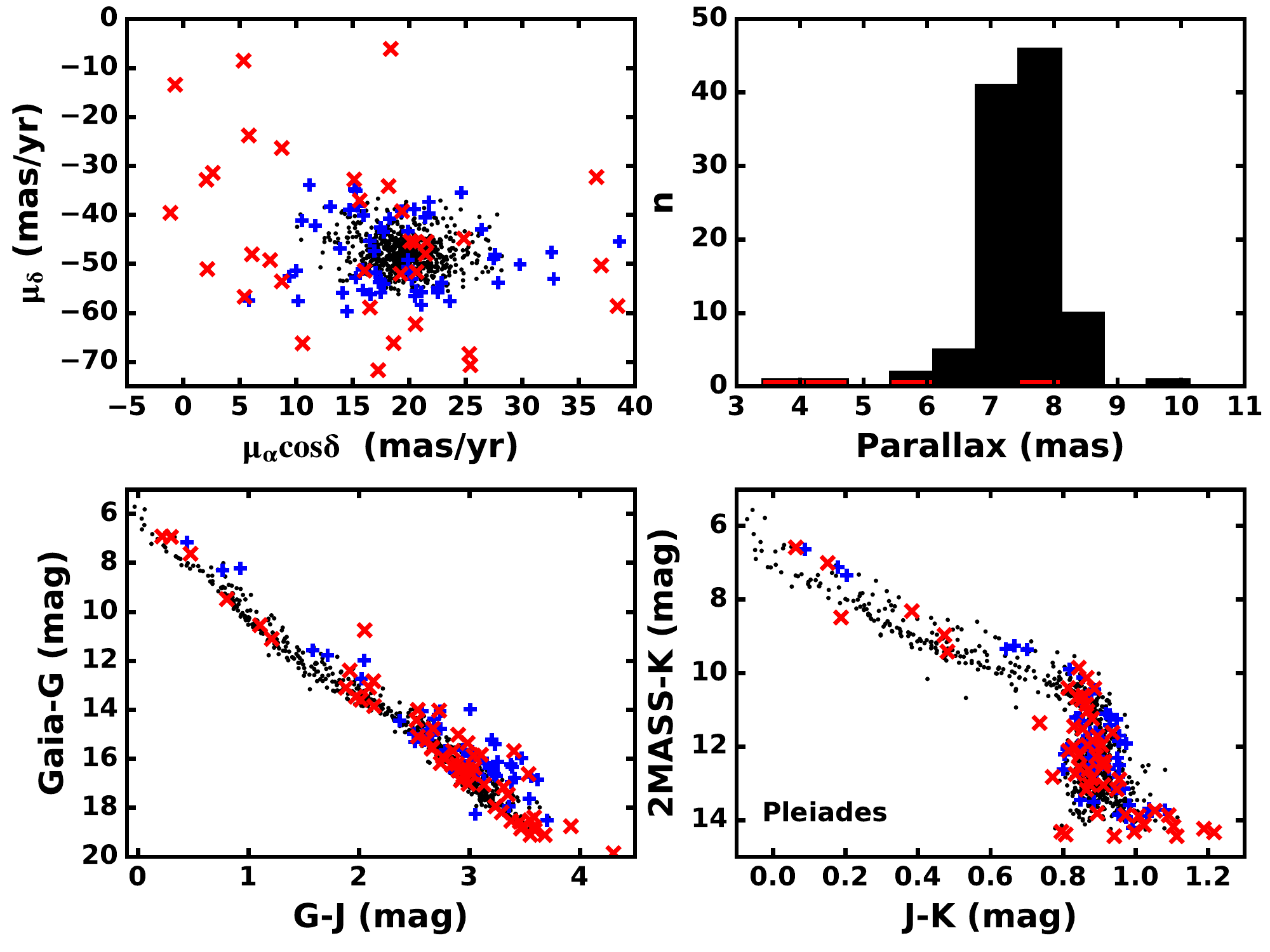}\\
\includegraphics[trim={8mm 0mm 15mm 13mm},clip,width=0.365\textwidth]{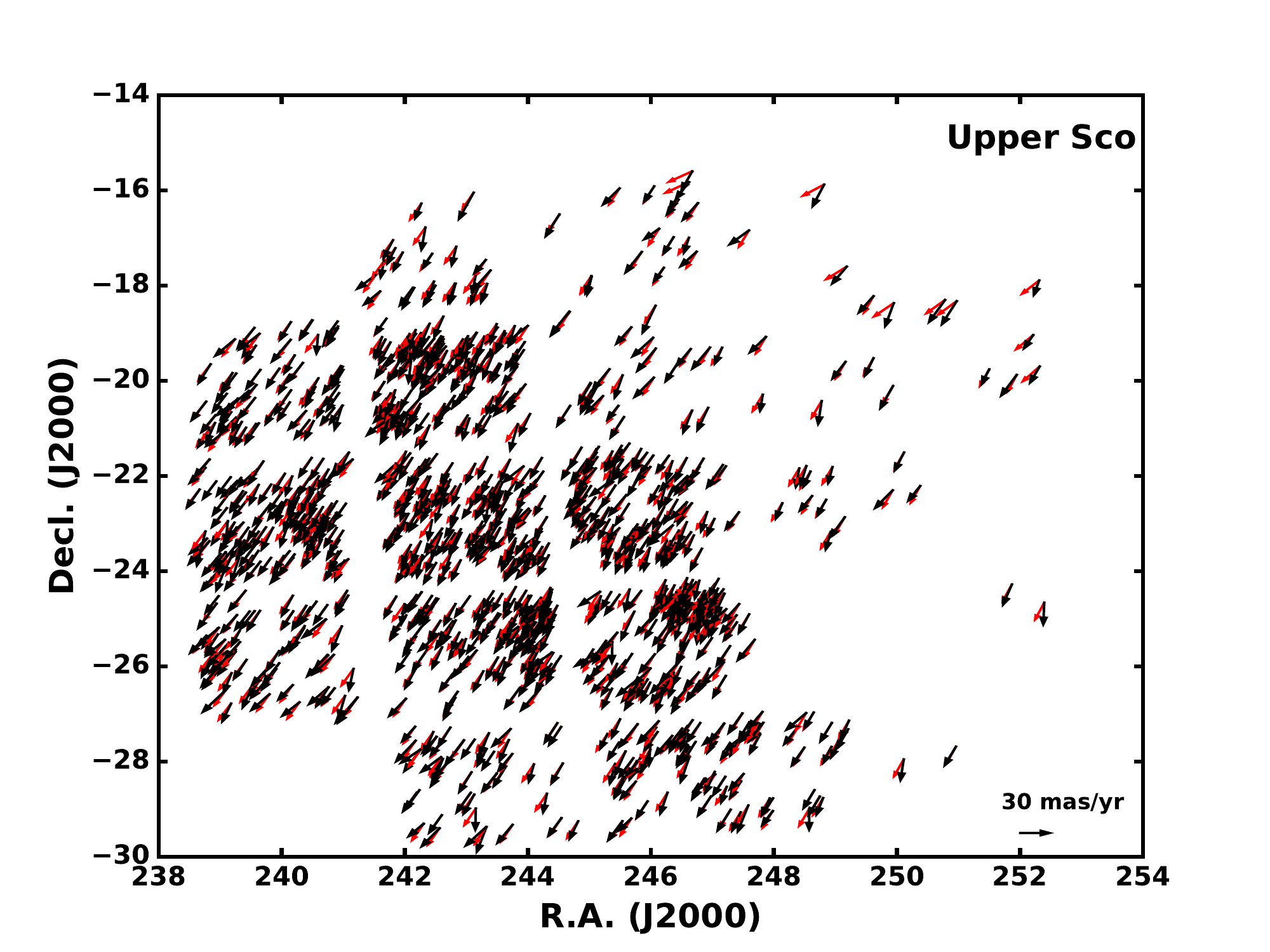}
\includegraphics[width=0.38\textwidth]{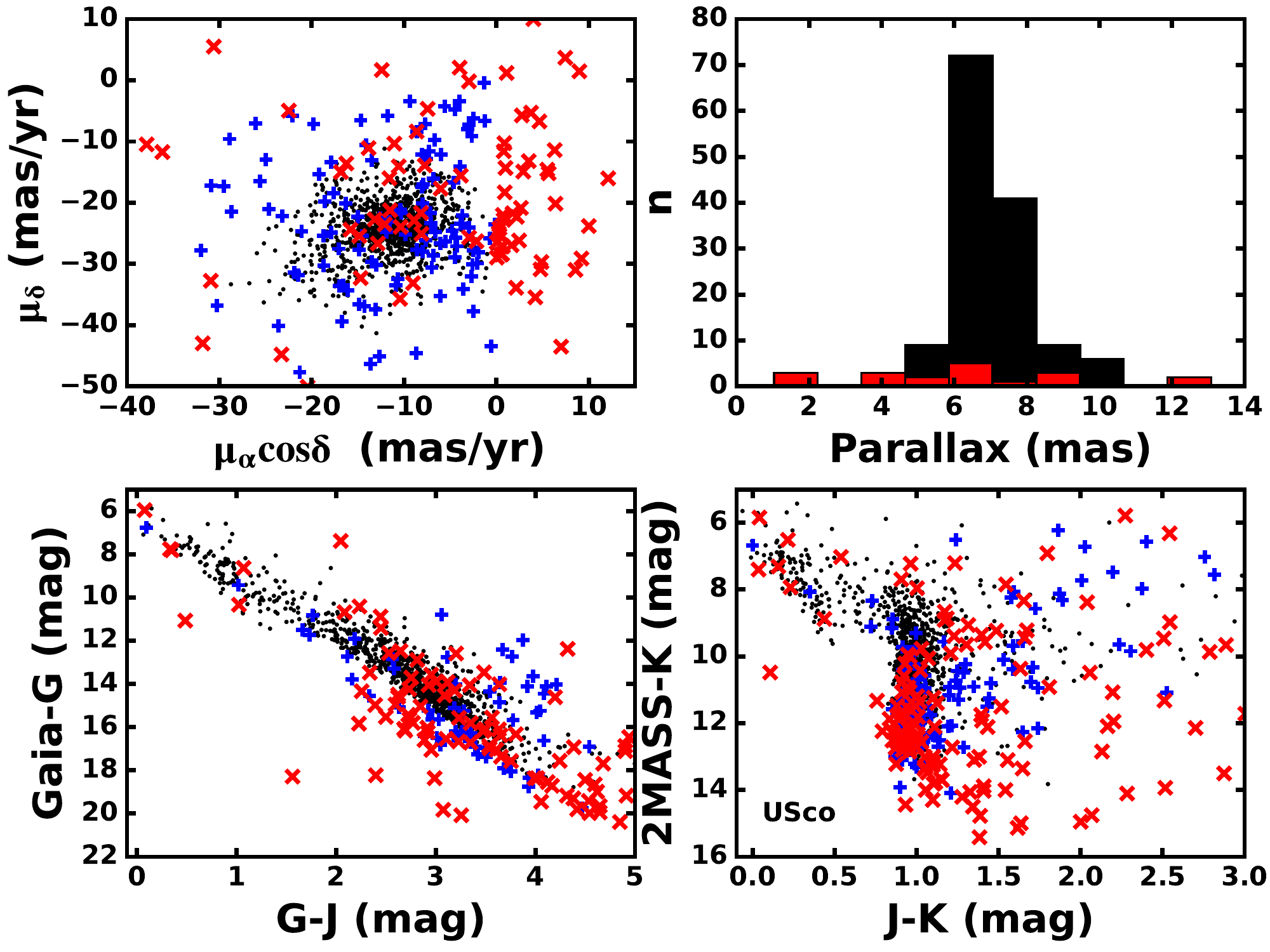}
\caption{Proper motion vector plots and membership diagnostics for the four young clusters. In the arrow plots (\emph{left}) we show candidates with $P_\mathrm{mem}>$50\%,  red/black arrows indicating the model proper motions and measured proper motions respectively. 
On the right panels, we show proper motion diagrams (\emph{top-left}), Parallax histogram (\emph{top-right}), and (G-J, G) and (J-K, K) CMD's (\emph{bottom-left} and \emph{bottom-right}) for each cluster. Candidates with $P_\mathrm{mem}>$50\% are shown in black in all panels. Previously suggested members from the literature (see text for discussion) with $P_\mathrm{mem}<$30\% (blue plusses), or that failed to pass our initial coarse cut (red-crosses) are also shown. Upper Sco member that are spectroscopically conferment were still searched for planets, and should be treated as members for statistical analyses. Proper motion outliers might be binary systems showing photocenter motion from variability of individual components, while CMD outliers might be heavily embedded objects where extinction complicates photometric distance estimates.}
\label{selectionplots}
\end{figure*}

The remaining members are then treated as in \citet{myfirstpaper,wifes1_2015}. We calculate photometric distances (in the absence of a true parallax) for each candidate from model isochrones in different colors. The majority of candidates possessed both 2MASS and Gaia-G band magnitudes ($>$90\%) and so we  prioritized the G-J CMD in the interest of uniformity. In the absence of Gaia-G magnitudes, we used, in order of availability, the V-J, R-J, B-V or J-K CMD to calculate photometric distances, with magnitudes sourced as described above. Figure \ref{cmdspline} shows the (G-J, G) CMD for the known literature members of each cluster compared to the model isochrones. We prioritize the combination of optical and near-IR photometry because it is more sensitive to temperature for K/M dwarfs than purely optical or IR photometry. From inspection of the (G-J, J) CMD for each cluster, we found that both the BHAC \citep{bhac15} and Padova PARSEC 1.2s \citep{chen14_padova} isochrones differ from the empirical sequence of members significantly for the K-M dwarfs.  Using the model isochrones to determine photometric distances will introduce systematics in our membership selection as a function of star color, however the availability of Gaia-G magnitudes for the majority of the \emph{K2} targets makes using it a necessity. To deal with the discrepancy we instead fit the known cluster members using a spline model. We iteratively fit a spline model, with knots every 0.4\,mag in color. On each iteration to a maximum of three, we remove potential binary stars by clipping members brighter than the fit by more than 1.5 times the RMS scatter about the fitted spline. The spline models were then used to interpolate photometric distances for all candidate cluster members without parallax measurements. We then use the photometric distances and proper motions for each candidate and calculate membership probabilities as described in \citet{myfirstpaper}. For Upper Scorpius, we also list all members with spectroscopic confirmation of youth observed by \emph{K2} (e.g., Li 6708\AA;~\citealt{preibisch01}).

\subsection{Comparison to Previous Membership Work}
 
In total, we identified 78, 850, and 841 candidates with membership probabilities $>$50\% in Hyades, Praesepe, and Pleiades respectively, with $>$90\% recovery rate of members listed in the literature in each cluster. Table \ref{shortmemtab} lists the candidate members identified by our selection. Figure \ref{selectionplots} displays the results of our membership selection in both the cluster CMDs and the position and velocity of the members.

For the three open clusters, our membership selection broadly agrees with the previous membership work in the literature. We reject 6 Hyades members from \citet{roeser11} on the basis of discrepant proper motions in the initial cuts. A further three Hyades members from \citet{roeser11} were also rejected following the Bayesian selection, where they were assigned probabilities of membership of $<$10\%. All three of these objects appear to sit significantly above the cluster sequence in either the (G-J,G) or (J-K, K) CMDs, and so we expect the low membership probabilities to be robust. 

For Praesepe, we find an overlap of $\sim$94\% with \citet{kraushillenbrand_comaber}. We reject 7 members on the basis of updated parallaxes in the TGAS catalog that disagree with the expected cluster distance of $\sim$181\,pc, and a further 19 with discrepant proper motions in the new catalogs. We also reject 6 members that sit below the main sequence in the (J-K,K) CMD. Of the remaining \citet{kraushillenbrand_comaber} members that passed our coarse cuts, 30 were assigned low membership probabilities. None of these objects have measured parallaxes, and all are found away from the core of the Praesepe proper motion diagram (Figure \ref{selectionplots}). The proper motions for 27 of these objects were sourced from the HSOY catalog, with the remaining three from PPMXL. Significant discrepancies in astrometry, and hence proper motions in comparison of PPMXL data to other catalogs have been observed (e.g., \citealt{urat1}). \citet{kraushillenbrand_comaber} compiled the proper motions for their membership from the USNO-B1.0, 2MASS, and SDSS \citep{sloan_dr12} astrometry, and applied sigma-clipping when computing proper motions. The HSOY catalog, which draws on 2MASS, PPMXL and new Gaia DR1 astrometry, did not implement any data quality assessment in their determination of proper motions. As such, we expect that these objects with discrepant proper motions that were identified as members in \citet{kraushillenbrand_comaber} may have spurious proper motions in the newer catalogs. We suggest that these objects may be tentatively included as members of Praesepe until reassessment with parallaxes from the upcoming Gaia DR2 is possible. 

We also compare our membership for Pleiades to the sample presented by \citet{dance1}. We reject 2 of the \citet{dance1} members on the basis of discrepant parallaxes, 2 which sit below the cluster sequence in the (J-K,K) CMD. We also reject 37 previously suggested members which have highly discrepant proper motions compared to the cluster space velocity (Figure \ref{selectionplots}). We then find 21 members that have low probabilities of membership, many of which sit slightly above the expected cluster sequence in the (G-J,G) CMD, and all but one of which has proper motion significantly different to the expected cluster core in the proper motion diagram.

\begin{deluxetable*}{cccccccccccc}
\tabletypesize{\scriptsize} 
 \tablecaption{Candidate members in the young clusters and associations observed by \emph{K2}.} 
\tablehead{ \colhead{EPIC} & \colhead{R.A.}    & \colhead{Decl.}   & \colhead{G}     & \colhead{K}    & \colhead{PMRA}     & \colhead{PMDE}    & \colhead{$\pi$}  & \colhead{PMRA$_c$} & \colhead{PMDE$_c$} & \colhead{P$_{m}$} & \colhead{P$_R$} \\ 
 \colhead{}     & \colhead{(J2000)} & \colhead{(J2000)} & \colhead{(mag)} & \colhead{(mag)}& \colhead{(mas/yr)} & \colhead{(mas/yr)} & \colhead{(mas)} & \colhead{(mas/yr)} & \colhead{(mas/yr)}  &  \colhead{(\%)}    & \colhead{(days)} \\ } 
 \startdata 
\multicolumn{12}{l}{\bf Hyades Candidate Members} \\
210317378 & 03 52 34.34 & +11 15 38.8 & 12.42 & 9.01 & 179.36$\pm$1.64 & -5.5$\pm$1.64 & 29.7$\pm$4.1$^{(1)}$ & 136.6 & -1.9 & 89 &  \\ 
210359769 & 03 55 01.44 & +12 29 08.1 & 9.68 & 7.57 & 131.22$\pm$0.17 & -10.25$\pm$0.08 & 21.8$\pm$0.2$^{(0)}$ & 136.8 & -5.6 & 100 &  \\ 
210365286 & 03 58 14.38 & +12 37 40.9 & 14.87 & 10.99 & 129.36$\pm$2.16 & -12.5$\pm$2.16 & 24.9$\pm$3.4$^{(1)}$ & 132.8 & -6.3 & 99 & 0.8686\\ 
\enddata 
 \tablecomments{Membership information for all the cluster candidates with probabilities of membership $>$10\% and all spectroscopically confirmed members of Upper Scorpius. The complete table is available online. G-band magnitudes are taken from Gaia DR1 \citep{gaiadr1}, and K-band magnitudes from 2MASS \citep{2mass}. Proper motions are taken in order of availability from the Tycho-Gaia Astrometric Solution \citep{gaiadr1}, \emph{Hipparcos} \citep{perrymantycho, leeuwen07}, UCAC4 \citep{ucac4} and PPMXL \citep{ppmxl}. We indicated the source of the parallax measurement, with (0) indicating a true measurement from either TGAS or \emph{Hipparcos}, and larger numbers indicating photometric distances based on the cluster sequences: (1) (G,G-J) photometry, (2) - (r,r-J) photometry, (3) (V,V-K) photometry, (4) (K, J-K) photometry. We also list the expected proper motions if the candidate shares the model space velocity vector of the cluster model. Finally, rotation period measurements from Covey et al., (private communication) for Upper Scorpius, \citet{rebull16_plei} for Pleiades, \citet{douglas16} for Hyades, and \citet{douglas17} for Praesepe candidates are listed if available in the literature. Candidates not considered in these studies were visually inspected for rotation periods smaller that 2\,days, which would impact the choice of detrending window. These objects are marked as ``$<$2'' in the table.}
 \label{shortmemtab} 
 \end{deluxetable*}

In Upper Scorpius we identify 381 candidates with probability of membership $>$50\% that are not identified as members in the literature. Our membership selection recovers $>$95\% of the high-mass membership from \citet{myfirstpaper}. We reject 8 objects with discrepant parallaxes that were suggested to be members in the wider literature (Table \ref{shortmemtab}). Of the lower-mass members from the literature, 108 lack either 2MASS photometry or proper motion measurements. The majority of these are in the most embedded regions of the Ophiuchius cloud cores ($(\alpha,\delta)\sim(246.5,-24.5)$) and are thus most likely members of the Ophiuchus star-forming region ($\sim$1\,Myr).  In addition to these objects, 96 members have significantly discrepant proper motions, approximately half of which are also found in the Ophiuchius cloud core. Of the remaining members which passed the coarse cuts, $\sim$86\% had probabilities of membership $>$50\%, which is consistent with the spectroscopic confirmation survey of \citet{wifes1_2015} for candidates without parallax measurements. Identification of members on the basis of kinematics when parallax measurements are not available can be considerably more difficult for young OB association subgroups such as Upper Scorpius compared to the older open clusters. Significant differential extinction makes estimation of photometric distances problematic for the most extincted objects. For those literature members with parallax measurements, 97\% have membership probabilities above 50\% indicating that our membership analysis is functioning when a reliable distance is available. We list all spectroscopically confirmed, and disk-bearing members of Sco-Cen and Ophiuchus observed by \emph{K2} in the membership tables (982 members in total), with appropriate references for completeness, regardless of kinematic membership probabilities, as well as new candidates identified in our selection.

\section{Transit Detections Among the Young Cluster Members}
\label{sec:detections}

We applied our detrending and transit search pipelines described above on the  candidate members (Section \ref{sec:membership}) of the four young clusters in order to identify young transiting exoplanets. Table \ref{table:dets} lists the detected periodic transit signals and indicates whether they are planet-like, eclipsing binary systems, or periodic variability caused by other occulting material (e.g., material at the corotation radius of 11\,Myr old Upper Scorpius members as suggested by \citealt{stauffer_plage}). Figures \ref{figs:det4}$-$\ref{figs:det1} display the phased transits. 

Our pipeline identifies all 10 known transiting exoplanet candidates in the four young clusters, including the exoplanet K2-75\,b transiting a star determined to be an interloper in the Pleiades region \citep{zeit2}. This alone is a significant improvement over existing pipelines that only locate a subset of the known exoplanets, or require multiple versions with hand-adjusted parameters to locate the full sample. We identify 17 eclipsing binary systems on the basis large transit depths, secondary eclipses and eclipse shapes, many of which have been previously identified (e.g., \citealt{kraususcoctio5,davidusco5,david16_eb}), and  two of which are previously unreported in the literature (EPIC 211705654 and 204432860). We also identify five systems with transit-like signals associated with stars determined to not be members of the young clusters upon validation with additional data. Two of these systems are not previously reported in the literature (EPIC~210696763 and 210736056).  We also detect periodic signatures that have inconsistent or variable shape associated with six Upper Scorpius members, that have previously been identified as non-planetary in nature \citep{stauffer_plage}, and one additional such candidate (EPIC 204882444).  All EB and planet candidate systems with stellar rotation periods $<$3\,days detected by the notch-filter pipeline were also recovered with LOCoR, excepting the EB EPIC 204432860, for which the orbital period very closely matches the stellar rotation period, and hence the system might be tidally locked. We will now comment on systems that appear planetary in nature but require some inspection, excluding the known exoplanets from earlier papers in this series.

\subsubsection*{EPIC~210696763 (Campaign 4)}
We detect two strong transiting exoplanet signals associated with EPIC~210696763 (periods of 3.64 and 7.98\,days). We also detect a possible third transit signal with period of 5.76\,days, though this third transiting signal is marginal (SNR$\sim$7). We have obtained a near-IR spectrum of this star with the Immersion Grating Infrared spectrograph (IGRINS) instrument \citep{park14_igrins,greg_igrins16} at the 2.7\,m Harlan J. Smith telescope at the McDonald Observatory. Extraction of the spectrum and calculation of the stellar radial velocity was performed as in \citet{zeit1} and \citet{igrins_plp_github,greg_rv_poster}. We determine the radial velocity of EPIC~210696763 to be 22.59$\pm$0.16\,km/s, which is inconsistent with the expected radial velocity of $\sim$5.7\,km/s \citep{leeuwen09} making it a highly unlikely member of the Pleiades cluster. However, this field star does indeed appear to host a probable 2- or 3-planet system.

\subsubsection*{EPIC~210736056 (Campaign 4)} 
We detect a $\sim$4\,mmag signal with period of 29.05\,days which includes three transits. EPIC~210736056 has proper motion highly consistent with the Pleiades space velocity, however on the (G-J,G) CMD for the cluster sequence, the target sits just below the main population, at the edge of what passes our coarse membership cuts. The combination of the photometric distance and proper motion yields a membership probability of  (P$_\mathrm{mem}=62$\%). EPIC~210736056 also sits slightly outside the core of the main Pleiades population ($\alpha$,$\delta$=$\sim$54.4$^\circ$,18.9$^\circ$), though some other known Pleiades members can be found in this region. Inspection of the K2SFF lightcurve (uncorrected for rotational variability) reveals a rotation period of $\gtrsim$30\,days, which is significantly slower than the expected rotational period for a $\sim$110\,Myr old M-type star (P$_\mathrm{Rot}<$20\,days). We also observed EPIC~210736056 with IGRINS, and determine the system radial velocity to be  7.5$\pm$0.2\,km/s. This is consistent with the expected radial velocity for Pleiades at the position of  EPIC~210736056  ($\sim$7.8\,km/s). As such we cannot confidently rule-out or confirm  the membership of EPIC~210736056 in the Pleiades. Further information, including a parallax measurement from the future Gaia DR2, an improved spectral type, and identification of other youth diagnostics are required to robustly determine membership.

\subsubsection*{EPIC~211804579 (Campaign 5)}
EPIC~211804579 is a high-probability Praesepe member ($>$95\%), and the depth ($\sim$1\,mmag) and shape of the periodic transit signal suggests the possibility of a planetary companion with orbital period of 1.5235\,days. Upon further inspection, a contaminating source, 2MASS J08361560+1722542, separated by $\sim$9.5'' at a position angle of  $\sim$270$^\circ$ was identified in the \emph{K2} pixel data for this target. Removing the pixels associated with 2MASS J08361560+1722542 and re-reducing the data with our pipeline resulted in a non-detection, implying that the transit signal is associated with the contaminant source. The color and magnitude (J-K=0.124, K=12.42\,mag) of   2MASS~J08361560+1722542 suggests that it is a background star at  d$>$1\,kpc with a spectral type of late-A or early-F, and is clearly not associated with EPIC~211804579. The signal therefore suggests that the background source is an eclipsing binary system

\subsubsection*{EPIC~203823381 (Campaign 2)}
The transit signal detection (P=8.278\,days) associated with EPIC~203823381 appears flat bottomed and is a depth consistent with an exoplanet transit (d$\sim$4\,mmag). \citet{vanderburg16} and \citet{crossfield16} both also identify a transiting signal at the same period and with similar depth. However, the K2SFF pointing-corrected \emph{K2} lightcurve for EPIC~203823381 does not show the $\sim$day timescale rotational variability that is often characteristic of Upper-Scorpius members. Our membership analysis of this object only moderately support membership in Upper Scorpius (P$_\mathrm{mem}$=80\%). The uncertainties on the proper motions are too large to be deterministic ((-5.7$\pm$2.1,-19.8$\pm$2.1)\,mas/yr). The Ecliptic Plane Input Catalog (EPIC) of \citep{huber_epic} suggests that EPIC~203823381 is a highly reddened background giant, however this is a common misclassification for nearby bright, inflated dwarfs and so is not necessarily a reliable estimate. We obtain an IGRINS spectrum of EPIC~203823381 from the McDonald Observatory 2.7\,m telescope. The IR spectrum lacks many of the spectral features expected of a M-dwarf, including the CO band-head. A combination of both broad and narrow lines are also present in the spectrum. We expect that this is a blend of a reddened A-type star and  an M-type star.  If the M-type star is an Upper Scorpius member, then the A-type background star must dominate the Kepler-band flux in order to obscure the expected rotational signature. In this case, even if the transiting object is associated with the M-type members it is highly diluted in the \emph{K2} lightcurve and is thus not an exoplanet. If the M-star provides the majority of the flux in the Kepler-band, then the lack of rotation indicates that it is unlikely to be a member of Upper Scorpius.

\subsubsection*{EPIC~204750116 (Campaign 2)}
We detected a transit signal with period of 23.43\,days and depth of 0.2\,mmag that is flat-bottomed within the precision of the data. \citet{vanderburg16} also identified this planet. EPIC~204750116 has proper motions highly consistent with membership in Upper Scorpius, however, spectroscopic follow-up from \citet{vanderburg16} give a radial velocity of 39.15$\pm$0.1\,km/s. Given the expected Upper Scorpius radial velocity of $\sim$1\,km/s, this star is unlikely to be part of Upper Scorpius. 

\subsection*{EPIC~204276894 (Campaign 2)}
We detect a marginal 0.5\,mmag transit signal with period of 1.957\,days with duration of 3-4\,hours that is flat-bottomed at the precision level of the detrended \emph{K2} lightcurve. The proper motion of the host star is consistent with membership in Upper Scorpius, however it does not show the expected rotational variability associated with 10\,Myr old stars. The Radial Velocity Experiment observations list a radial velocity of 47.2\,km/s \citep{rave5} which is highly inconsistent with Upper Scorpius membership, and a $\log{g}=1.0$, indicating that it is a background giant.

\begin{deluxetable*}{ccccccccc}
\tabletypesize{\scriptsize} 
\tablewidth{0pt}
 \tablecaption{Transit signal detections among the young cluster members from the notch-filter pipeline.} 
\tablehead{ 
\colhead{EPIC} & \colhead{Cluster} & \colhead{R.A.} & \colhead{Decl.} & \colhead{P} & \colhead{T$_0$} & \colhead{{D}epth} & \colhead{P$_\mathrm{rot}$} &\colhead{Note} \\
\colhead{}& \colhead{}& \colhead{(J2000)} & \colhead{(J2000)} & \colhead{(days)} & \colhead{(MJD)} & \colhead{(mmag)} &\colhead{(days)} &  \colhead{}\\
}
\startdata
\input{detection_table_nsfw.tex}
\enddata
\tablecomments{Periods, transit times and depths from periodic transit-like signals detected by our notch-filter pipeline. In the final column we list the discovery publication reporting known cluster planets from K2, EB when that the periodic signal is produced by an eclipsing binary system, ``Non-mem'' for transit signals that look planetary but appear to be associated with stars unlikely to be members. The transits associated with EPIC 211804579 was determined to be due to a nearby object and is marked ``NS''.} We also note some non-planetary periodic signals reported to be produced by material at the corotation radius around three Upper Sco stars (``NP'') reported in \citep{stauffer_plage}.
\label{table:dets}
\end{deluxetable*}

\begin{deluxetable*}{ccccccccc}
\tabletypesize{\scriptsize} 
\tablewidth{0pt}
 \tablecaption{Transit signal detections among the young cluster members from the LOCoR pipeline.} 
\tablehead{ 
\colhead{EPIC} & \colhead{Cluster} & \colhead{R.A.} & \colhead{Decl.} & \colhead{P} & \colhead{T$_0$} & \colhead{{D}epth} & \colhead{P$_\mathrm{rot}$}&\colhead{Note} \\
\colhead{}& \colhead{}& \colhead{(J2000)} & \colhead{(J2000)} & \colhead{(days)} & \colhead{(MJD)} & \colhead{(mmag)} & \colhead{days}& \colhead{}\\
}
\startdata
\input{detection_table_lcr.tex}
\enddata
\tablecomments{Periods, transit times and depths from periodic transit-like signals detected by the linear combination of rotations pipeline. Columns are as for Table \ref{table:dets}.}
\label{table:dets2}
\end{deluxetable*}

\subsubsection*{EPIC~205046529 (Campaign 2)}
EPIC~205046529 is a spectroscopically confirmed member of the Upper Scorpius moving group \citep{preibisch02}. The lightcurve shows transient flux-dips at a period of 1.836\,days that vary in depth over the course of K2 Campaign 2 (1-15\,mmag), almost disappearing in the second half of the dataset. Furthermore, the shape of the transit-like features are inconsistent with a transiting exoplanet. Follow-up observations by \citet{stauffer_plage} identify a close companion and attribute the dips to clouds of particulate material in Keplerian orbits near the corotation radius of the star and is unlikely to be an exoplanet.

\section{Limits on Additional Planets in the ZEIT Systems}\label{sec:limits}

The presence and frequency of multiple planet systems in young clusters can also provide a constraint on the dynamical state of a planetary system, or populations of planetary systems. Of the multitude of transiting exoplanet systems discovered in the original Kepler mission \citep{batalha13}, 20\% have been shown to host at least two transiting exoplanets \citep{fabrycky14}. The fraction and architecture of multiple systems may also inform the formation mechanisms responsible for creating the transiting exoplanet distribution (e.g., \citealt{lissauer11,morton14,ballard16}). Young ($<$1\,Gyr) exoplanet systems may provide an avenue for assessing the mechanisms responsible for creating the older systems observed in the \emph{Kepler} sample: dynamical mechanisms that may scatter tightly packed multiple exoplanet systems to mutually inclined orbits (e.g., Lidov-Kozai interaction and planet-planet scattering, \citealt{naoz12,chatterjee08})  operate over timescales of hundreds of millions of years to billions of years and so may not have yet produced the observed population split seen in the \emph{Kepler} samples.


To place limits on the presence of additional planets in the ZEIT systems, we perform injection-recovery testing using the detrending an transit search pipelines that we have developed. Because the details of the rotational stellar variability and the effects of the pointing instability are significantly different for each star, and one region of exoplanet parameter space is not necessarily predictive of limits in other regions of parameter space (e.g., at or near the rotation period), the entire range of parameter space of interest must be uniformly tested for each star.

We inject 2000 transiting exoplanet signals into the raw data for each planet host on random orbits using the BAsic Transit Model cAlculatioN code (BATMAN, \citealt{batman}), adopting the host star parameters of \citet{zeit1,zeit3,zeit4} and \citet{zeit2}. We sampled uniformly in orbital period in the range of 1$-$30\,days, and on impact parameter and orbital phase. For the injected planet radii, we chose radii spanning 0.5$-$10\,R$_\oplus$, with half of the points taken from a uniform distribution, and half taken from a $\beta$-distribution with coefficients $\alpha=2$ and $\beta=6$. This ensured sufficient coverage at all radii of interest, while simultaneously providing finer sampling at the key radii where sensitivity is expected to vary (1-5\,R$_\oplus$). 

We fix eccentricity to $e=0$ in our injection testing, because including eccentricity requires two additional dimensions in the injection recovery testing (eccentricity and argument of periapsis), while not producing a significant change in detectability at the range of periods of interest at the \emph{K2} $\sim$30\,min cadence. 

For each simulated planet, we perform the injection-recovery testing on the entire transit-search pipeline using the Lonestar5 cluster at the Texas Advanced Computing Center by executing the following steps:

\begin{itemize}
\item Start with extracted pixel aperture photometry from K2SFF \citep{vanderburgk2} and inject the transit model using BATMAN \citep{batman}). 

\item Apply pointing systematic correction using K2SFF \citep{vanderburgk2}.

\item Apply the notch-filter detrending pipeline.

\item Identify periodic transit signals using BLS and calculate power-spectrum peak significances as described above.

\item  In the case of identified transit signals with SNR $>$7, check if the detected period matches the injected period to within 1\%, and that the measured transit phase is consistent to within 1\% of the injected phase.
\end{itemize}

\begin{figure*}
\centering
\includegraphics[trim={0 0 5mm 0},clip,width=0.48\textwidth]{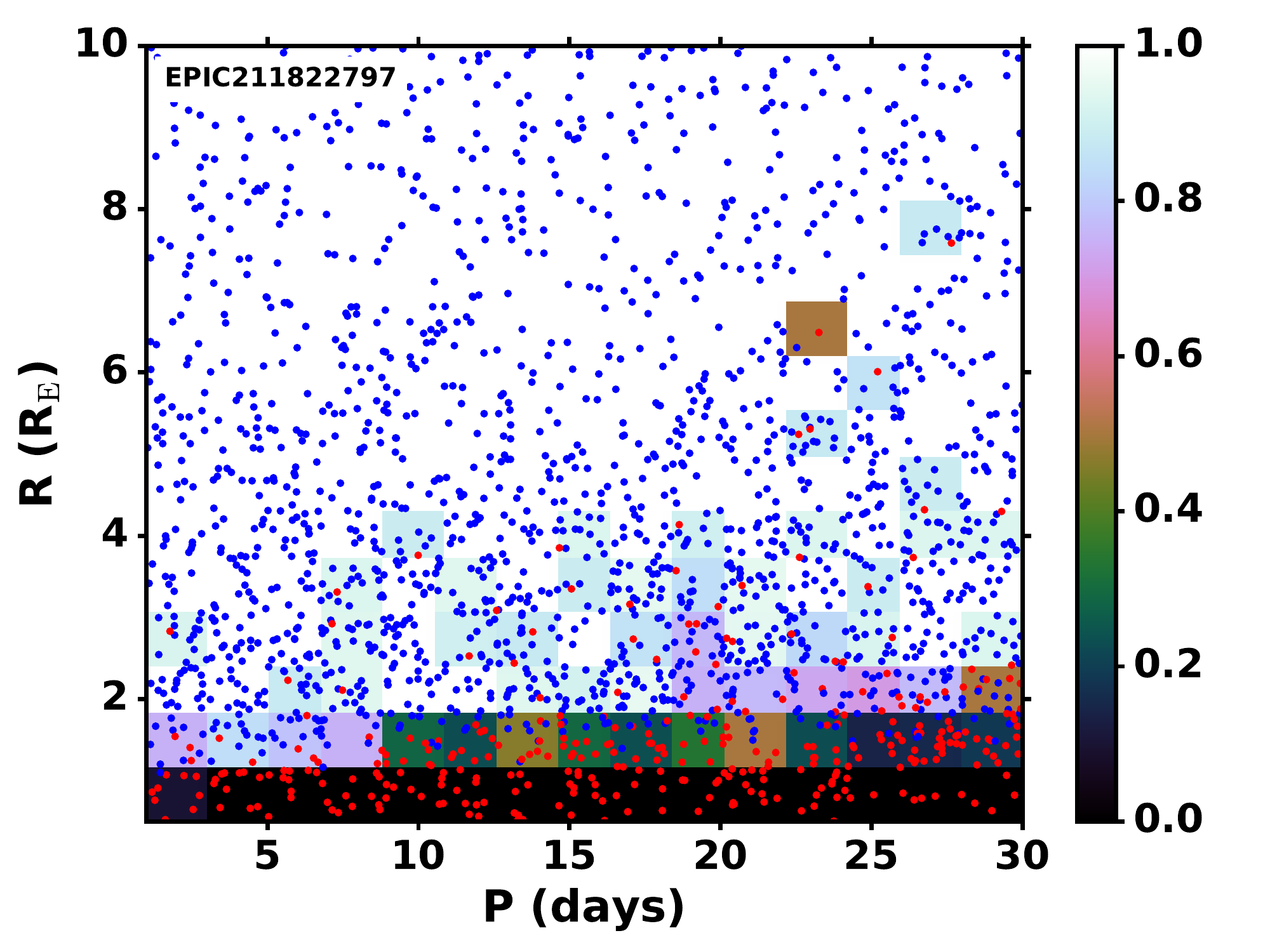}
\includegraphics[trim={0 0 5mm 0},clip,width=0.48\textwidth]{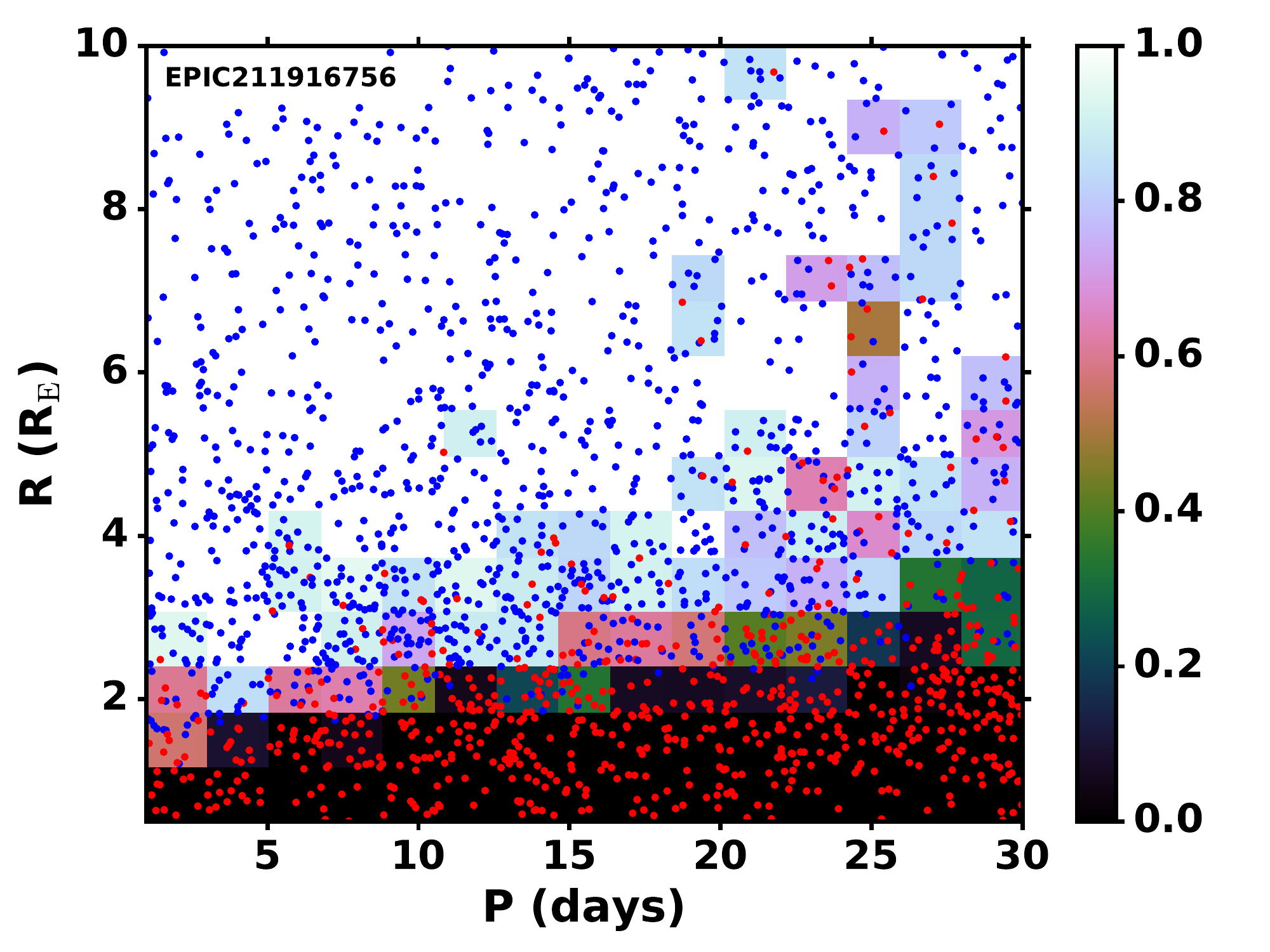}\\
\includegraphics[trim={0 0 5mm 0},clip,width=0.48\textwidth]{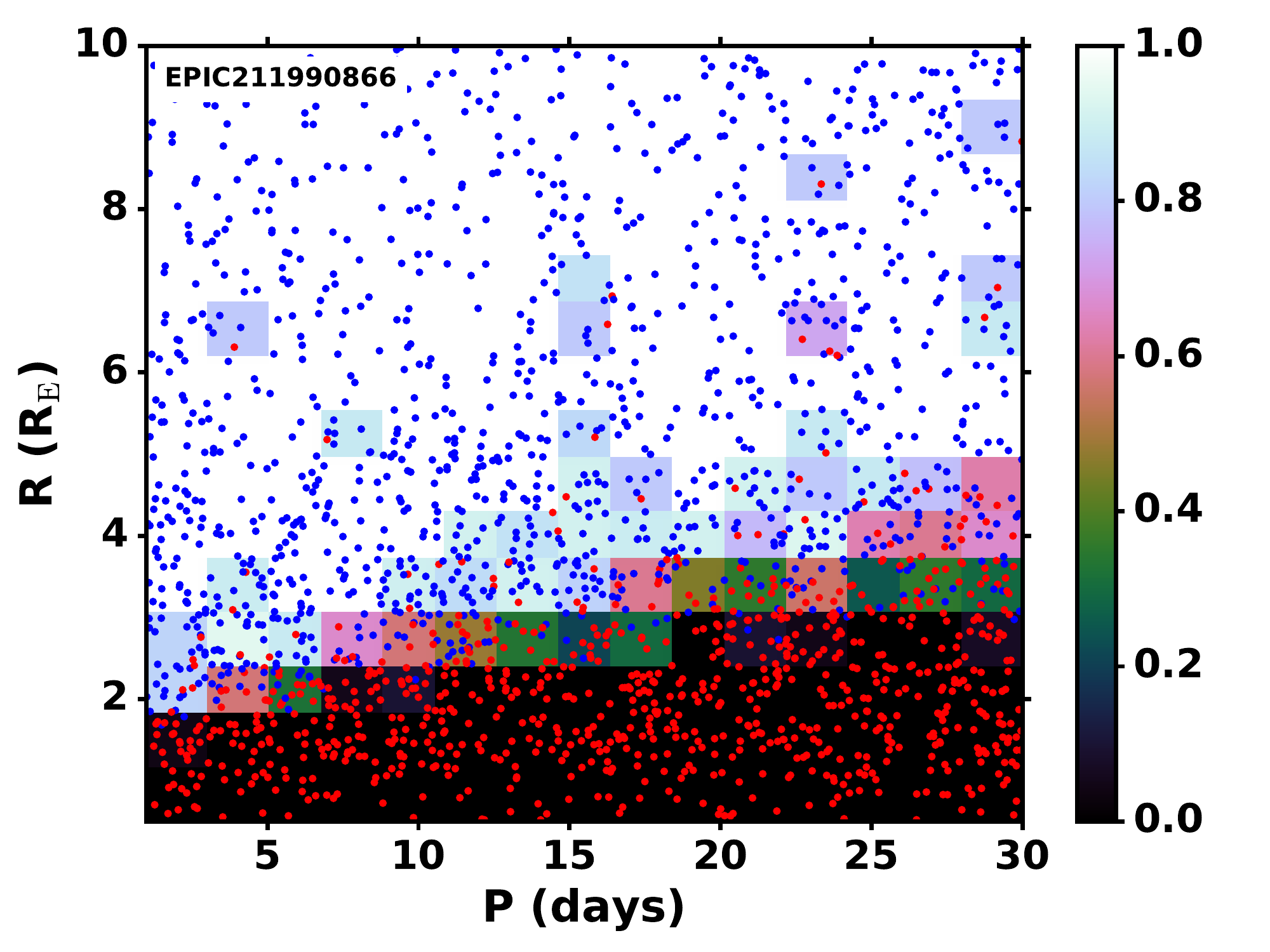}
\includegraphics[trim={0 0 5mm 0},clip,width=0.48\textwidth]{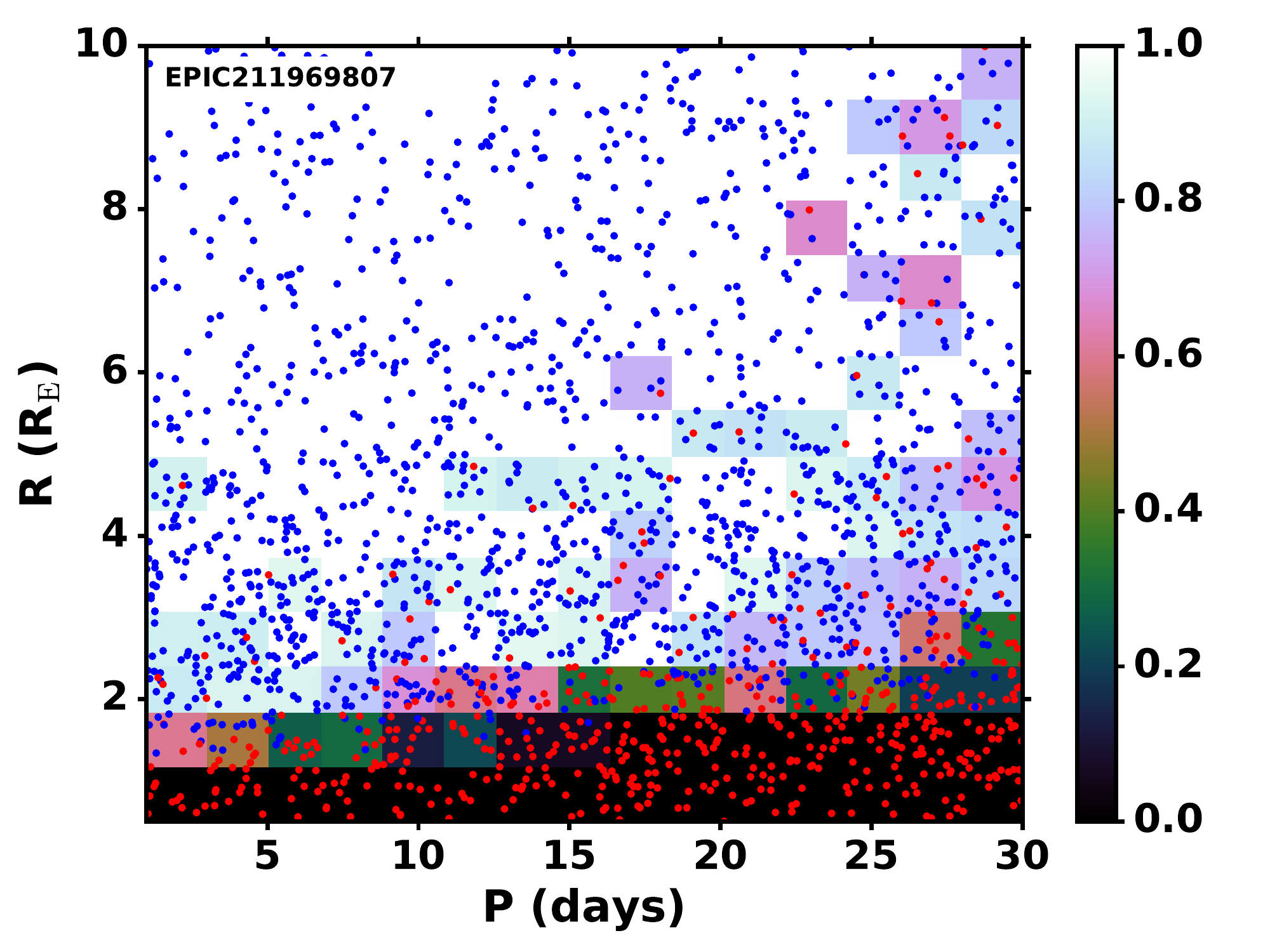}\\
\includegraphics[trim={0 0 5mm 0},clip,width=0.48\textwidth]{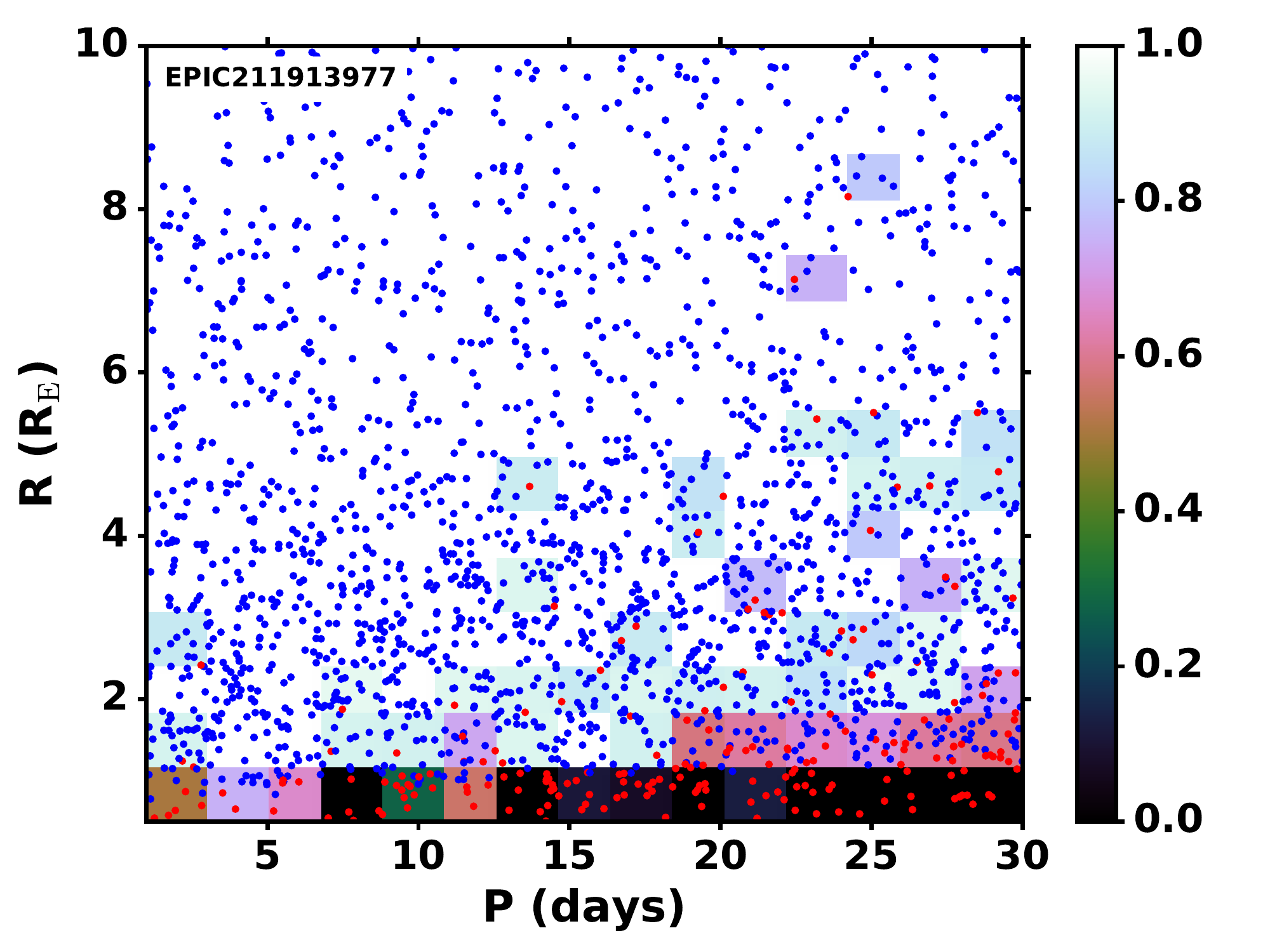}
\includegraphics[trim={0 0 5mm 0},clip,width=0.48\textwidth]{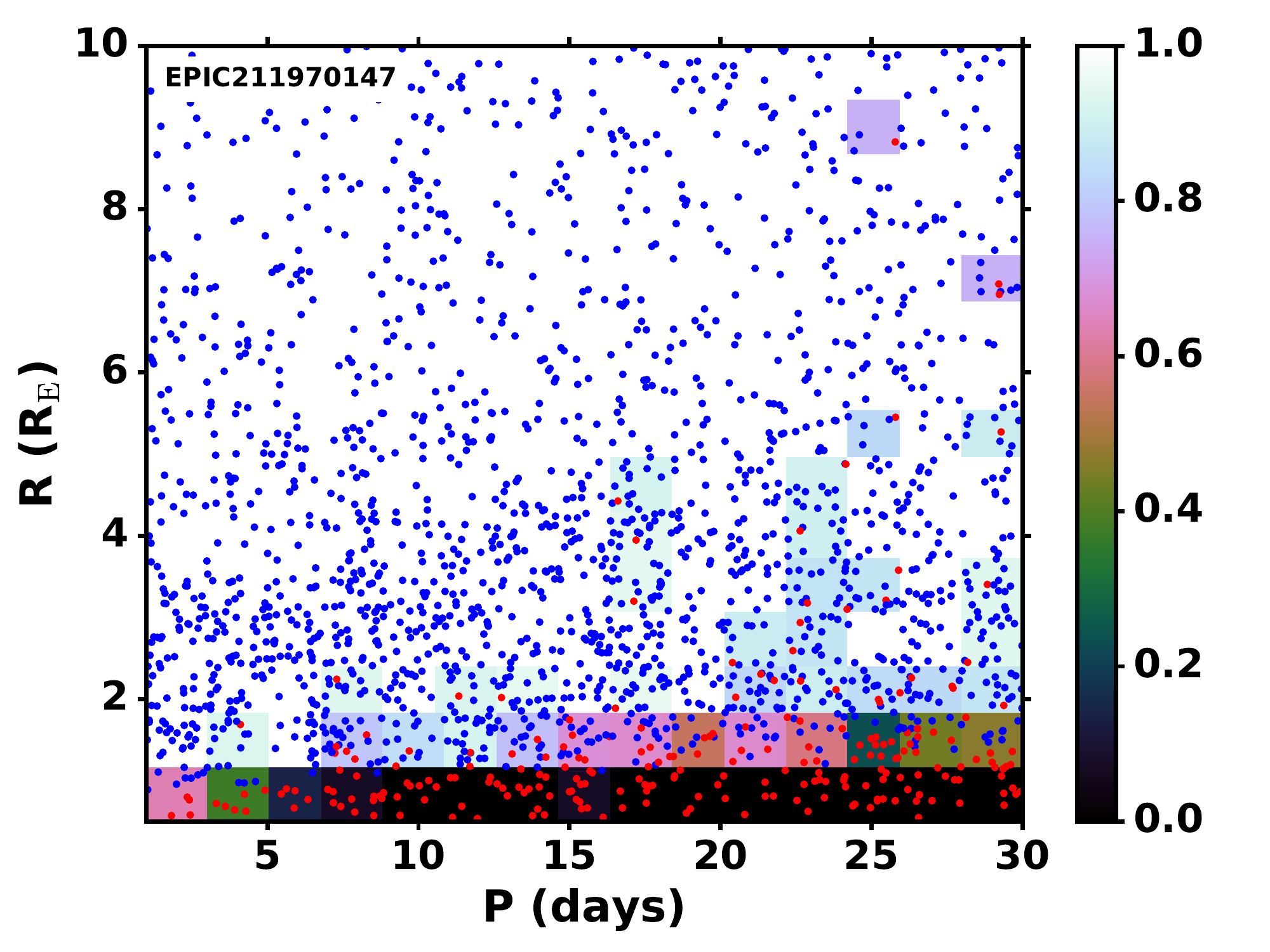}
\caption{Results of the injection-recovery tests for the six confirmed Praesepe exoplanet hosts. Recovered and missed injected planets are shown in blue and red respectively over a uniform grid interpolated from the injection tests shown in color. Planets with R$\gtrsim$2\,R$_\oplus$ are typically recovered over the range of periods tested.}
\label{injrec_plots}
\end{figure*}

\begin{figure*}
\centering
\includegraphics[trim={0 0 5mm 0},clip,width=0.48\textwidth]{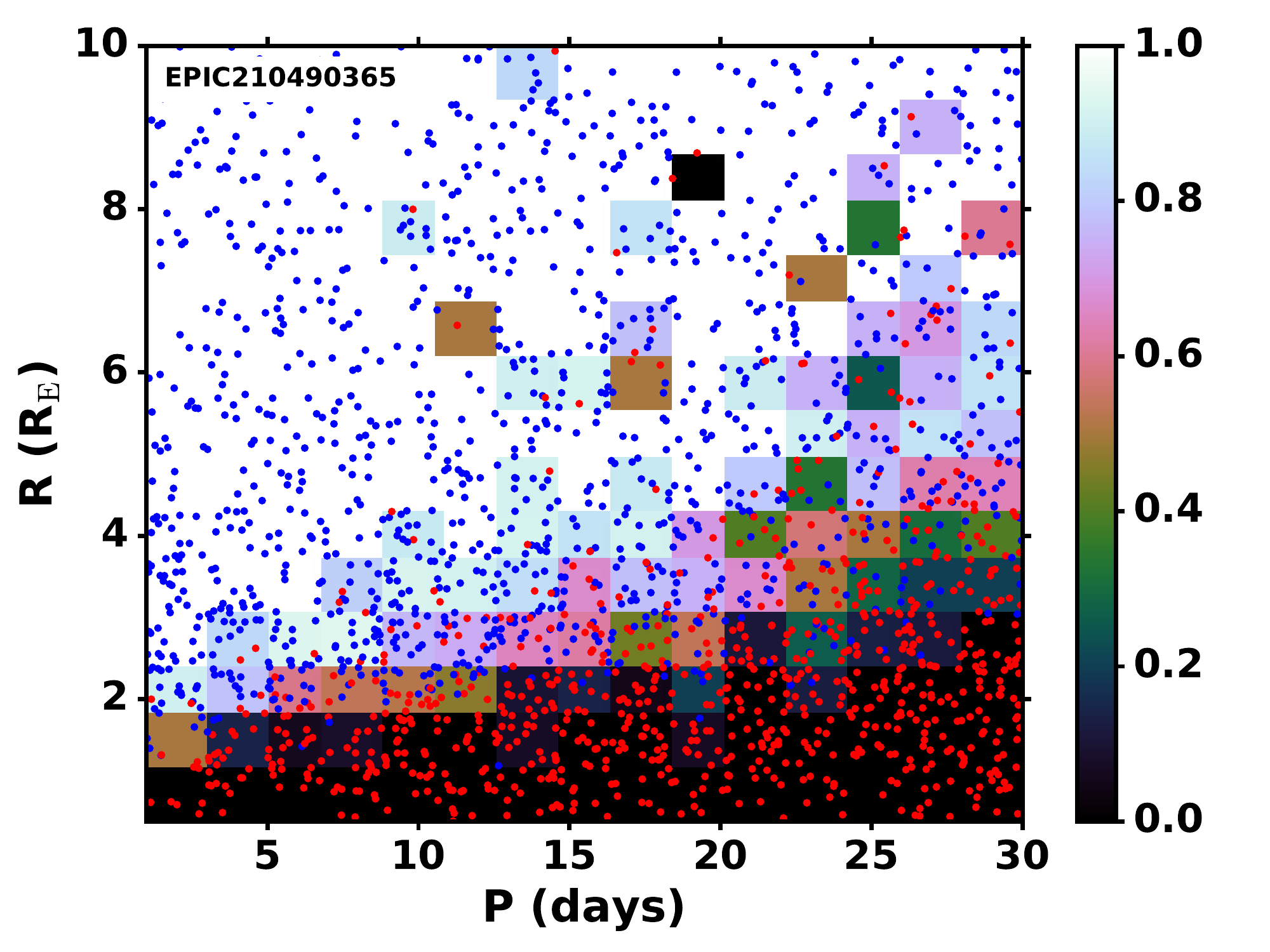}
\includegraphics[trim={0 0 5mm 0},clip,width=0.48\textwidth]{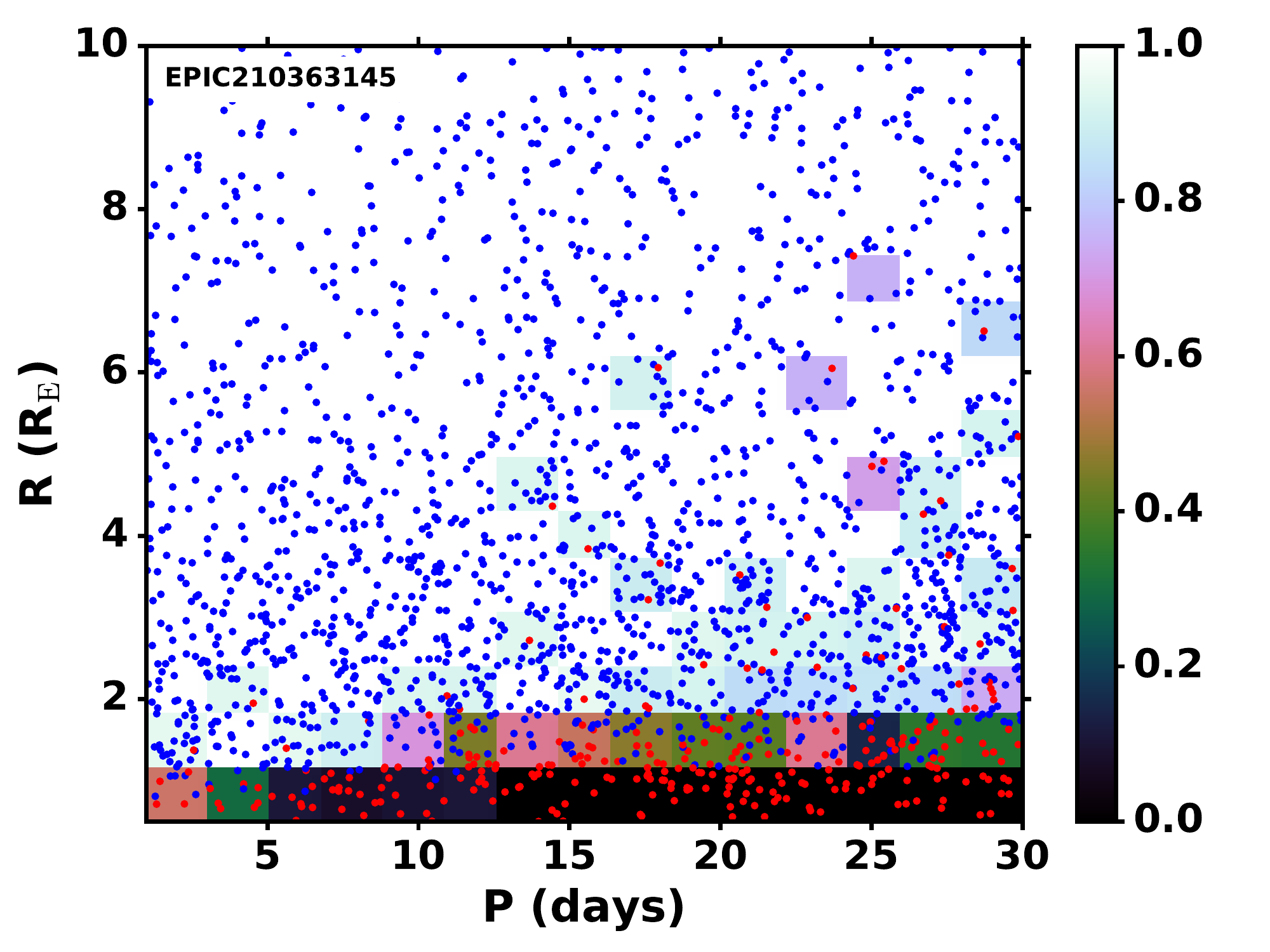}\\
\includegraphics[trim={0 0 5mm 0},clip,width=0.48\textwidth]{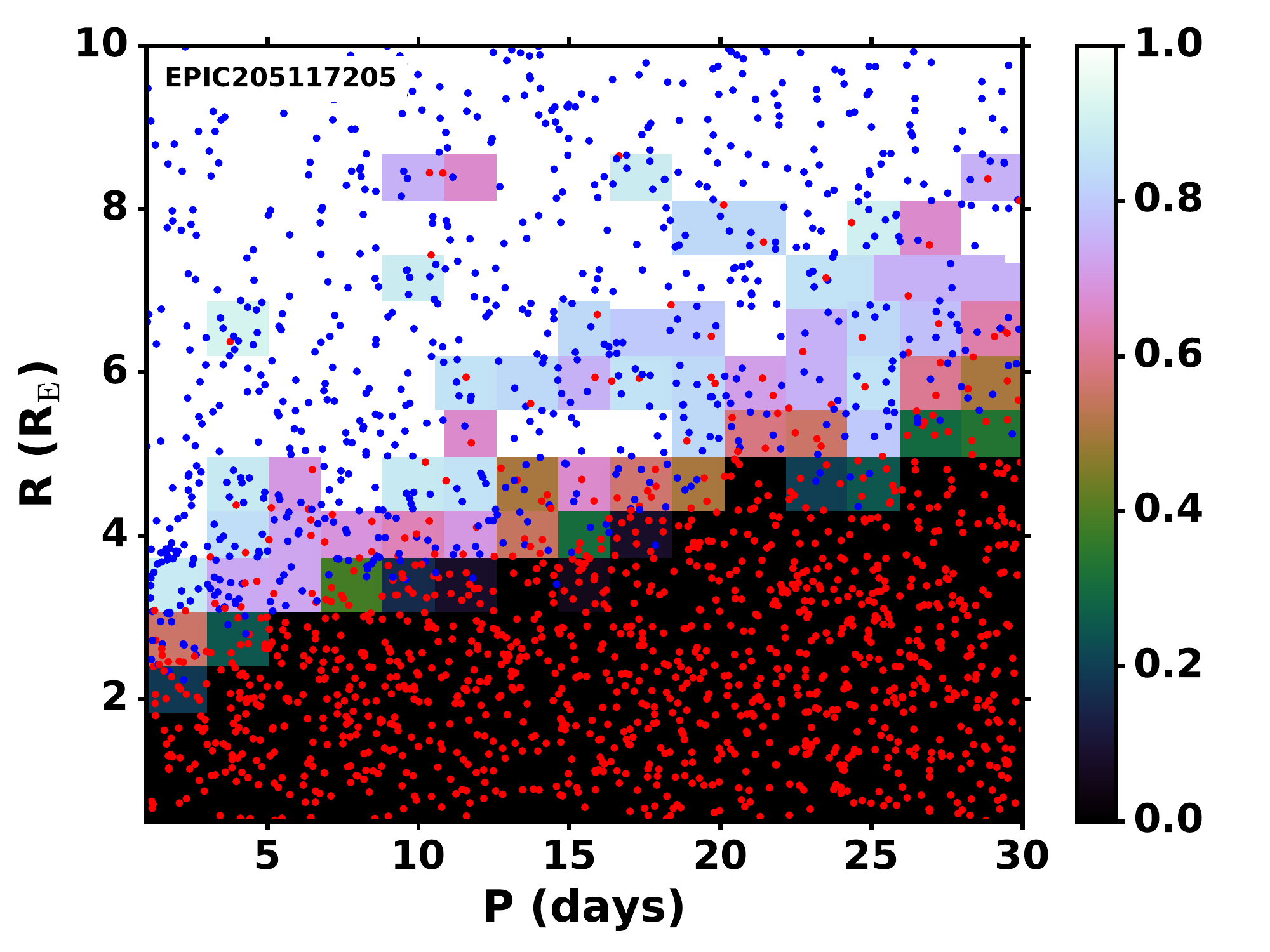}
\includegraphics[trim={0 0 5mm 0},clip,width=0.48\textwidth]{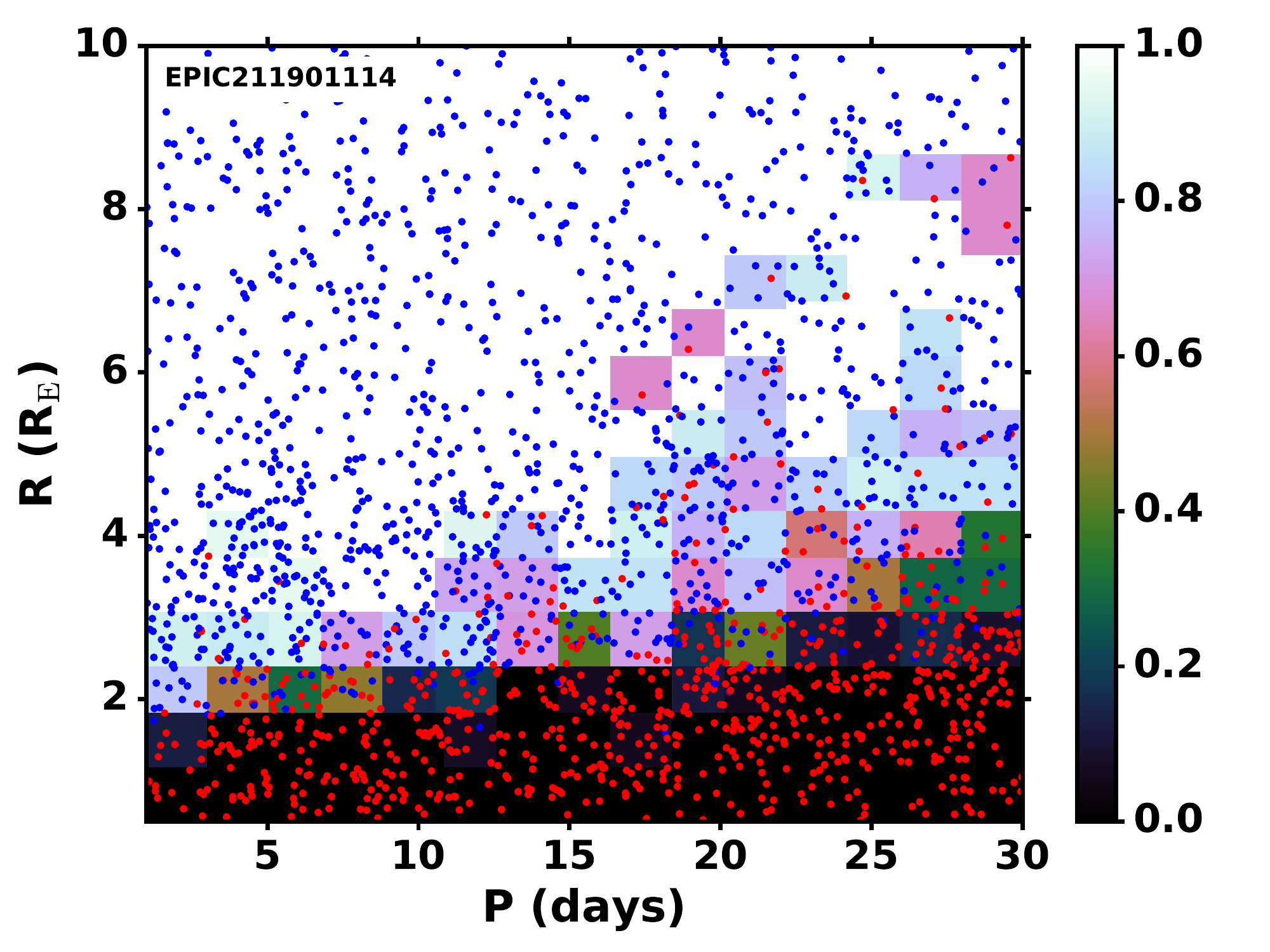}
\caption{Results of the injection-recovery tests for the Hyades (\emph{top left}), Pleiades non-member (\emph{top right}), Upper Sco (\emph{lower left}) and candidate Praesepe (\emph{lower right}) planet host. Recovered and missed injected planets are shown in blue and red respectively over a uniform grid interpolated from the injection tests shown in color. Additional planets with R$\gtrsim$2 and 4\,R$_\oplus$ are ruled out for EPIC 210490365 and 210363145 respectively. The youth, and hence PMS inflation of EPIC 205117205 accounts for the incompleteness below 5\,R$_\oplus$, though if young planets are also inflated, the observations may have similar sensitivity to the overall population of all planets.}
\vspace{5mm}

\label{injrec_plots2}
\end{figure*}

The 1\% period and phase tolerances were chosen such that for 1000 trial planets, $<$1 false alarm detections pass the tolerances. Figures \ref{injrec_plots} and \ref{injrec_plots2} show the detection efficiency maps for each ZEIT system as a function of injected orbital period and planet radius. We show both the individual injected trials, and the resulting binned completeness grid for each planet host.

For the eight exoplanet systems in the 600-800\,Myr old Praesepe and Hyades clusters, our pipeline is sensitive to additional planets with radii larger than $\gtrsim$1-2\,R$_\oplus$ for the M-type planet hosts, and $\gtrsim$3\,R$_\oplus$ for the larger F-type host EPIC211990866 (K2-100). For K2-33 (EPIC 205117205) our search is sensitive to larger transiting planets, with the boundary on detectability occurring in the range of 2-4\,R$_\oplus$, depending on orbital period.

To estimate the number of multiple systems we would expect in the young cluster observations given the sensitivity limits presented in Figures \ref{injrec_plots} and \ref{injrec_plots2} we take the archive of exoplanet detections from the original \emph{Kepler} mission \citep{nasa_exo_archive13}, and remove all systems with host star effective temperatures T$_\mathrm{eff}>6000$\,K, or surface gravities inconsistent with dwarf stars ($\log{g} < 3.5$). For the remaining systems we assign probability of detection as a single or multiple system for each exoplanet if it where hosted by each of the ZEIT planet hosts based on the injection-recovery results. We then count the number of detected systems and multiples. From this calculation we find that the expected number of multiple systems in the nine ZEIT planet hosting systems in 1.5$^{+1.2}_{-1.0}$, which is consistent with our detection of no multiple systems.

As mentioned in \citet{zeit4} the Praesepe planets detected in the \emph{K2} data appear to have larger radii, as a population, than the older late-type dwarf population from the original \emph{Kepler} mission. Similarly, the radius of the single planet detected in Upper Scorpius (K2-33\,b, 5.04\,R$_\oplus$) is highly uncommon for an $\sim$0.5\,M$_\odot$ host in the older populations. While the injection testing we present here for these systems cannot directly address the question of whether or not the young planet population features inflated radii, we can estimate the effect of a radius inflation on the expected number of multiples. For the extreme case of radii being twice as large on average than the \emph{Kepler} population, we find that the expected number of multiple transiting planets increases to 2$^{+1.5}_{-1.3}$ systems. This is still not significantly different from our null detection of additional planets. 

\section{Summary \& Conclusions}

We have presented two new methods for removing rotational stellar variability from lightcurves of active young stars that can be used in future to assess the occurrence of exoplanets at young ages. We reported the results of applying the methods to the members of the Upper Scorpius, Hyades, Pleiades and Praesepe clusters observed by the \emph{K2} mission in campaigns 2, 4 and 5. We also present a full kinematic membership selection for the four clusters in the \emph{K2} dataset for campaigns 2, 4 and 5, identifying members on the basis of proper motions, photometry, and parallaxes where appropriate.

We recover all known planets and eclipsing binary systems in our period-search range, as well as several non-planetary signals that were previously suggested to be created by orbiting clouds of material around Upper Scorpius members. We identify some additional candidate transiting exoplanets and show that these are either blended eclipsing binary systems or non-members in the clusters of interest. Two of these non-member planet candidates, the multi-planet system EPIC~210696763, and the 29.05\,day period EPIC~210736056, are previously unreported in the literature.

We also test the sensitivity of our pipeline to additional transiting exoplanets in the confirmed planet-hosting systems using an injection-recovery test. While we do not detect any additional transiting planets in these systems, we place strong limits on the range of radii and periods at which additional undetected transiting planets may still be identified. For systems in Praesepe and Hyades (600-800\,Myr) our method is sensitive to objects of 1-2\,R$_\oplus$ and larger,  and for the younger, more inflated Upper Scorpius planet host K2-33, we are sensitive to $\sim$3\,R$_\oplus$ transiting objects. We compare the completeness estimates to the \emph{Kepler}-prime planet distribution \citep{batalha13}, and find that the lack of any multiple systems among the nine planet hosts is consistent within the current completeness, even for the extreme case that young exoplanet radii are inflated by a factor of two compared to the population of exoplanets for older field-age stars. From the limited injection-recovery testing that we have applied to our pipeline, we achieve sensitivities that approach the peak of the exoplanet distributions for sun-like and cooler stars \citep{petigura13,dressing15}.

There are a number of avenues for improvement of the detrending pipeline we have presented here. Use of an actual transit model in place of the box-shaped notch that we employ would significantly improve the fit for cases where the ingress and egress of transits are resolved. In particular, the shorter cadence data that may be available from the future \emph{TESS} mission, or searches for longer period transits using return visits to cluster members in later \emph{K2} campaigns will benefit from a more suitable transit model. Furthermore, the Bayesian information criterion model selection can also be used to identify single transit events similar to the methods explored by \citet{foreman_mackey16} following the addition of a resolved transit model that includes ingress and egress points. Additionally, applying the notch-filter detrending as part of the phased transit search would also significantly improve sensitivity to small transiting planets, and allow simultaneous removal of the \emph{K2} roll and stellar variability signals by leveraging multiple detrending windows simultaneously.  This would also increase sensitivity to longer duration transits where degeneracy between the rotation and transit models in a single window becomes significant by fitting a common transit duration and depth across multiple period-phased windows.
 
Our pipeline will provide powerful utility for identifying transiting exoplanets at key ages in future \emph{K2} campaigns, \emph{TESS} observations, as well as future ground based transit searches (e.g. \citealt{yeti}). Data from \emph{K2} campaign 13, which targets the $\sim$2\,Myr old Taurus-Auriga star-forming region and additional Hyades members has already been downloaded and will soon be available, and future campaigns 15 and 16 will revisit Upper Scorpius and Praesepe, increasing the sample of observed members in each cluster by a factor of 50\%, and allowing searches for longer period transiting exoplanets for many revisited targets. \emph{TESS} will also observe the full population of nearby young moving-group members which will comprise some thousands of stars younger than $\sim$100\,Myr, many of which are yet to be identified.With these increases in sample size, we can more strongly address the prevalence of multi-planet systems in the young clusters. Additionally, both the \emph{K2} and \emph{TESS} mission will observe significant numbers of rotationally variable stars that are unassociated with any known cluster or association, but whose ages can be approximated through gyrochronology (e.g. \citealt{mamajek08}). This sample may provide a statistical population at ages not otherwise sampled by nearby young moving-groups or open clusters.

Extension of the injection-recovery testing described here to the full population of the young clusters and associations observed by \emph{K2} will directly inform key planet formation questions. With completeness information, it will be possible to assess whether the trend of large-radius exoplanets suggested in the Hyades and Praesepe planet samples of \citet{zeit1,zeit4} relative to the original \emph{Kepler} missions holds, and thus constrain the prevalence and rate of mass-loss in the first $\sim$Gyr of planetary evolution \citep{fultongap}.  Additionally, measurement of the occurrence rate of exoplanets as a function of cluster or stellar association age will provide the first direct measurement of the close-in exoplanet migration timescale, which will rule-out or confirm different mechanisms as the dominant method for exoplanet migration (e.g., disk-migration; \citealt{lubow_migration} or dynamical migration; \citealt{fabrycky07,chatterjee08}).

 \nocite{pecaut16,luhman12,aarnio08,preibisch98,chen11,martin04,gregorio-hetem92,meyer93,kunkel2000,martin98,bourvier92,kraushill09,lodieu06,erickson11,slesnik06,wilking05,alvesdeoliveira10,cieza10,luhman99,doppmann03,natta02,elias78oph,walter94,ardila2000,lodieu11,martin98_2,lodieu08,preibisch02,herbig88,vieira03}

\section*{Acknowledgments}
This paper includes data collected by the K2 mission. Funding for the K2 mission is provided by the NASA Science Mission directorate.
This research has made use of the Exoplanet Follow-up Observation Program website, which is operated by the California Institute of Technology, under contract with the National Aeronautics and Space Administration under the Exoplanet Exploration Program. This work used the Immersion Grating Infrared Spectrometer (IGRINS) that was developed under a collaboration between the University of Texas at Austin and the Korea Astronomy and Space Science Institute (KASI) with the financial support of the US National Science Foundation under grant AST-1229522, of the University of Texas at Austin, and of the Korean GMT Project of KASI. The authors acknowledge the Texas Advanced Computing Center (TACC) at The University of Texas at Austin for providing HPC resources that have contributed to the research results reported within this paper\footnote{http://www.tacc.utexas.edu}. This research has made use of the NASA Exoplanet Archive, which is operated by the California Institute of Technology, under contract with the National Aeronautics and Space Administration under the Exoplanet Exploration Program. ACR was supported [in part] by NASA K2 Guest Observer Cycle 4 grant NNX17AF71G. AWM was supported through NASA Hubble Fellowship grant 51364 awarded by the Space Telescope Science Institute, which is operated by the Association of Universities for Research in Astronomy, Inc., for NASA, under contract NAS 5-26555. This work was performed [in part] under contract with the California Institute of Technology (Caltech)/Jet Propulsion Laboratory (JPL) funded by NASA through the Sagan Fellowship Program executed by the NASA Exoplanet Science Institute. We would also like to thank the anonymous referee for helping to improve this publication.

\software{K2SFF (Vanderburg \& Johnson 2014), BATMAN (Kreidberg 2015)}
\facilities{Kepler, Smith (IGRINS), Texas Advanced Computing Center}

\bibliographystyle{apj}
\bibliography{master_reference}

\input{detfigurecode_nsfw.tex}

\input{detfigurecode_lcr.tex}

\end{document}

%% file: detection_table_nsfw.tex
210490365 & Hy & 04 13 05.62 & +15 14 51.9 & 3.4843 & 57062.58461 & 7.2 &1.88 & K2-25, ZEIT-I \\  
210827030 & Hy & 04 07 03.25 & +20 16 50.9 & 0.5111 & 57063.518630 & 7.7 & & EB \\ 
211705654 & Pr & 08 45 05.67 & +15 59 23.9 & 2.5331 & 57138.444062 & 1.8 & & EB \\ 
211804579 & Pr & 08 36 16.27 & +17 22 54.0 & 1.5235 & 57139.628845 & 0.6 & & NS \\ 
211822797 & Pr & 08 41 38.49 & +17 38 24.0 & 21.1595 & 57123.305645 & 0.7 & 14.66& K2-103, ZEIT-IV\\  
211901114 & Pr & 08 41 35.69 & +18 44 35.0 & 1.6488 & 57139.182137 & 1.9 &8.61 & Zeit-IV \\ 
211913977 & Pr & 08 41 22.58 & +18 56 02.0 & 14.6843 & 57137.980878 & 0.5 & 10.62 &K2-101, ZEIT-IV\\  
211916756 & Pr & 08 37 27.06 & +18 58 36.1 & 10.1338 & 57130.619456 & 4.8 &23.86 & K2-95, ZEIT-IV\\ 
211946007 & Pr & 08 42 39.44 & +19 24 52.0 & 1.9828 & 57139.152776 & 106.3&2.25 & EB \\ 
211969807 & Pr & 08 38 32.82 & +19 46 25.8 & 1.9741 & 57140.372898 & 1.1 & 17.08&K2-104, ZEIT-IV \\ 
211970147 & Pr & 08 40 13.45 & +19 46 43.7 & 9.9147 & 57139.658287 & 0.2 &11.60& K2-102, ZEIT-IV\\ 
211972086 & Pr & 08 50 49.84 & +19 48 36.5 & 6.0162 & 57142.867 & 179.3 &7.49& EB\\  
211990866 & Pr & 08 38 024.3 & +20 06 21.8 & 1.6739 & 57139.027323 & 0.6 & 4.26&K2-100, ZEIT-IV\\  
212002525 & Pr & 08 39 42.03 & +20 17 45.1 & 11.6158 & 57135.991382 & 100.4 &12.63& EB\\  
212009427 & Pr & 08 31 29.87 & +20 24 37.5 & 0.77845 & 57141.040820 & 4.2 & 1.56&EB\\ 
210363145 & Pl & 03 40 54.82 & +12 34 21.4 & 8.2008 & 57062.6123 & 0.7 && K2-75, ZEIT-II\\  %
210696763 & Pl & 03 31 020.3 & +18 18 58.4 & 3.6494 & 57061.51111 & 1.1 && Non-mem\\ 
210696763 & Pl & 03 31 020.3 & +18 18 58.4 & 5.7663 & 57062.461715 & 1.0 && Non-mem\\ 
210696763 & Pl & 03 31 020.3 & +18 18 58.4 & 7.98221 & 57059.963262 & 1.7 && Non-mem\\ 
210736056 & Pl & 03 37 35.45 & +18 53 47.1 & 29.0529 & 57044.194729 & 2.2 && Candidate Member\\ 
211082420 & Pl & 03 47 29.45 & +24 17 18.0 & 2.4616 & 57065.38 & 85.0 && EB\\ 
211093684 & Pl & 03 49 42.28 & +24 27 46.8 & 7.0502 & 57057.679115 & 40.6 &7.27& EB\\ 
203476597 & US & 16 25 57.92 & -26 00 37.4 & 1.4408 &56892.938242 & 17.9 &3.23& EB\\  
203710387 & US & 16 16 30.68 & -25 12 20.2 & 2.8090 & 56894.71338 & 52.6 & 2.56&EB\\  
203849738 & US & 16 09 52.88 & -24 41 53.5 & 0.6199 & 56894.6 & 27.7 &0.62& NP\\  
204165788 & US & 16 25 35.08 & -23 24 18.8 & 0.7474 &56893.140594 & 1.8 && EB\\ 
204143627 & US & 16 15 59.25 & -23 29 36.3 & 1.124 & 56893.726808 & 6.7 &1.12& NP\\  
204364515 & US & 16 01 21.56 & -22 37 26.5 & 1.4562 & 56894.298093 & 18.1 &1.46& NP\\ 
204432860 & US & 16 02 00.39 & -22 21 23.9 & 2.8742 & 56893.40870& 8.0 &2.87& EB\\ 
204506777 & US & 16 07 17.79 & -22 03 36.6 & 1.6281 & 56893.45074 & 16.3 && EB\\  
204760247 & US & 15 57 40.46 & -20 58 59.1 & 9.2048 & 56903.57 & 173.8 && EB\\ 
204882444 & US & 15 55 05.13 & -20 26 07.8 & 1.1488 & 56893.455574 & 10.7 &0.38& NP\\  
205046529 & US & 16 10 26.39 & -19 39 51.1 & 1.8361 & 56892.63254 & 6.0 &2.56& NP\\  
205117205 & US & 16 10 14.74 & -19 19 09.4 & 5.4224 & 56898.6705 & 2.0 &6.29& K2-33, ZEIT-III\\  
205483258 & US & 16 23 24.55 & -17 17 27.1 & 5.6683 & 56890.422916 & 72.9&5.66 & NP\\  
202963882 & US & 16 13 18.91 & -27 44 02.5 & 0.6308 & 56893.2333 & 61.6 && EB\\  
203692610 & US & 16 10 31.63 & -25 16 01.7 & 1.8204 & 56893.295249 & 5.4 &1.82& NP\\ 
203823381 & US & 16 28 32.89 & -24 47 54.9 & 8.2781 & 56893.602256 & 2.8 && Non-mem\\ 
203868608 & US & 16 17 18.99 & -24 37 18.7 & 4.5417 & 56893.768268 & 81.0 &5.66& EB\\  
204276894 & US & 16 34 39.22 & -22 58 14.8 & 1.957 & 56893.373134 & 0.1 && Non-mem\\ 
204750116 & US & 15 55 16.73 & -21 01 36.6 & 23.4302 & 56875.443931 & 0.5 && Non-mem\\ 

%% file: detection_table_lcr.tex
210490365 & Hy & 04 13 05.62 & +15 14 51.9 & 3.4843 & 57062.58461 & 7.2 &1.88& K2-25, ZEIT-I \\ 
211946007 & Pr & 08 42 39.44 & +19 24 52.0 & 1.9829 & 57139.1475 & 102.2 &2.25& EB \\ 
212009427 & Pr & 08 31 29.87 & +20 24 37.5 & 0.77845 &  57141.040852 & 4.2 &1.56& EB \\ 
203710387 & UsK & 16 16 30.68 & -25 12 20.2 & 2.809 & 56894.71339 & 51.2 &2.56& EB\\ 
205046529 & UsK & 16 10 26.39 & -19 39 51.1 & 1.8367 & 56892.627982 & 1.6 &2.56& NP\\ 

%% file: detfigurecode_nsfw.tex
\begin{figure*} 
\centering 
\includegraphics[width= 0.35\textwidth]{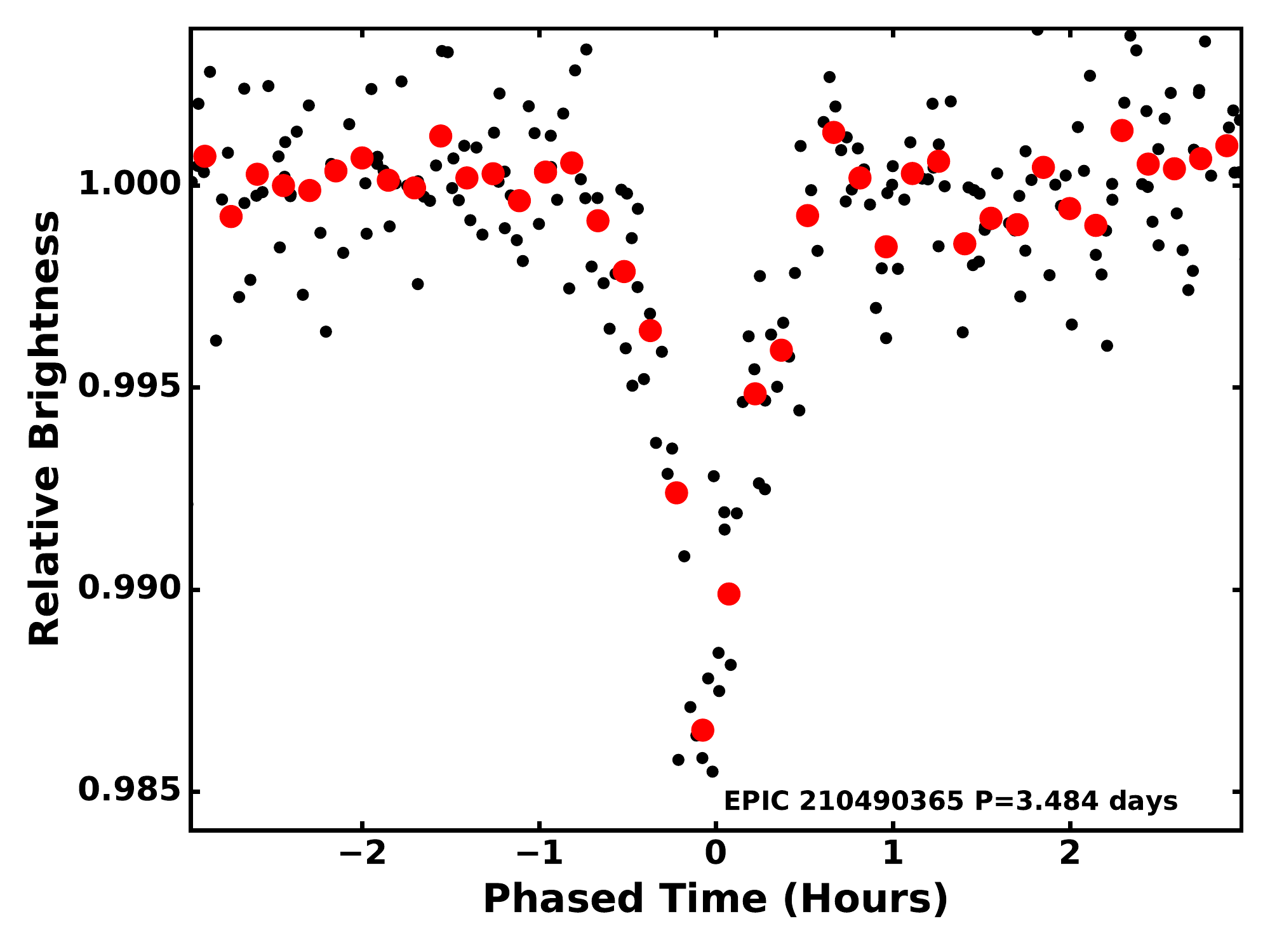} 
\includegraphics[width= 0.35\textwidth]{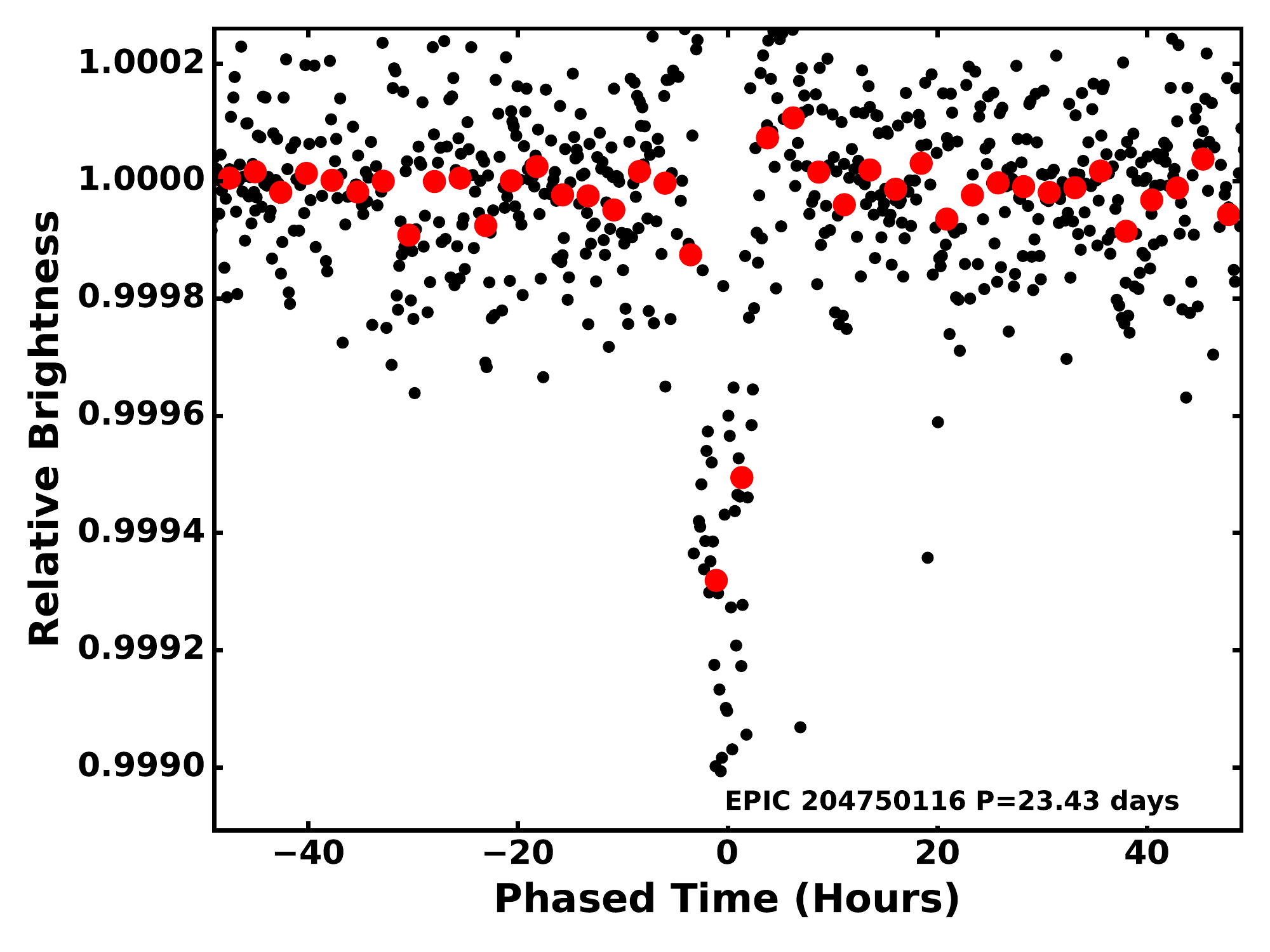} \\  
\includegraphics[width= 0.35\textwidth]{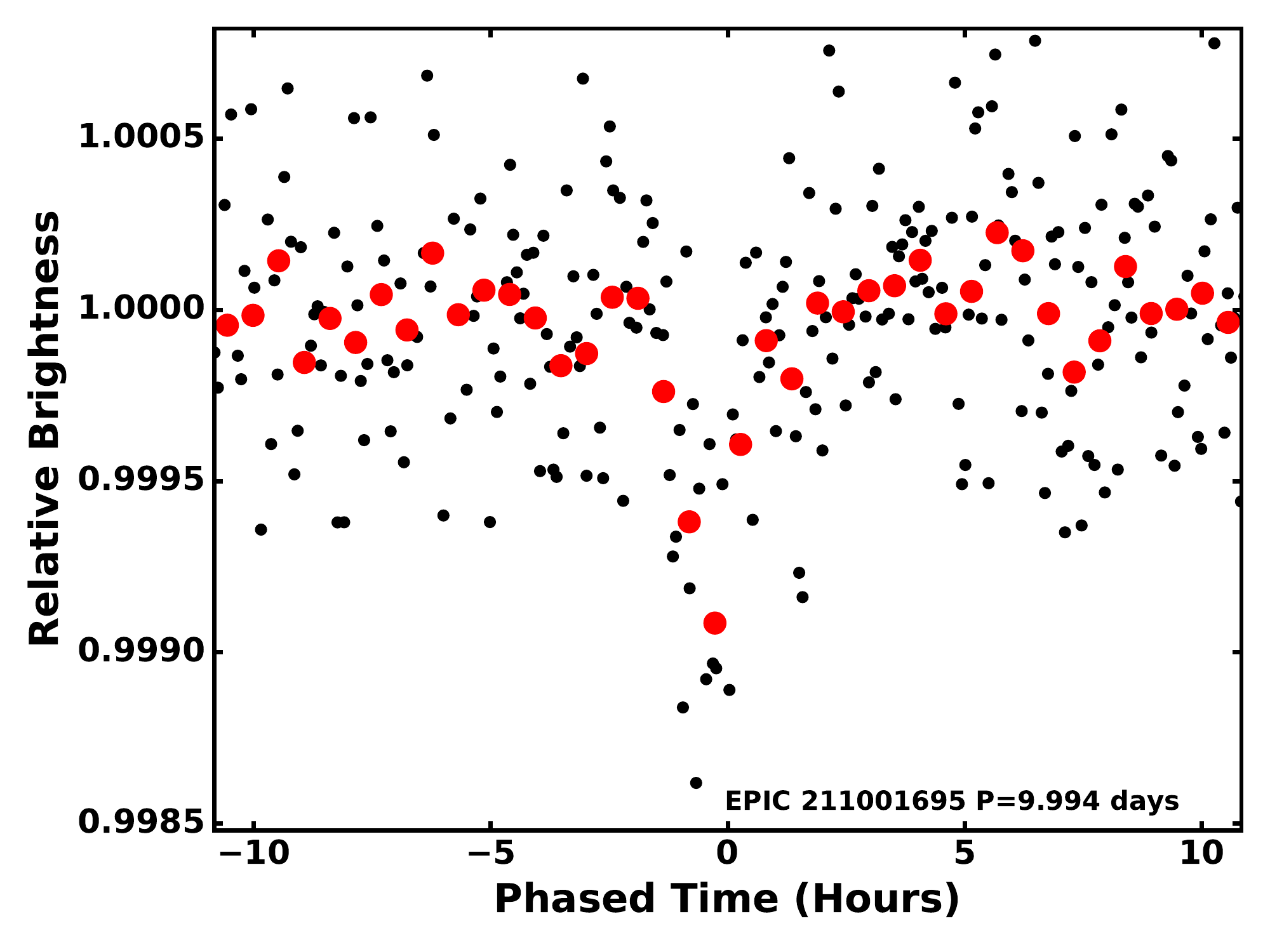} 
\includegraphics[width= 0.35\textwidth]{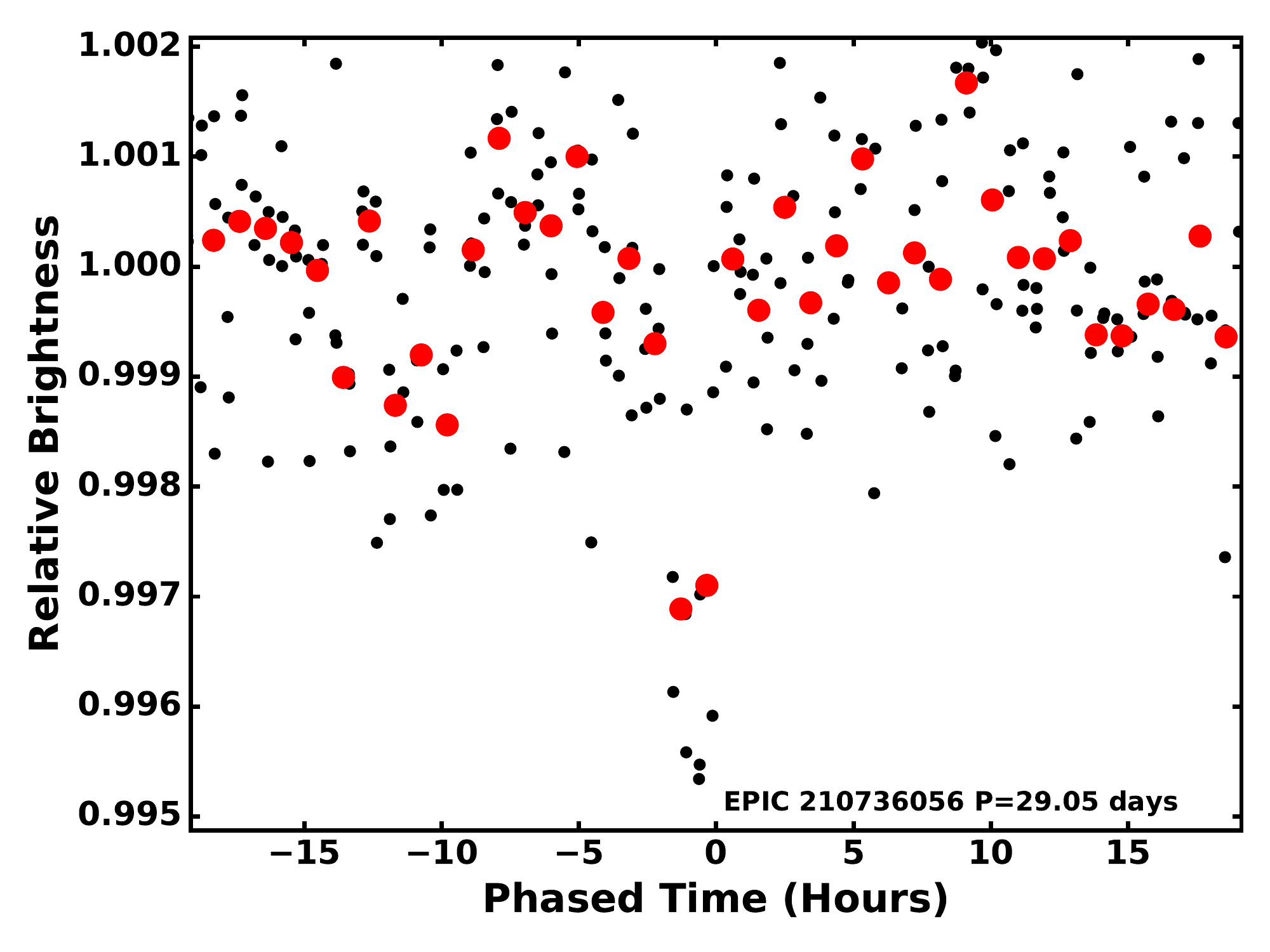} \\  
\includegraphics[width= 0.35\textwidth]{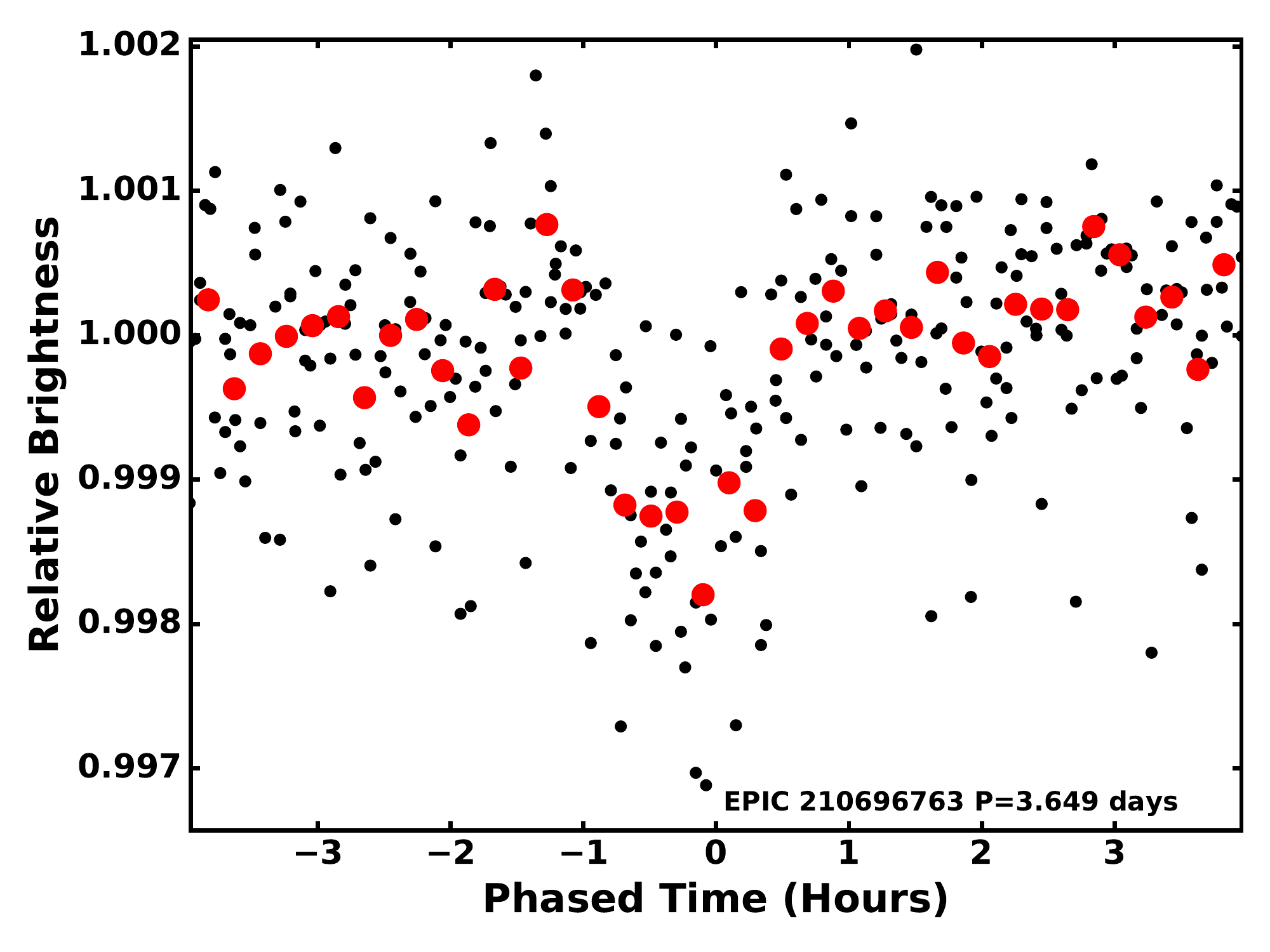} 
\includegraphics[width= 0.35\textwidth]{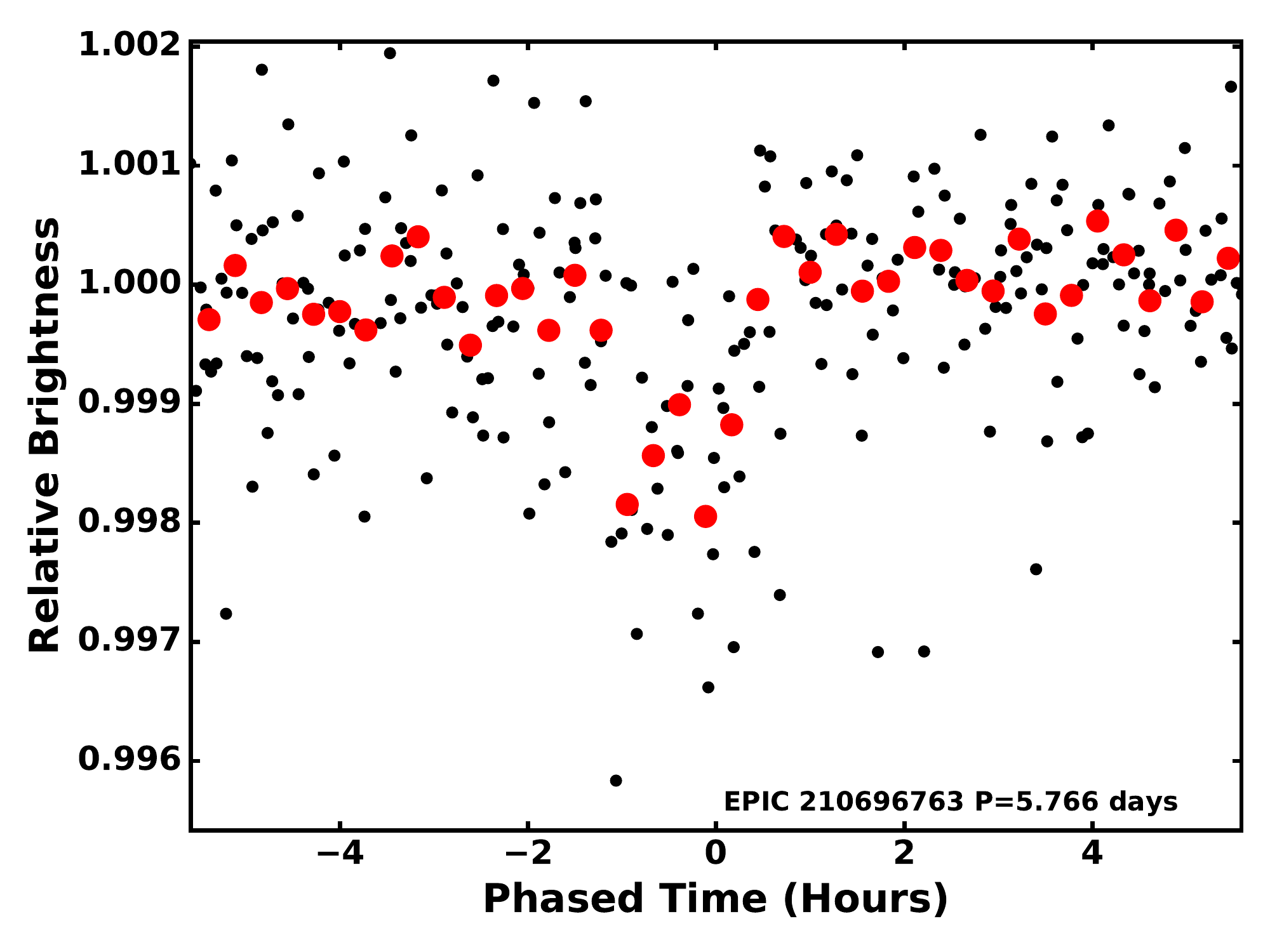} \\  
\includegraphics[width= 0.35\textwidth]{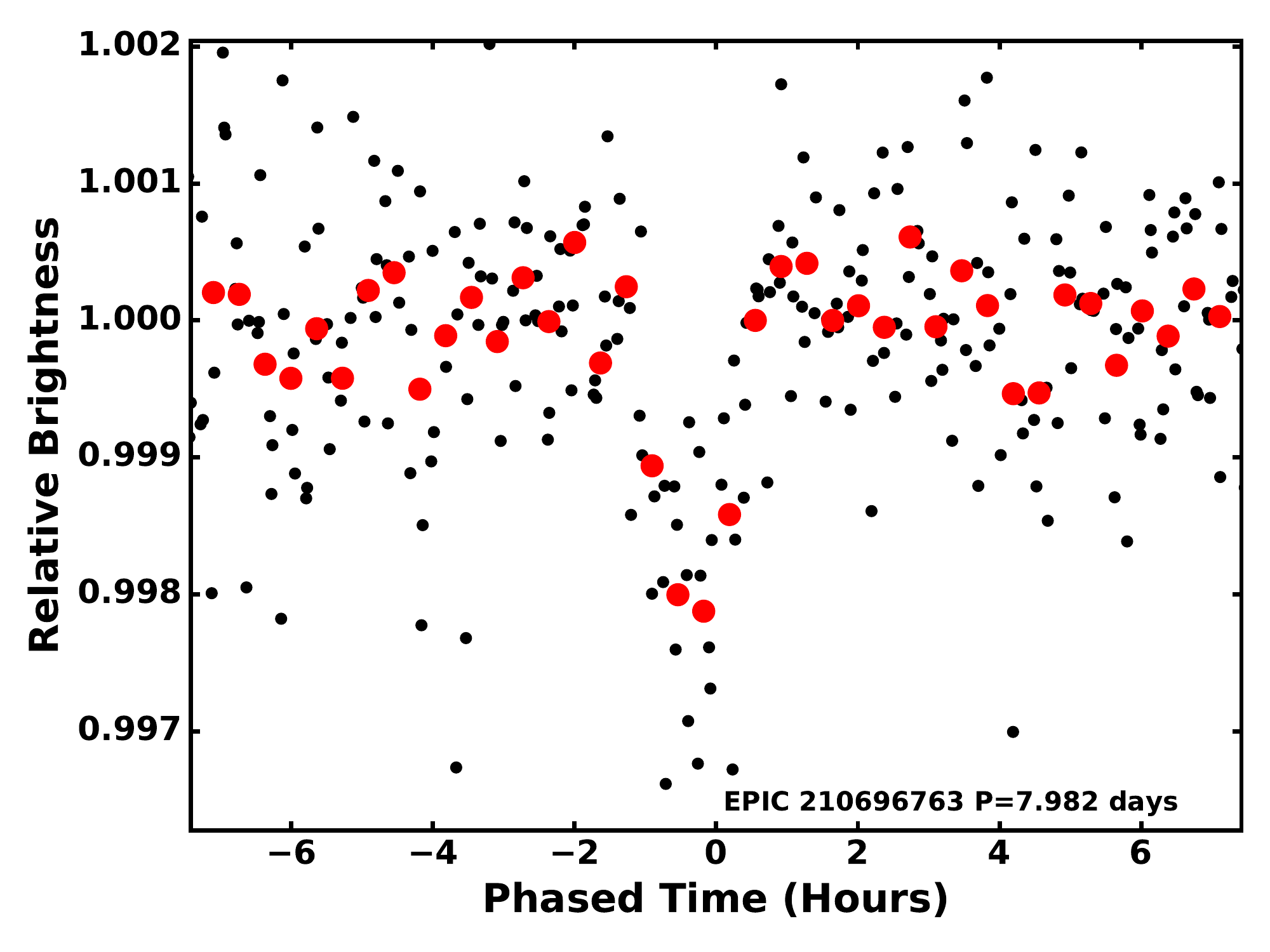} 
\includegraphics[width= 0.35\textwidth]{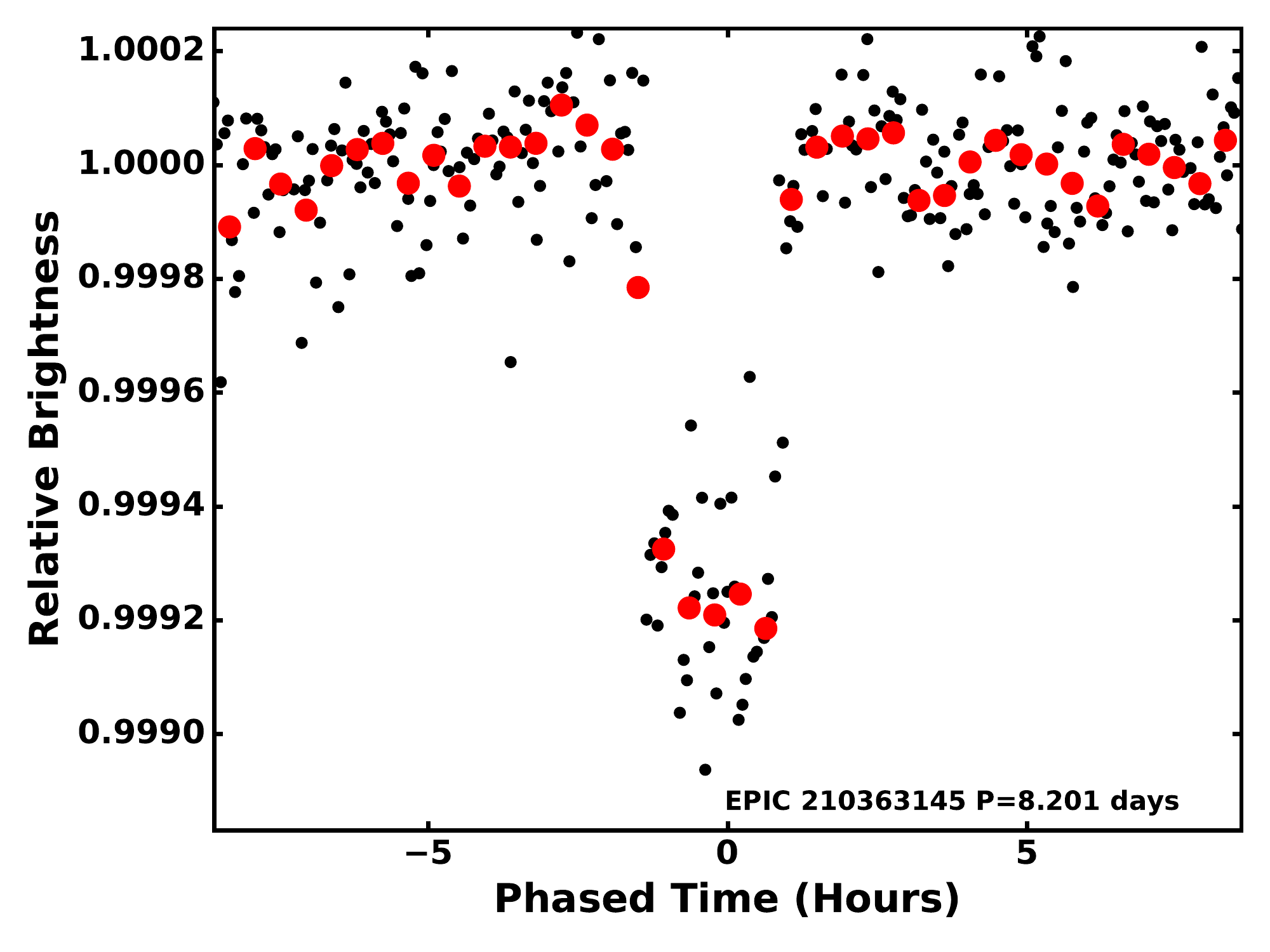} \\  
\includegraphics[width= 0.35\textwidth]{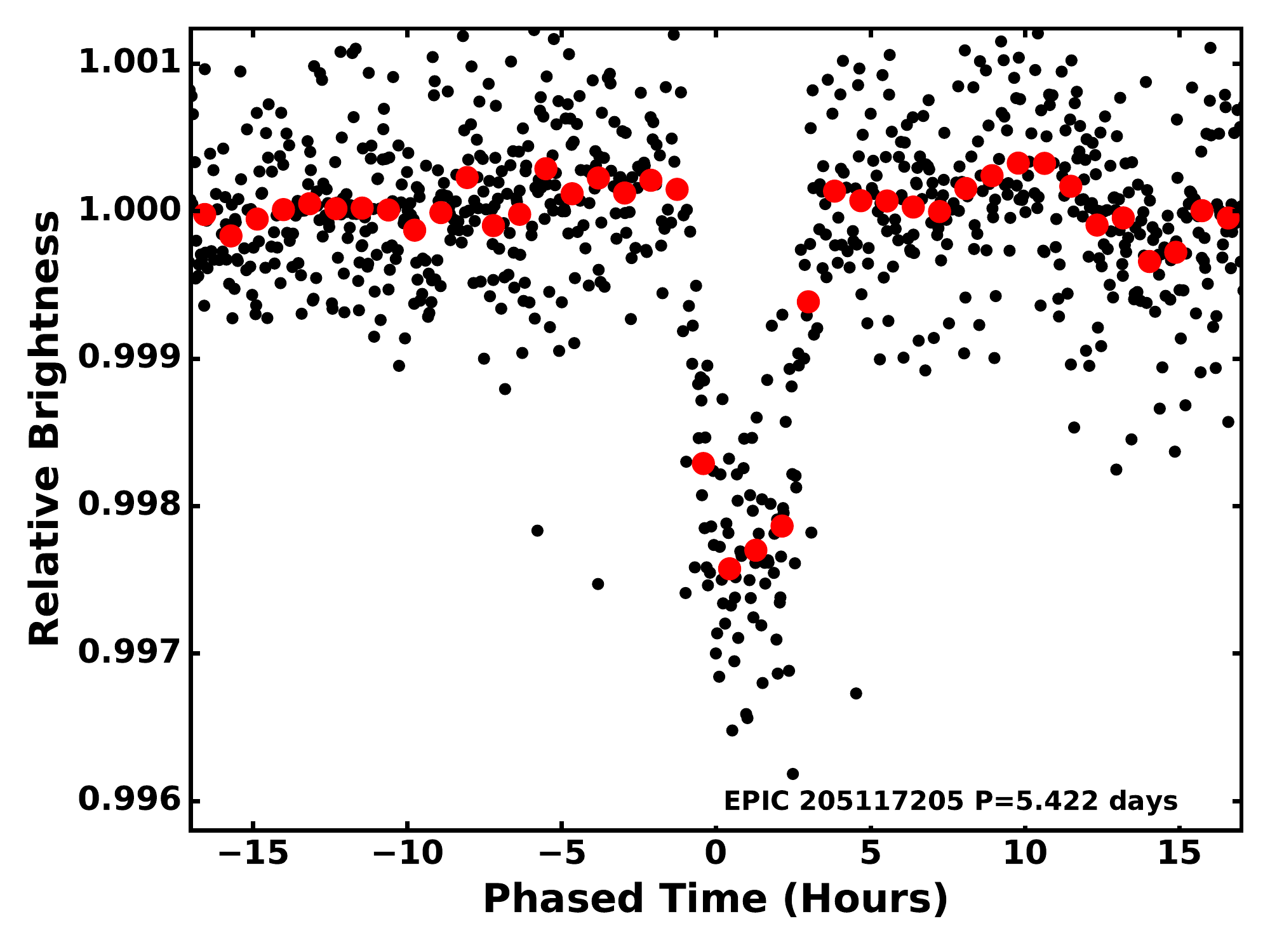} 
\includegraphics[width= 0.35\textwidth]{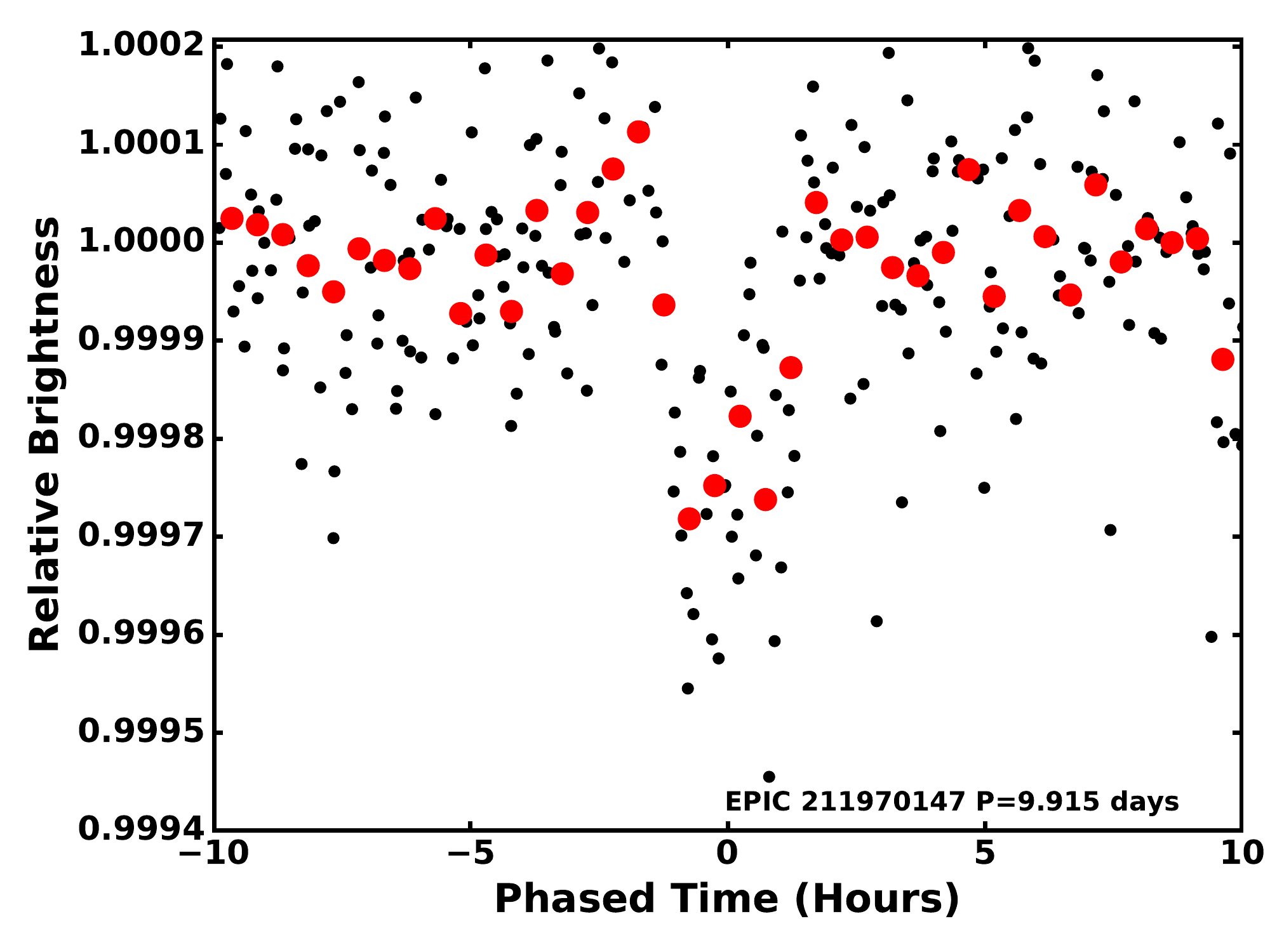} \\  
\caption{Periodic signals detected with the notch-filter pipeline that appear planetary, phased to the detected period. Black points are the 
detrended \emph{K2} observations, and red circles the binned lightcurve.} 
\label{figs:det1} 
\end{figure*}

\begin{figure*} 
\centering 
\includegraphics[width= 0.35\textwidth]{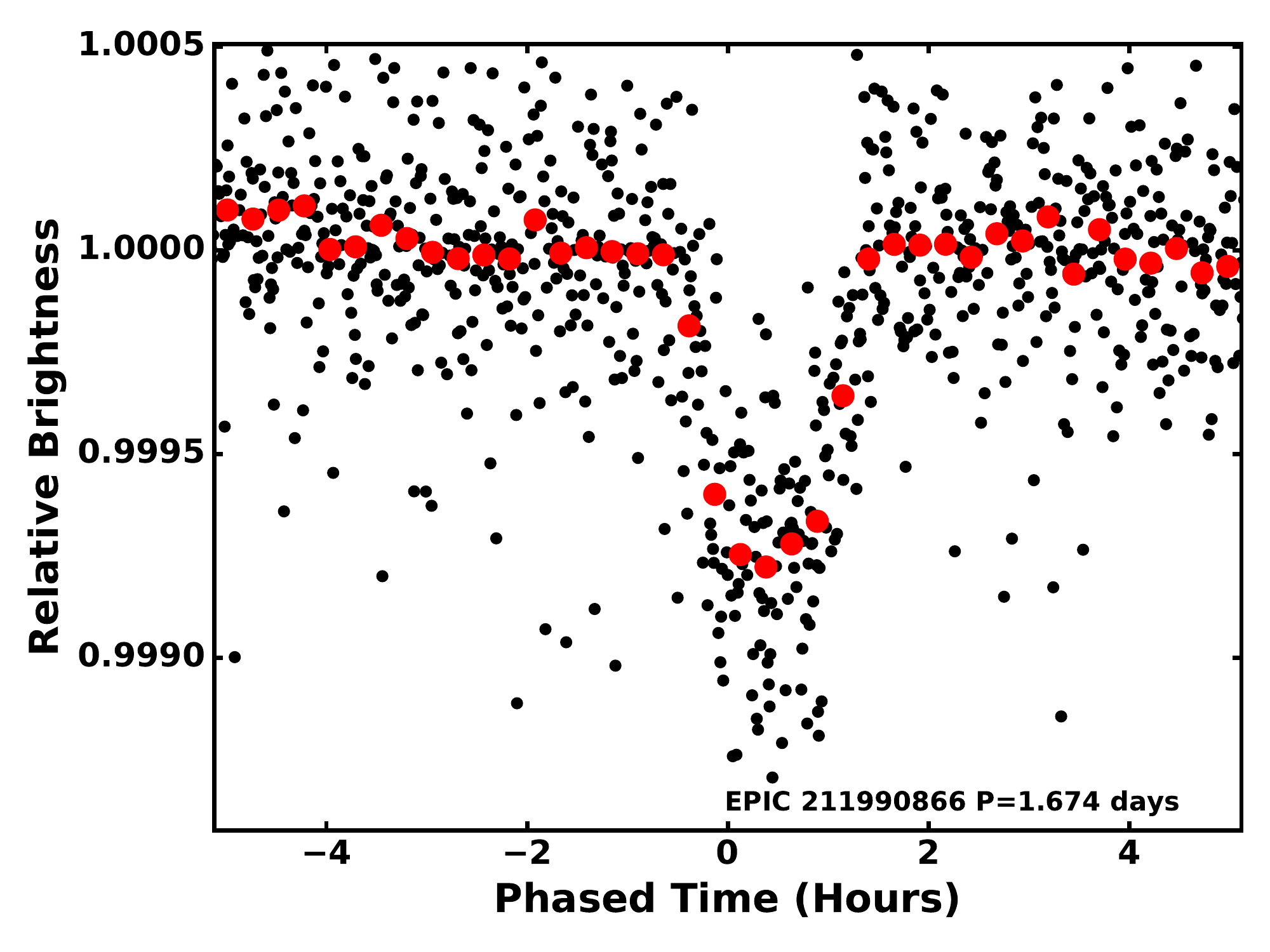} 
\includegraphics[width= 0.35\textwidth]{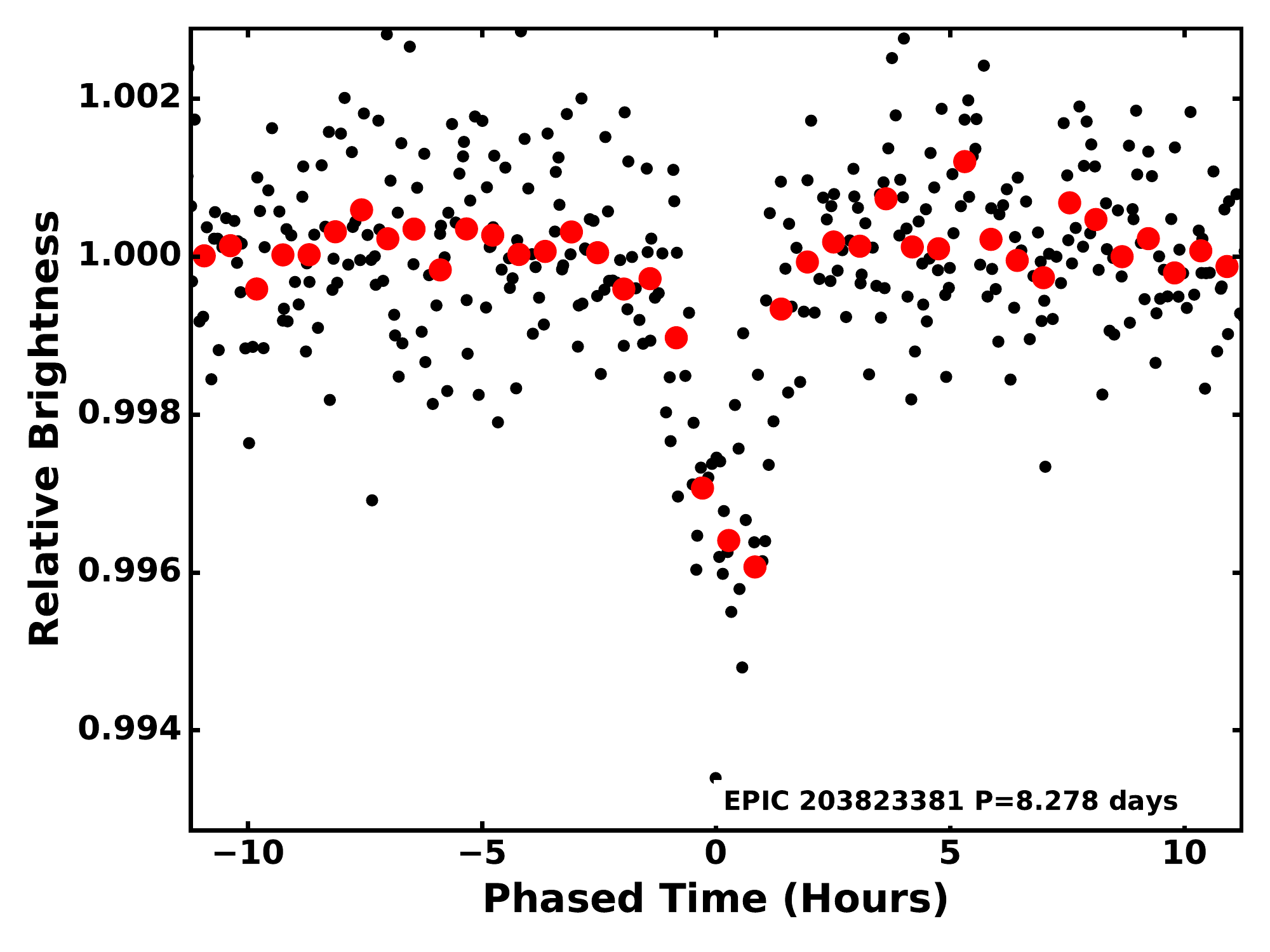} \\  
\includegraphics[width= 0.35\textwidth]{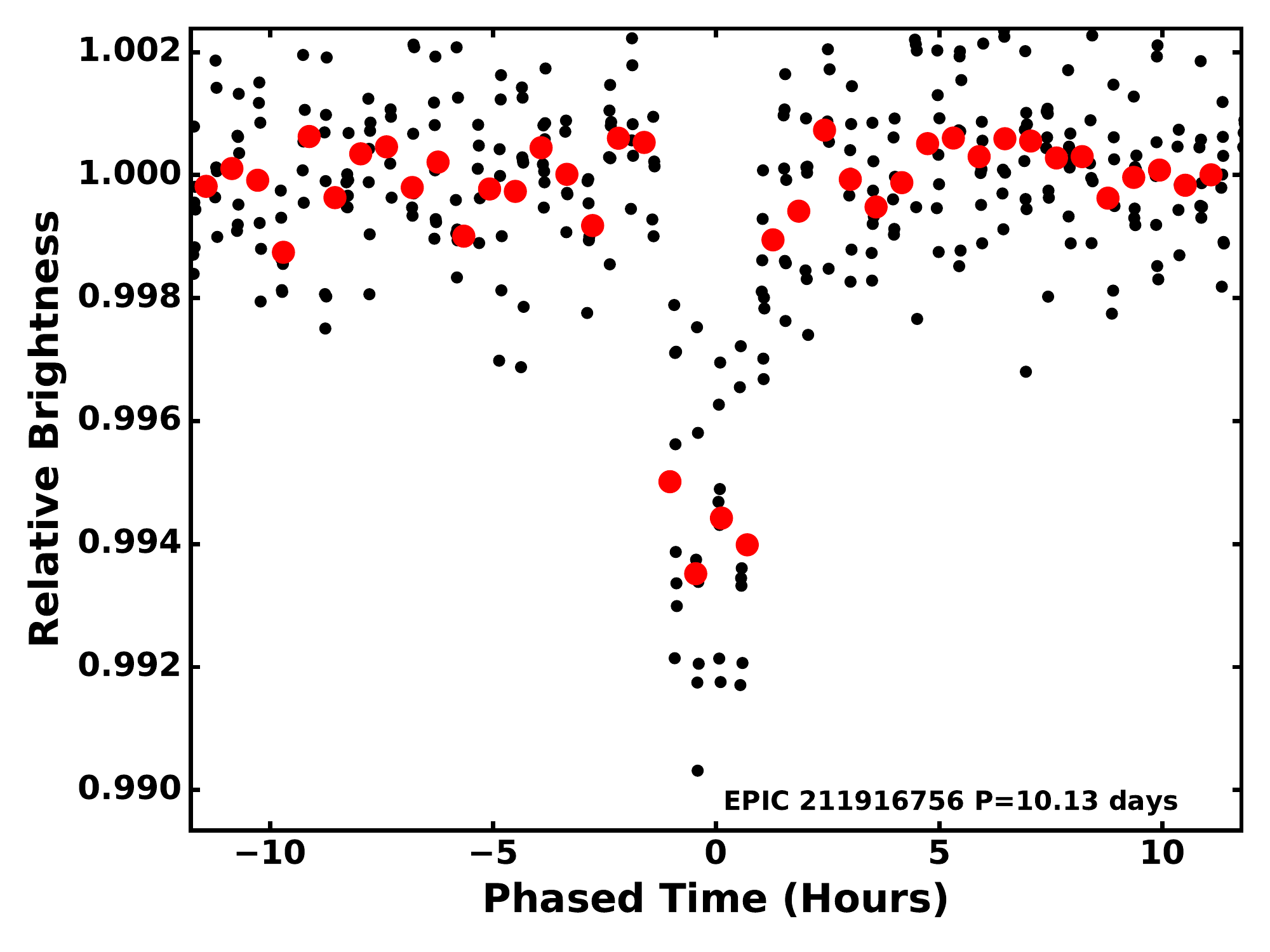} 
\includegraphics[width= 0.35\textwidth]{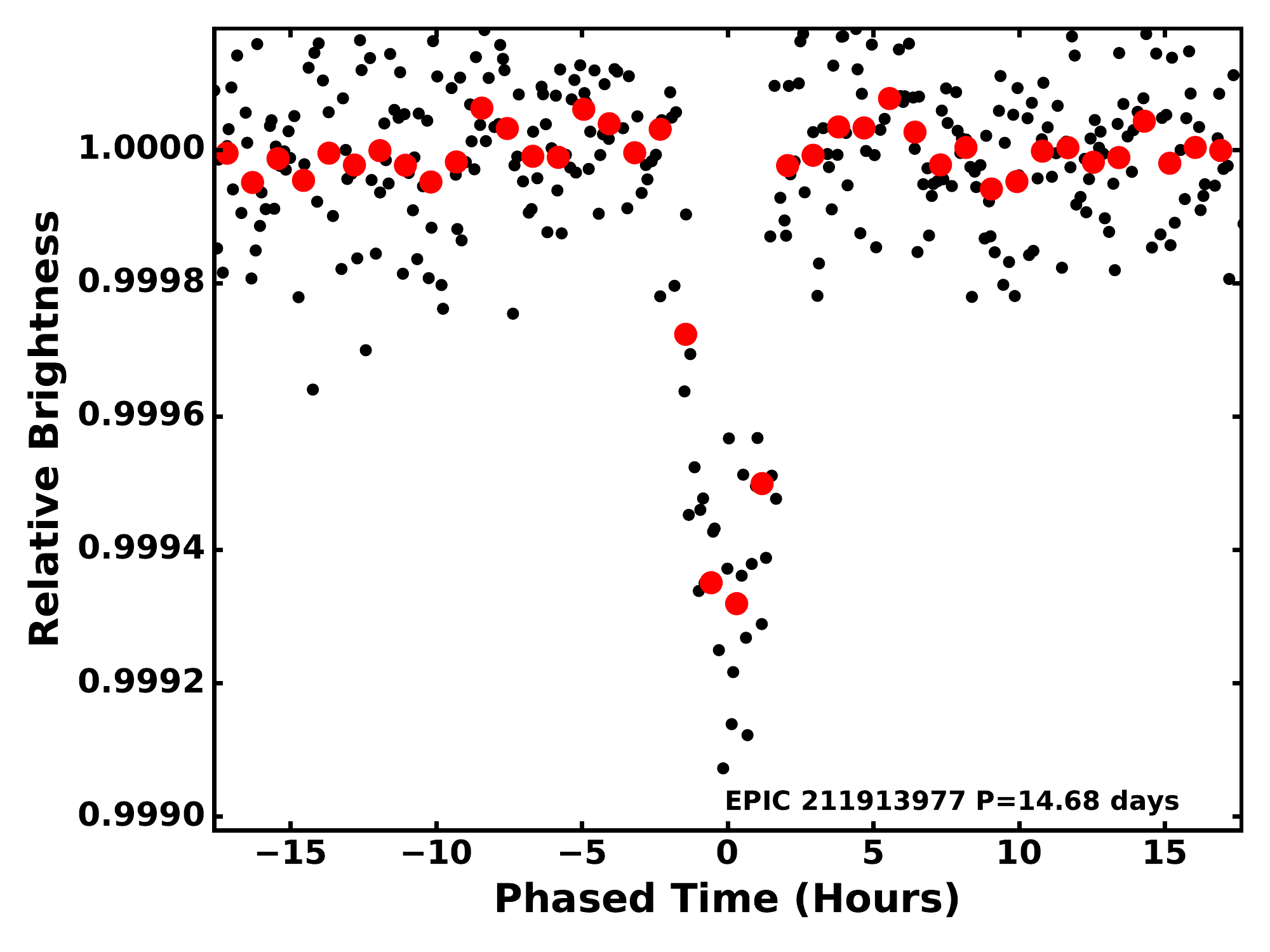} \\  
\includegraphics[width= 0.35\textwidth]{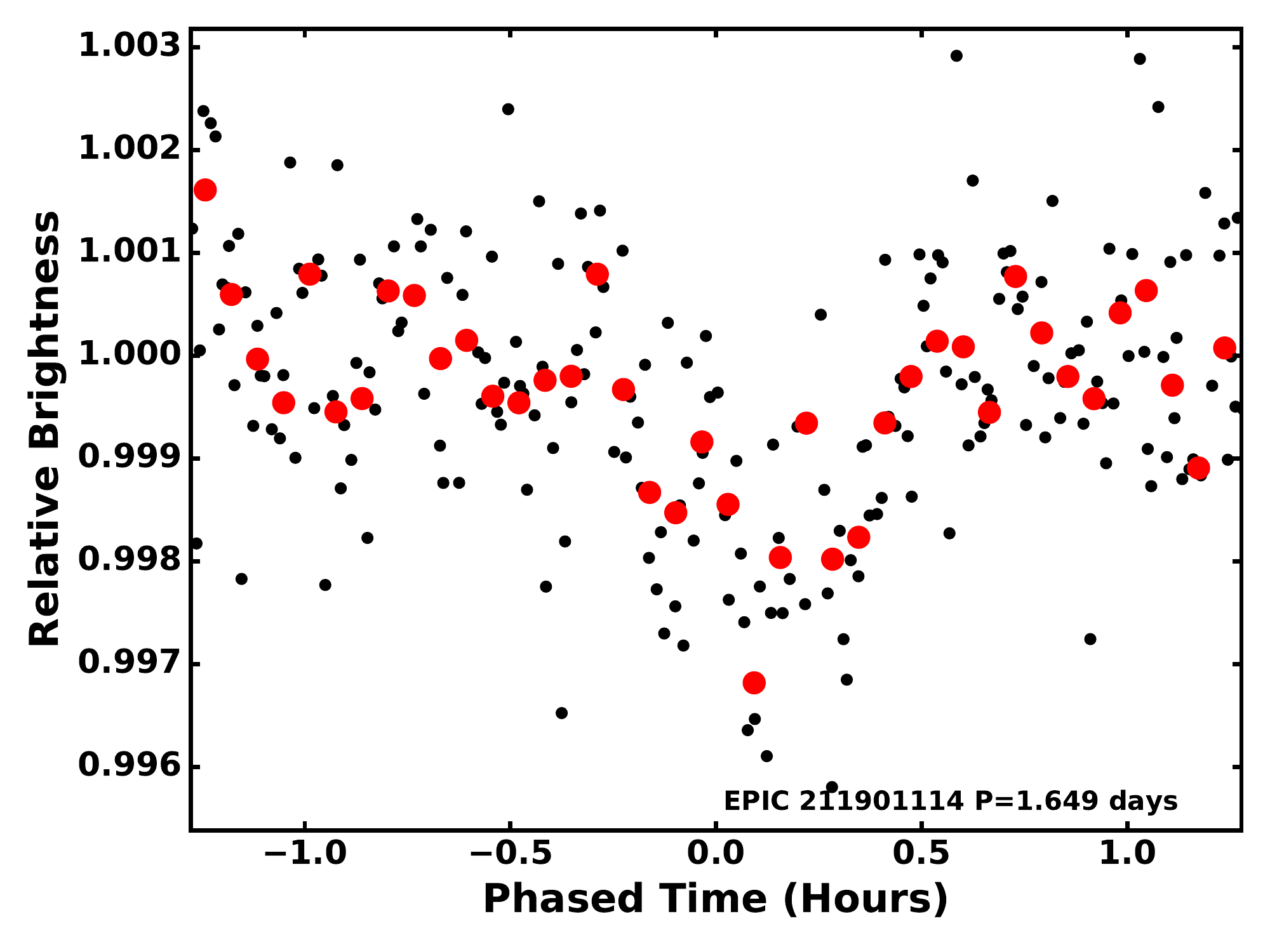} 
\includegraphics[width= 0.35\textwidth]{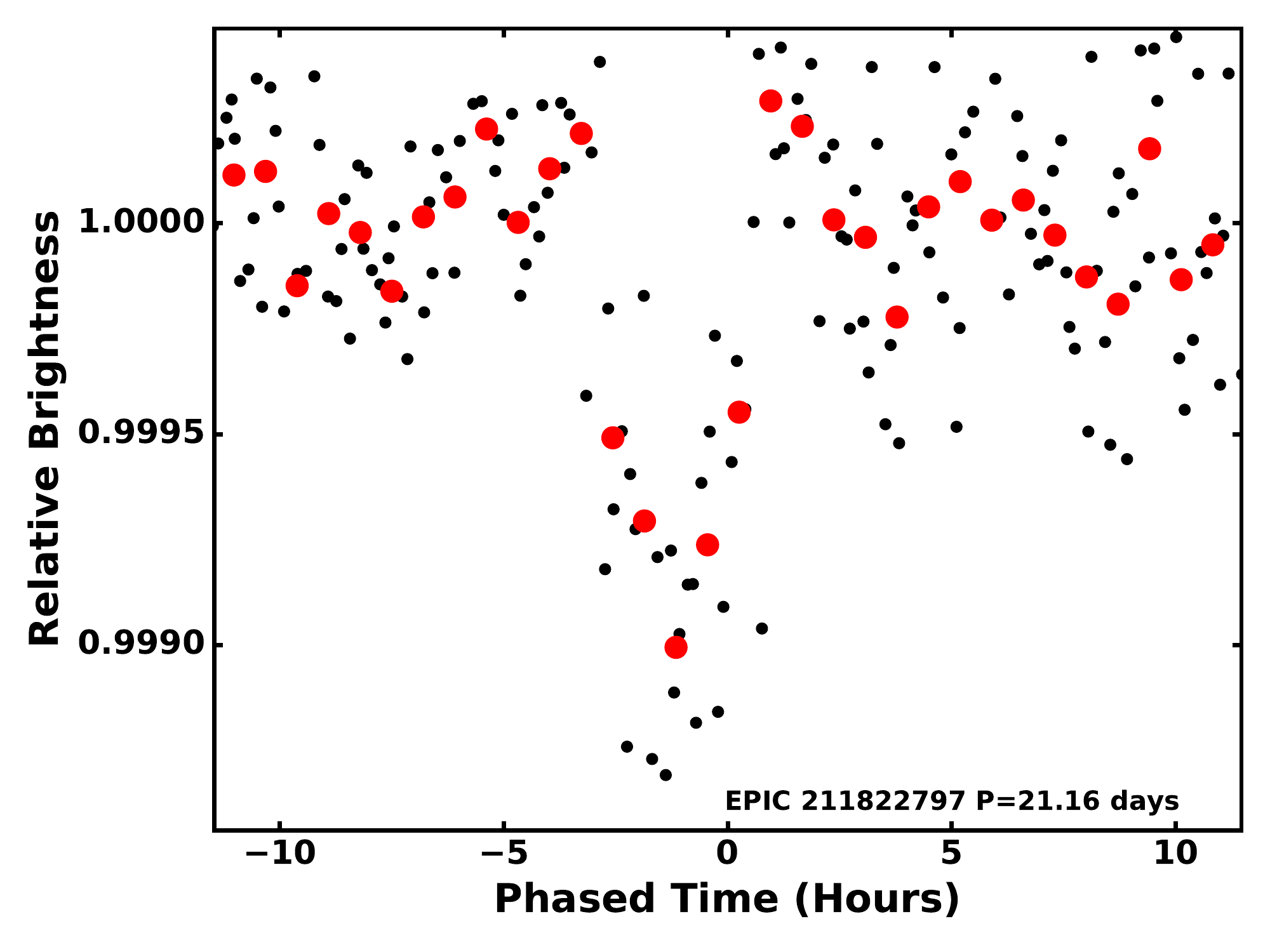} \\  
\includegraphics[width= 0.35\textwidth]{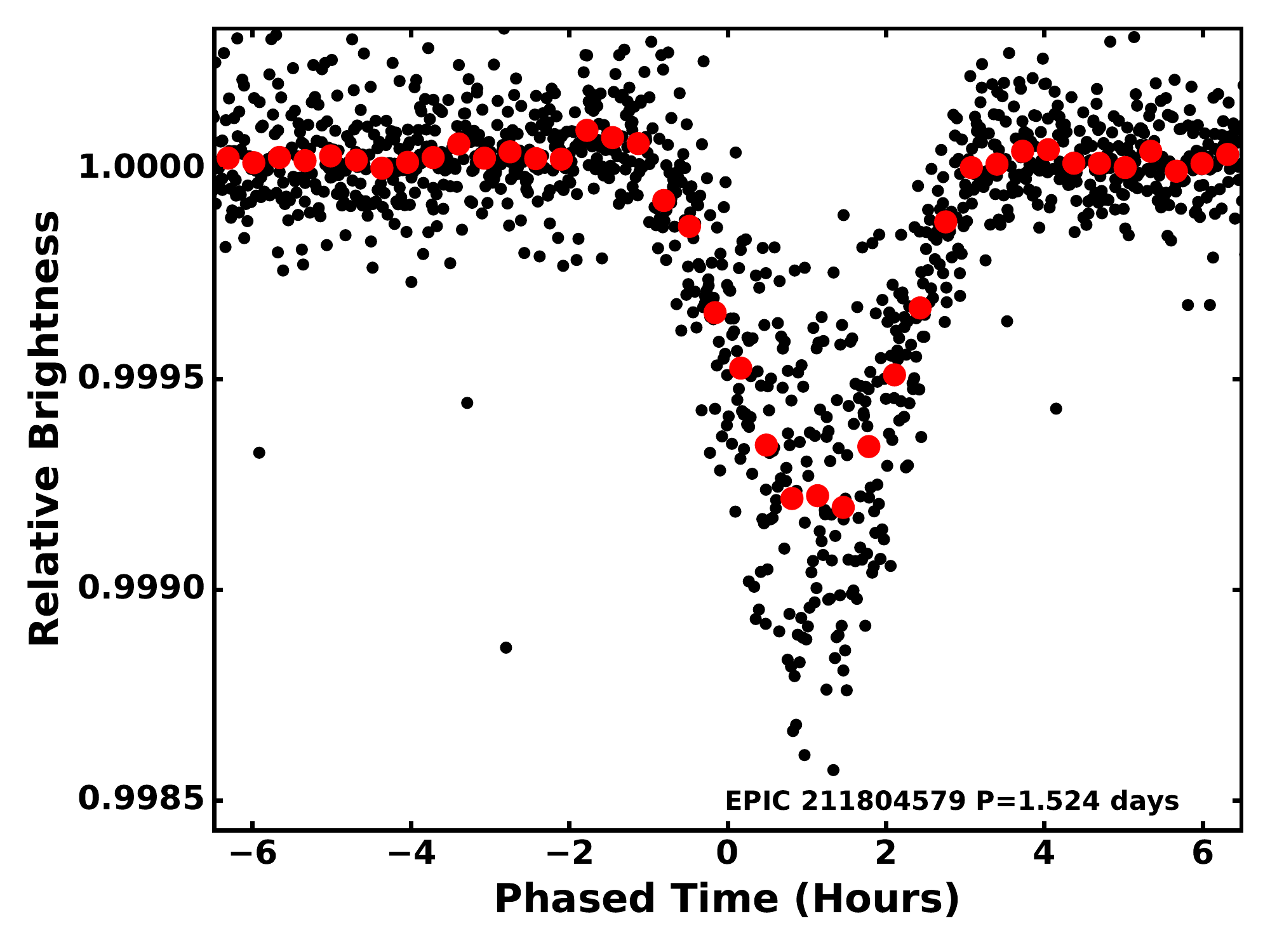} 
\includegraphics[width= 0.35\textwidth]{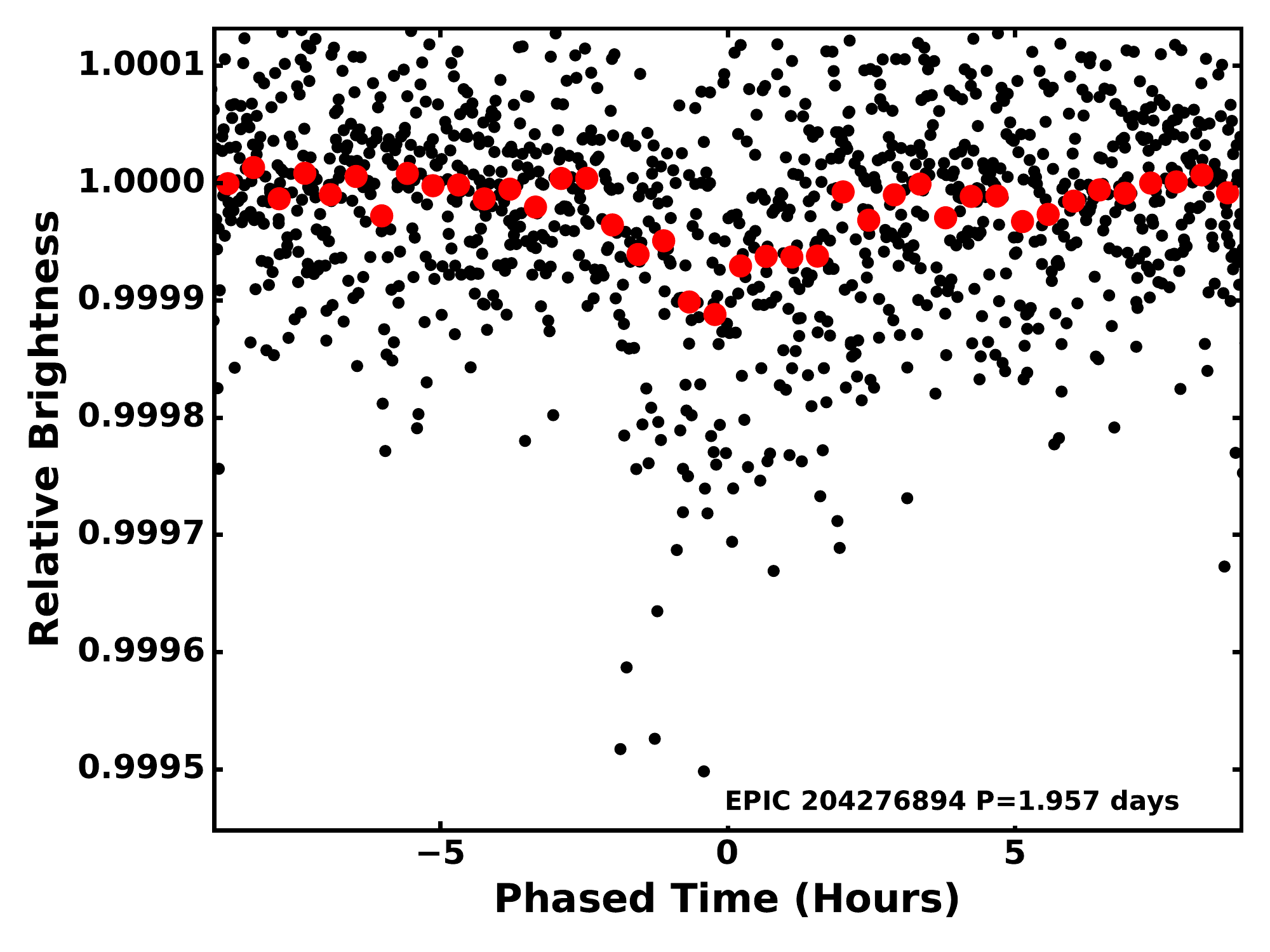} \\  
\includegraphics[width= 0.35\textwidth]{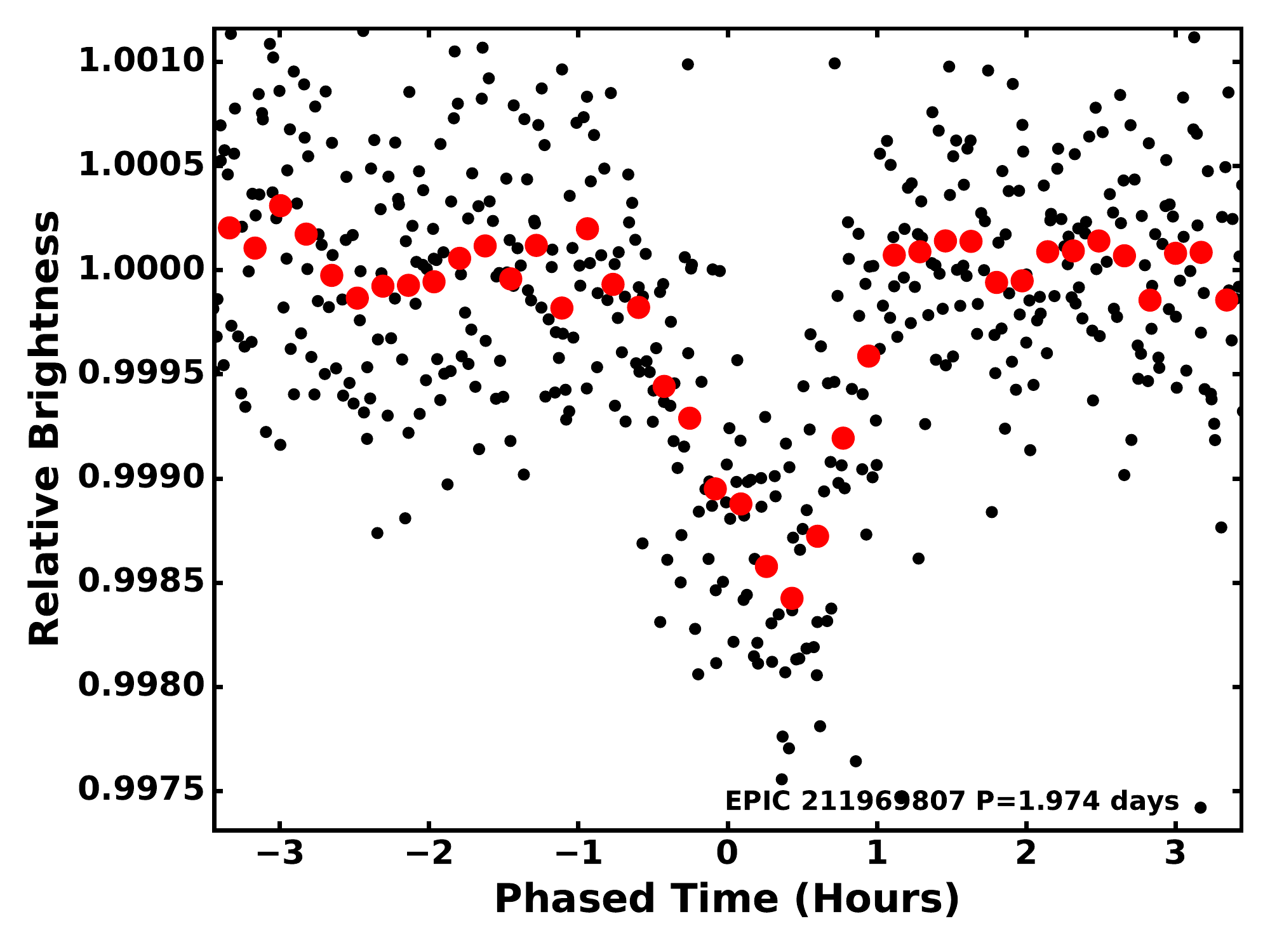} 
\caption{Periodic signals detected with the notch-filter pipeline that appear planetary, phased to the detected period.} 
\label{figs:det2} 
\end{figure*}

\begin{figure*} 
\centering 
\includegraphics[width= 0.35\textwidth]{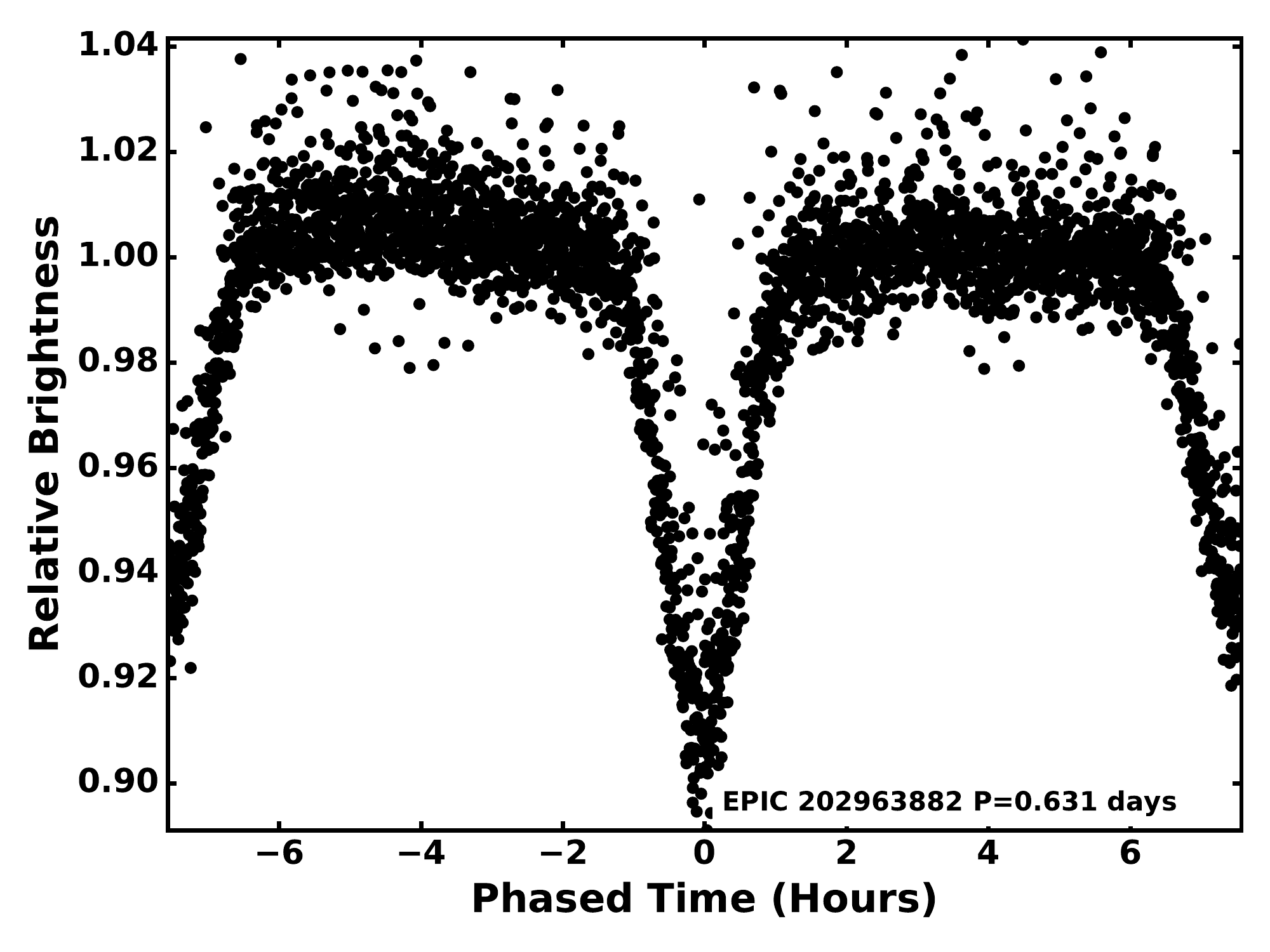} 
\includegraphics[width= 0.35\textwidth]{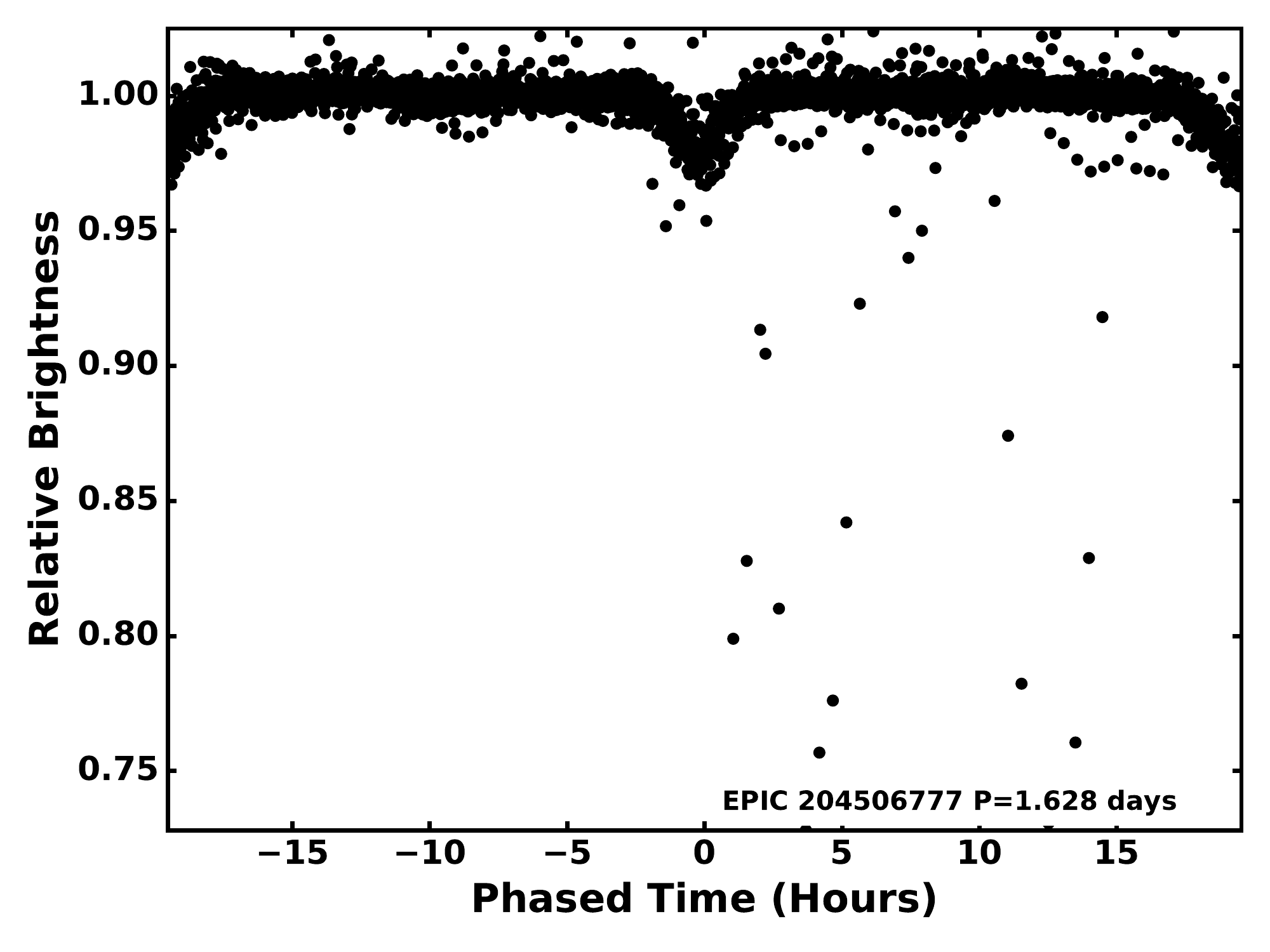} \\  
\includegraphics[width= 0.35\textwidth]{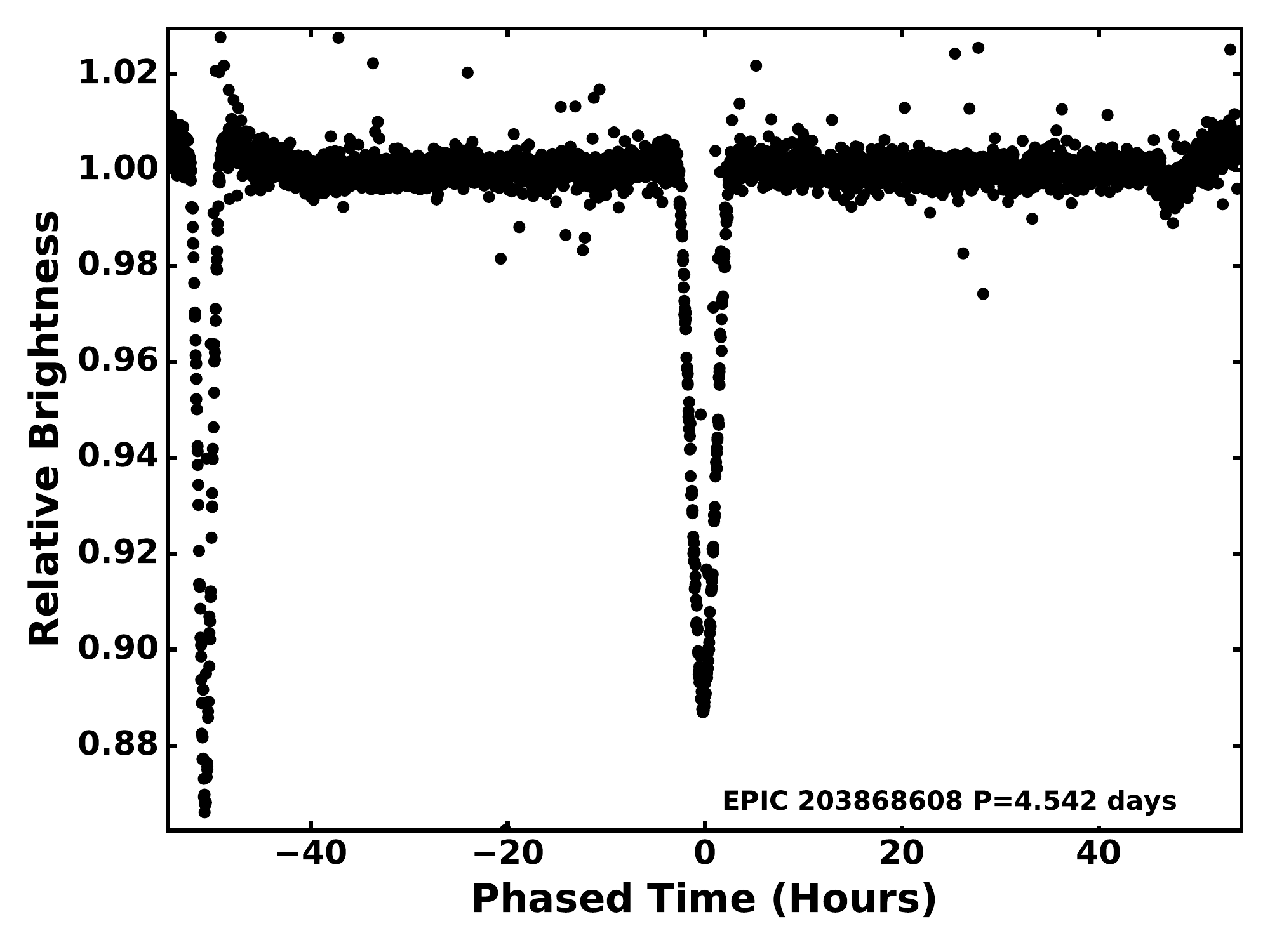}   
\includegraphics[width= 0.35\textwidth]{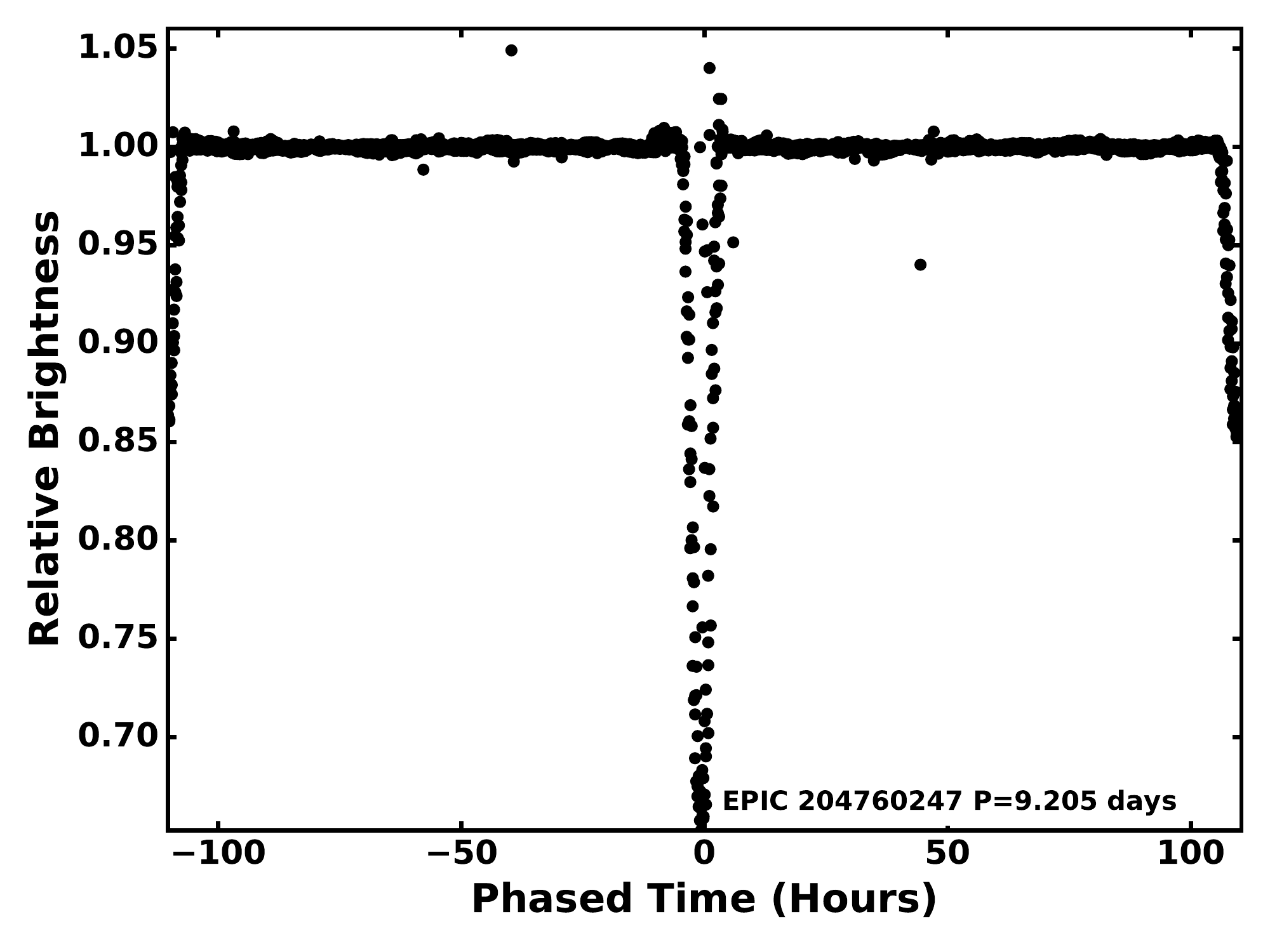} \\
\includegraphics[width= 0.35\textwidth]{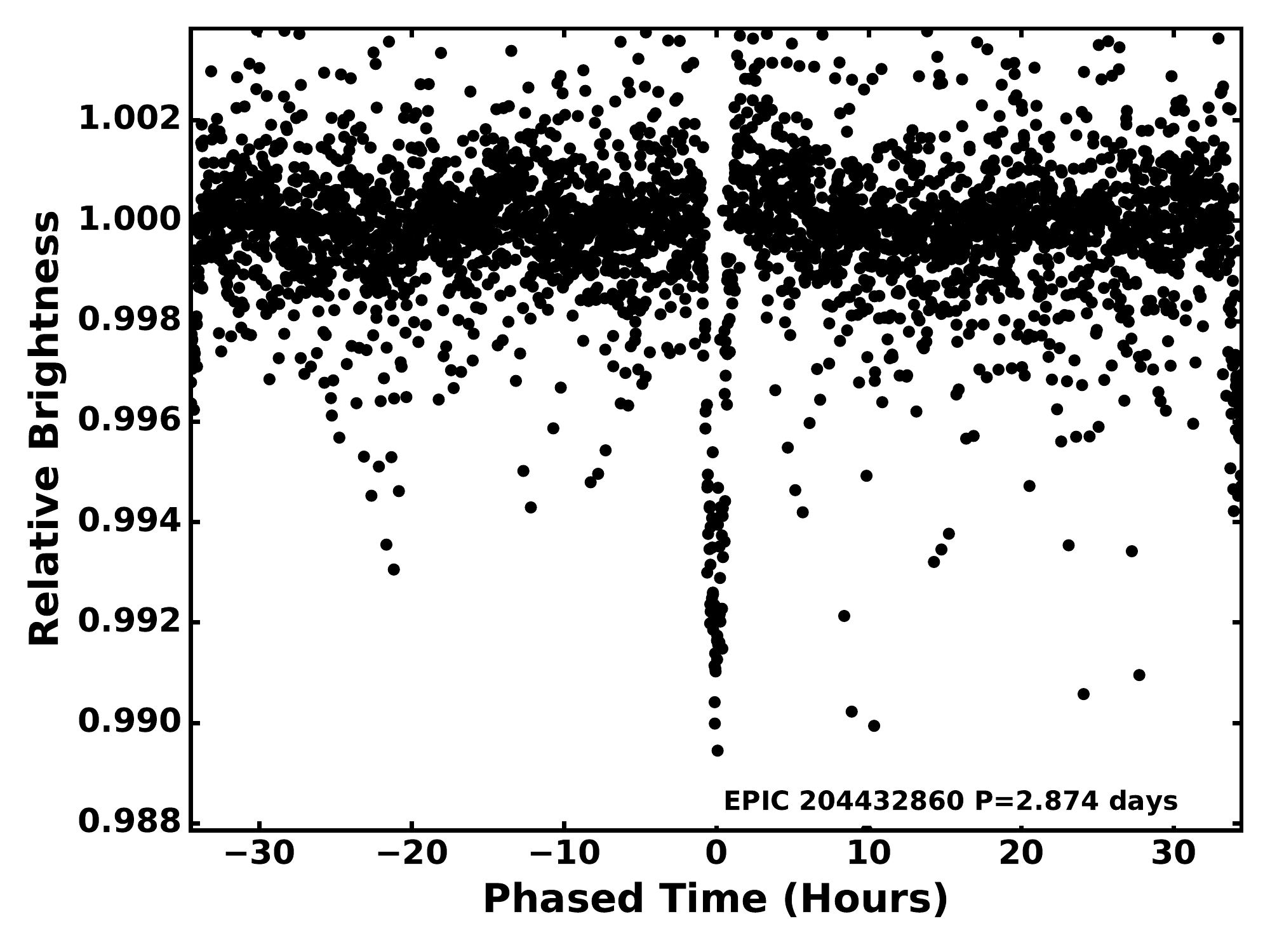}   
\includegraphics[width= 0.35\textwidth]{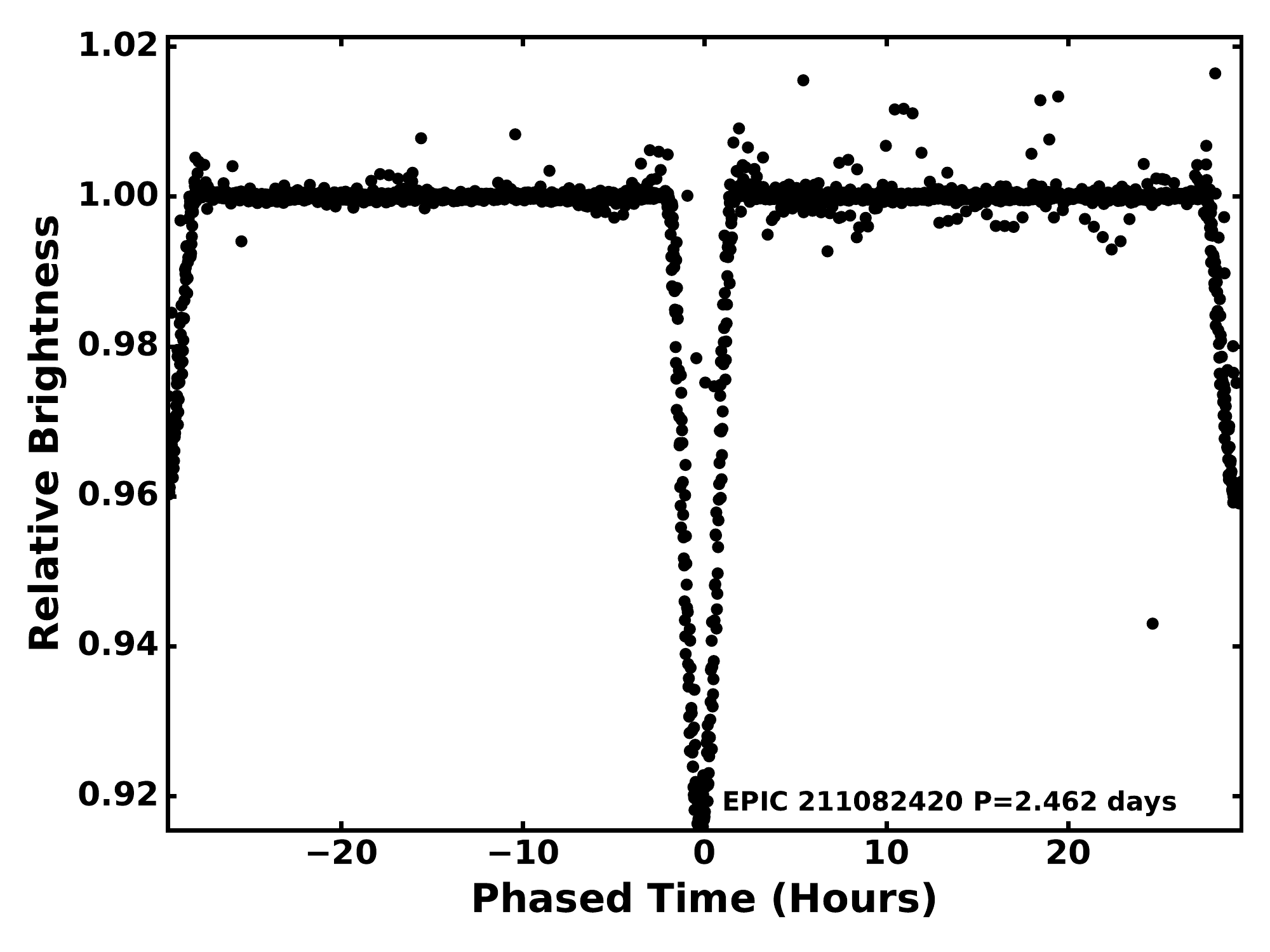} \\
\includegraphics[width= 0.35\textwidth]{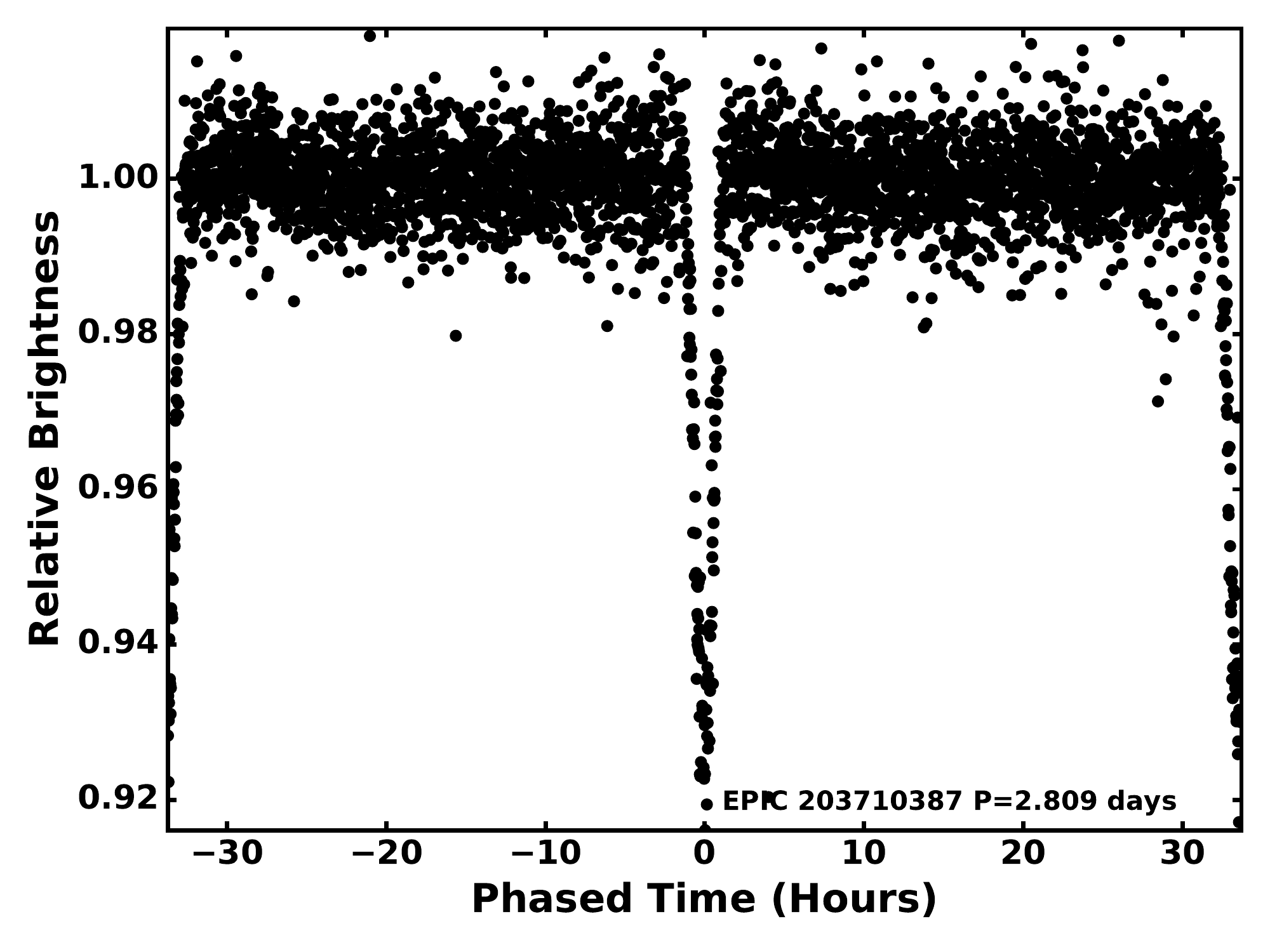}   
\includegraphics[width= 0.35\textwidth]{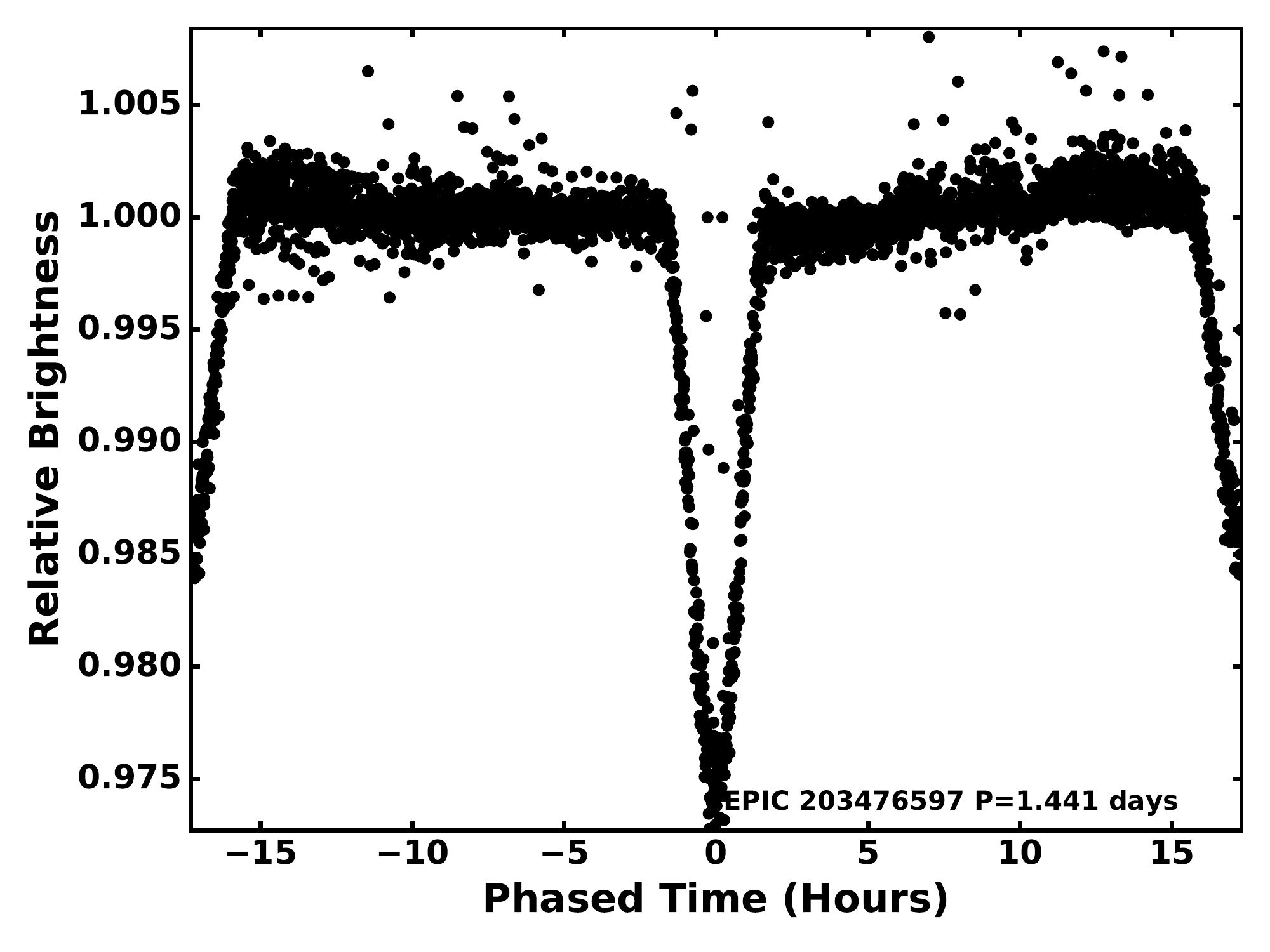} \\
\includegraphics[width= 0.35\textwidth]{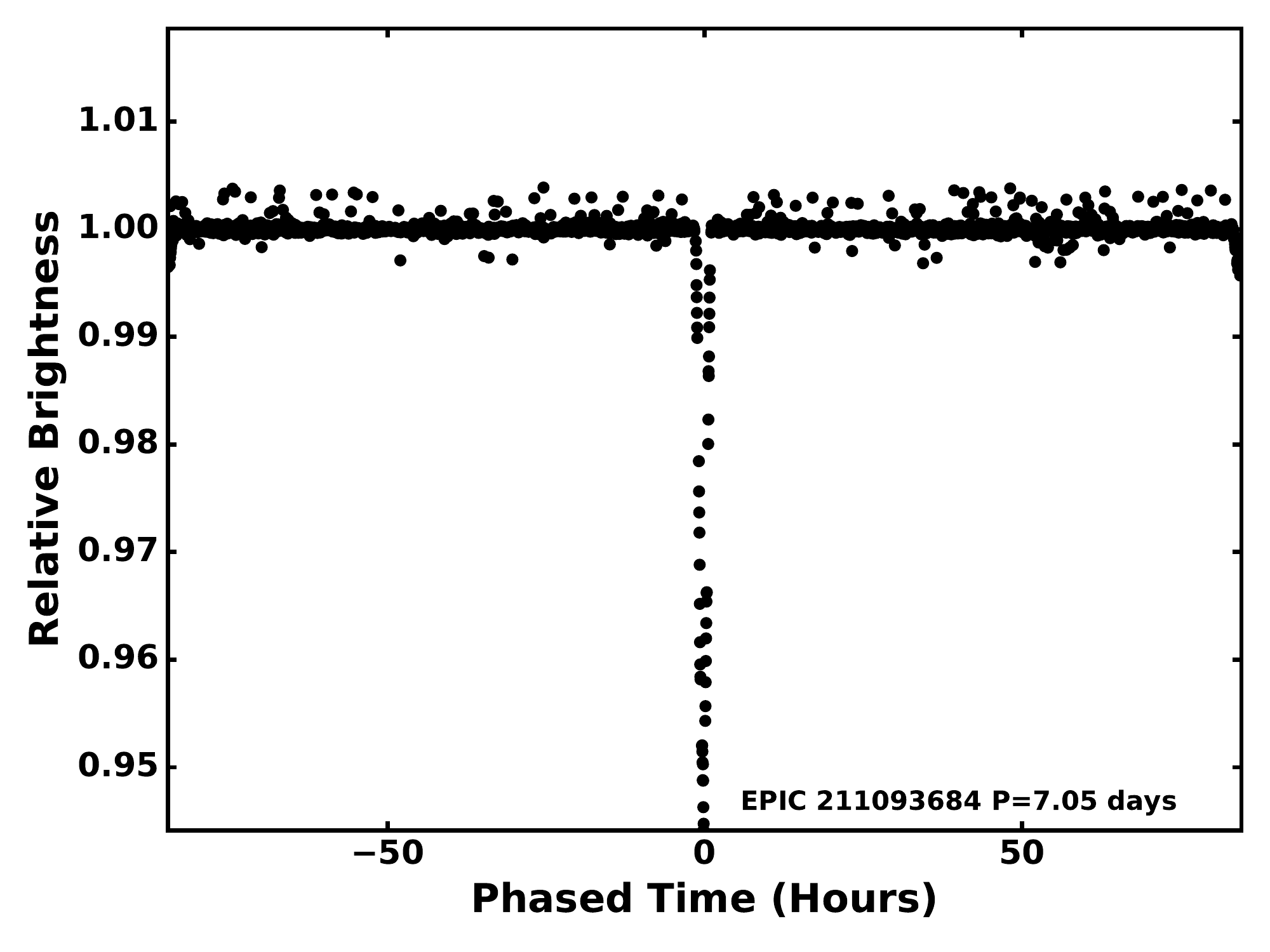} 
\includegraphics[width= 0.35\textwidth]{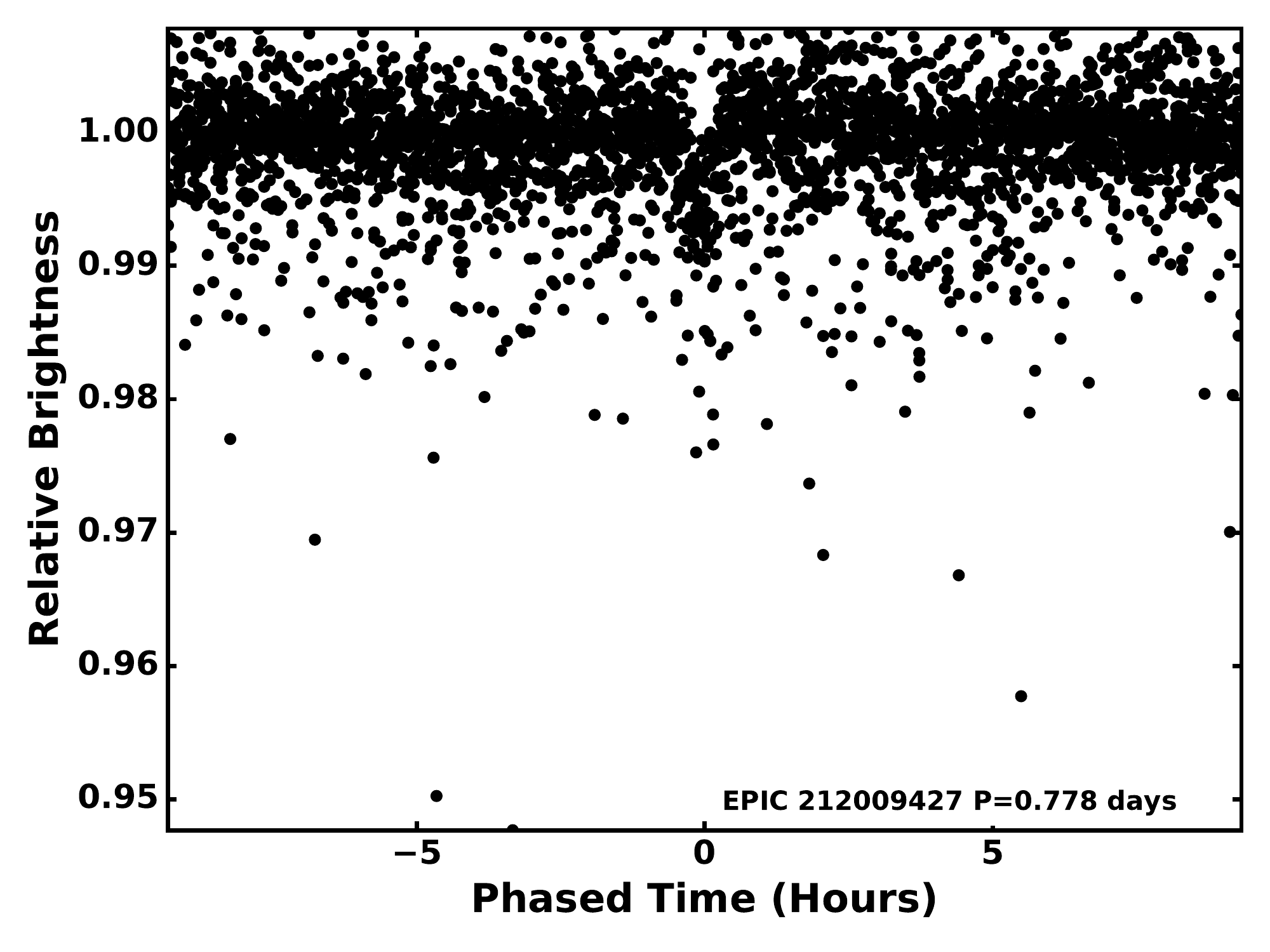} \\
\caption{Eclipsing binaries detected with the notch-filter pipeline.} 
\label{figs:det3} 
\end{figure*}

\begin{figure*} 
\centering 
\includegraphics[width= 0.35\textwidth]{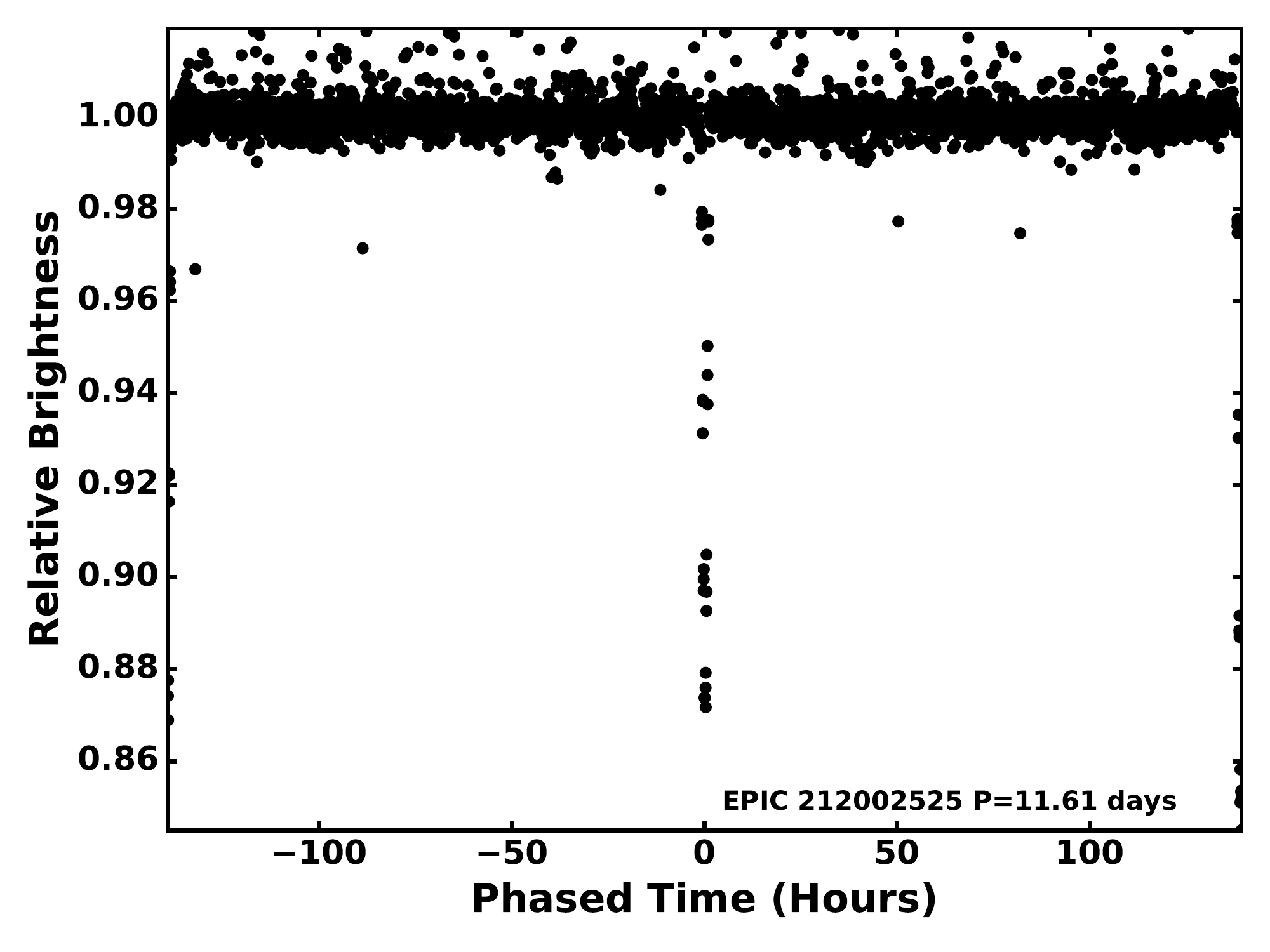} 
\includegraphics[width= 0.35\textwidth]{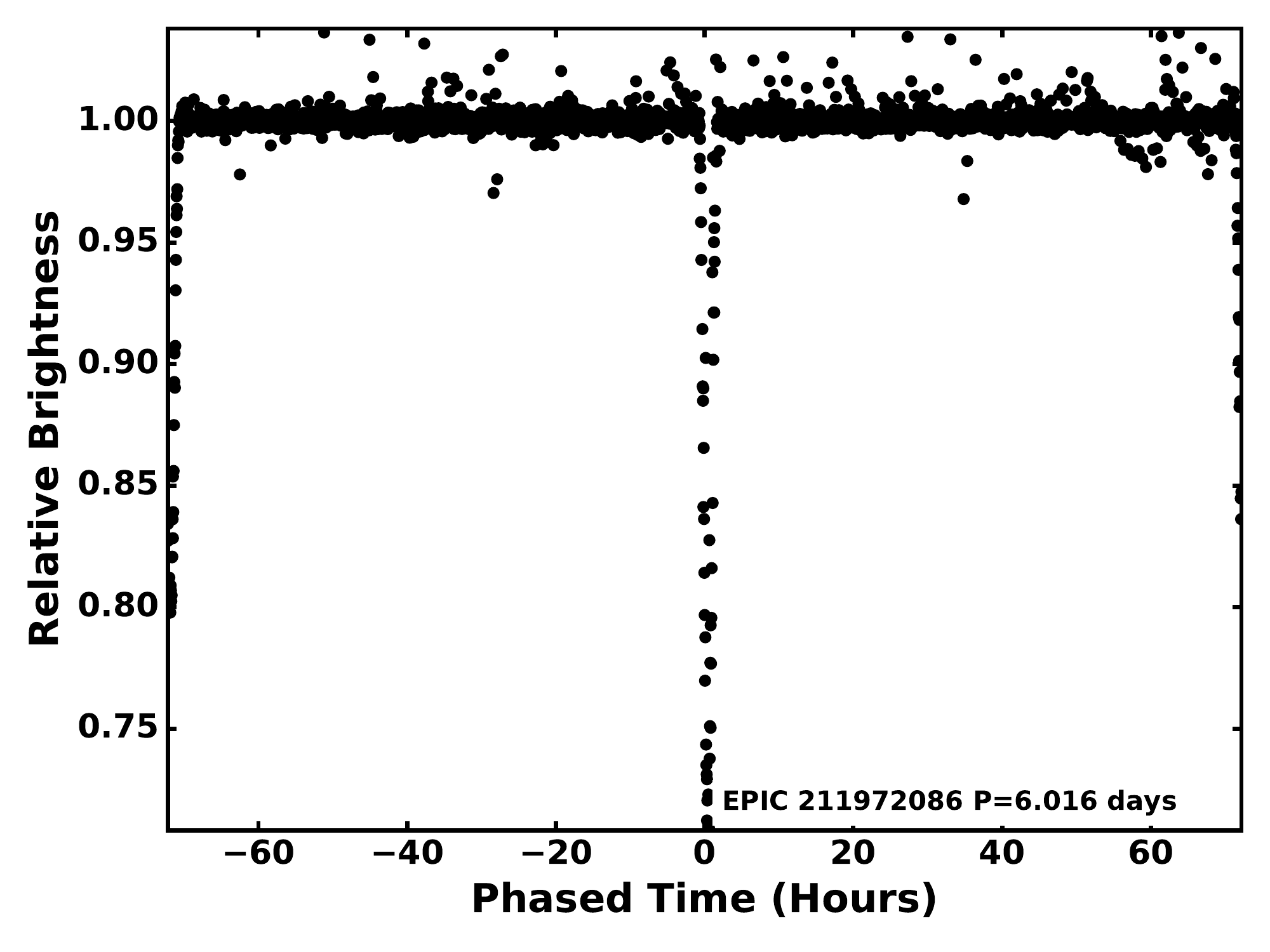} \\  
\includegraphics[width= 0.35\textwidth]{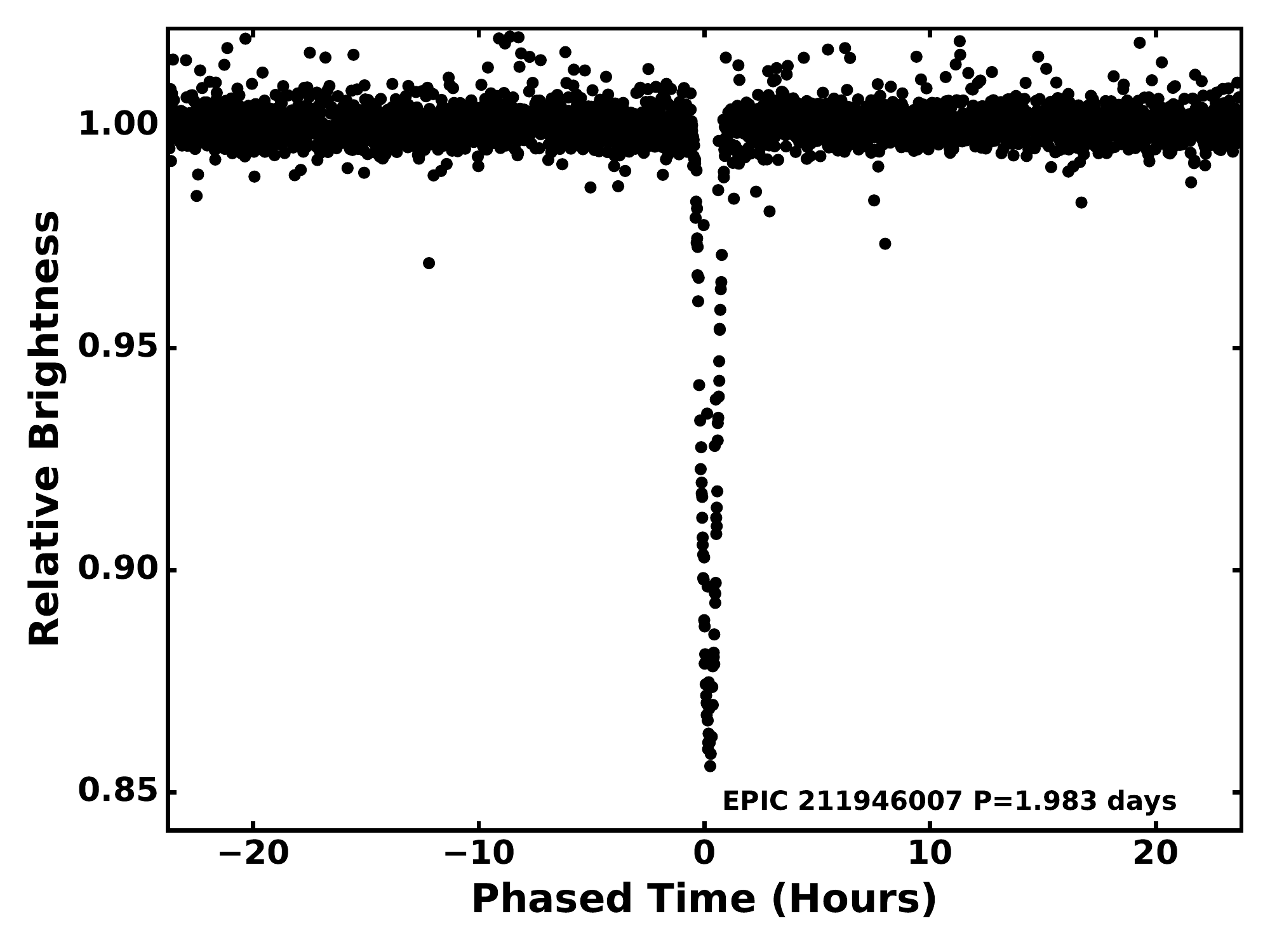} 
\includegraphics[width= 0.35\textwidth]{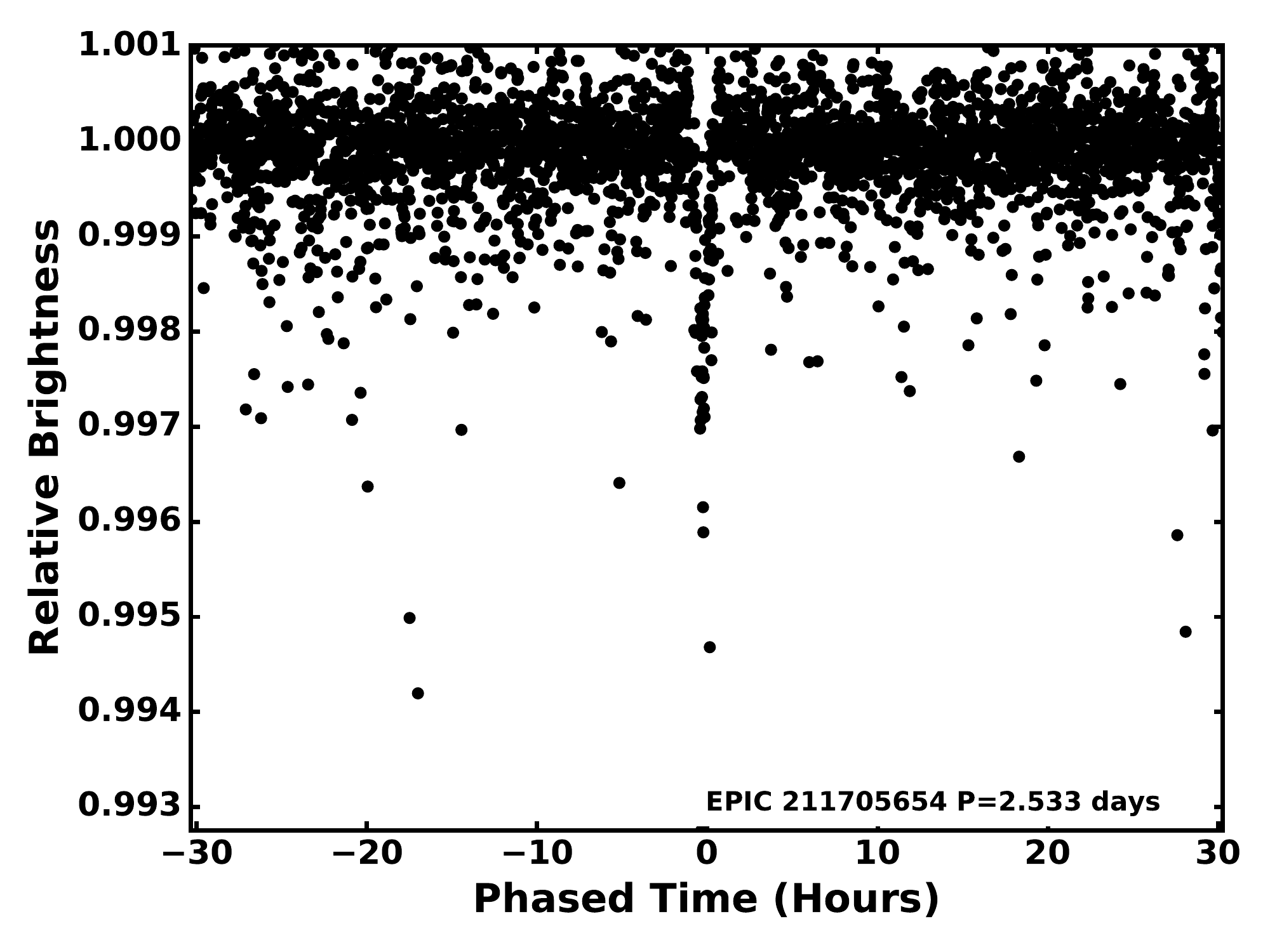} \\  
\includegraphics[width= 0.35\textwidth]{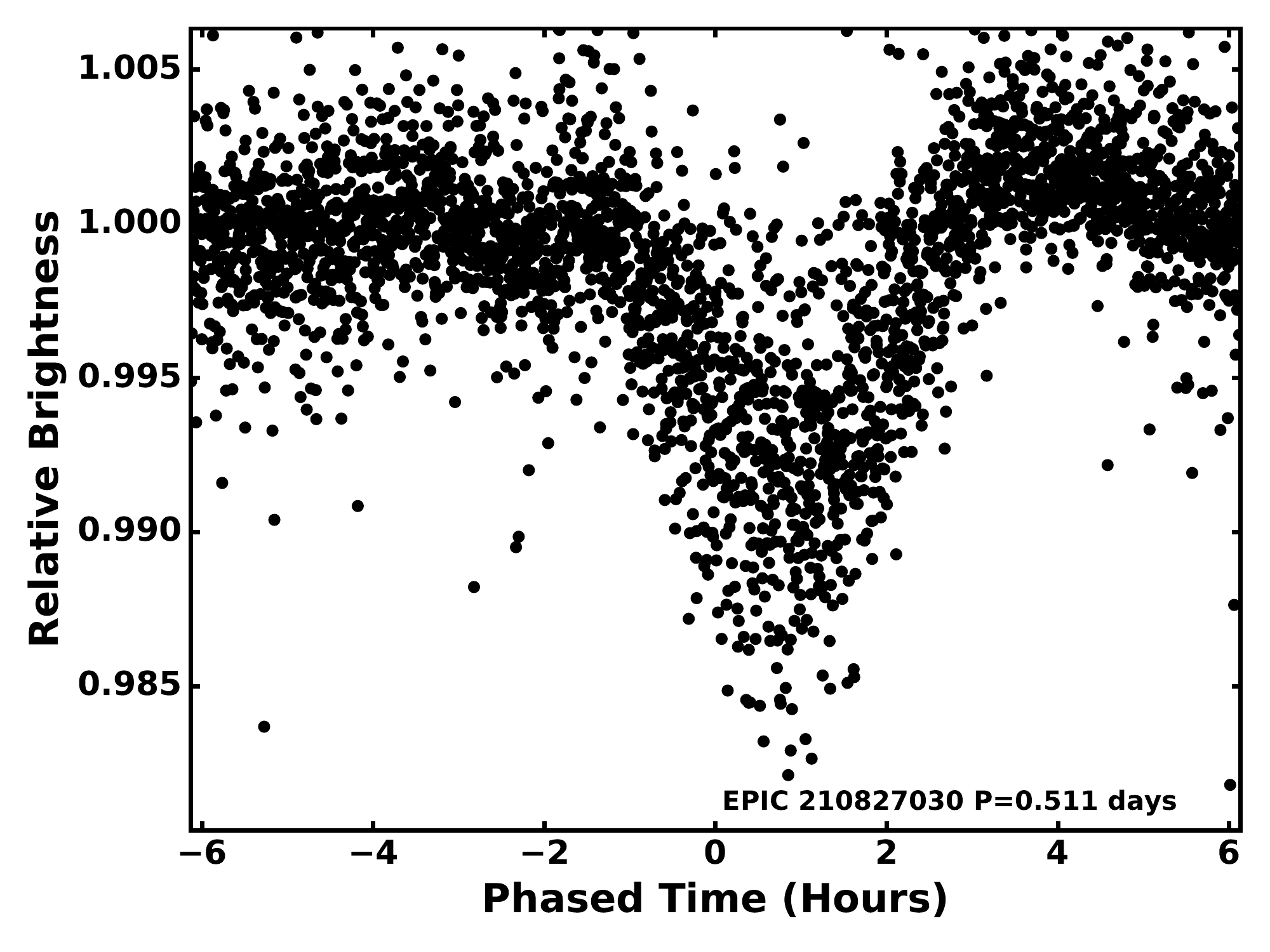} 
\includegraphics[width= 0.35\textwidth]{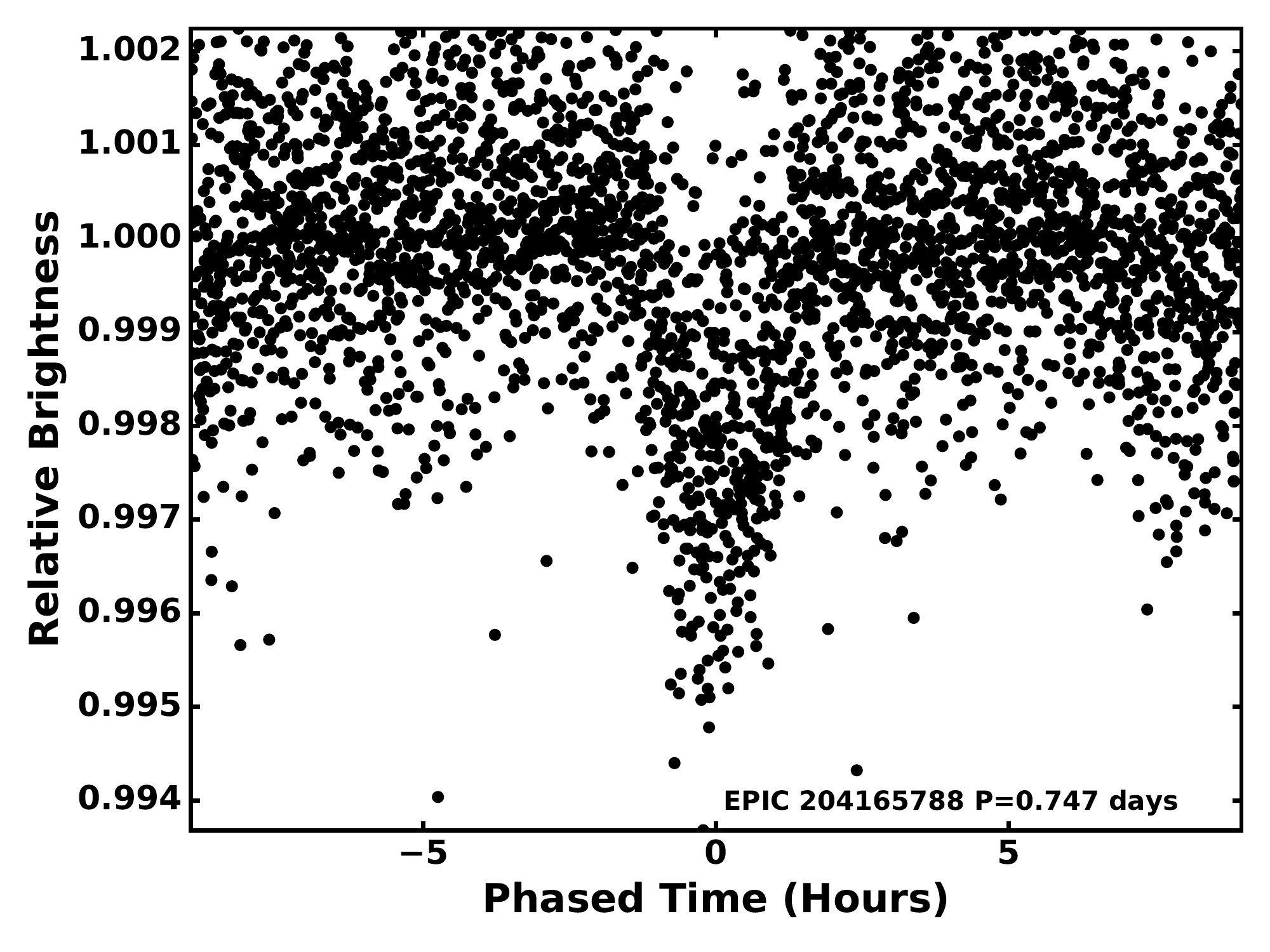} \\  
\caption{Eclipsing binaries detected with the notch-filter pipeline.} 
\label{figs:det4} 
\end{figure*}

\begin{figure*} 
\centering 
\includegraphics[width= 0.35\textwidth]{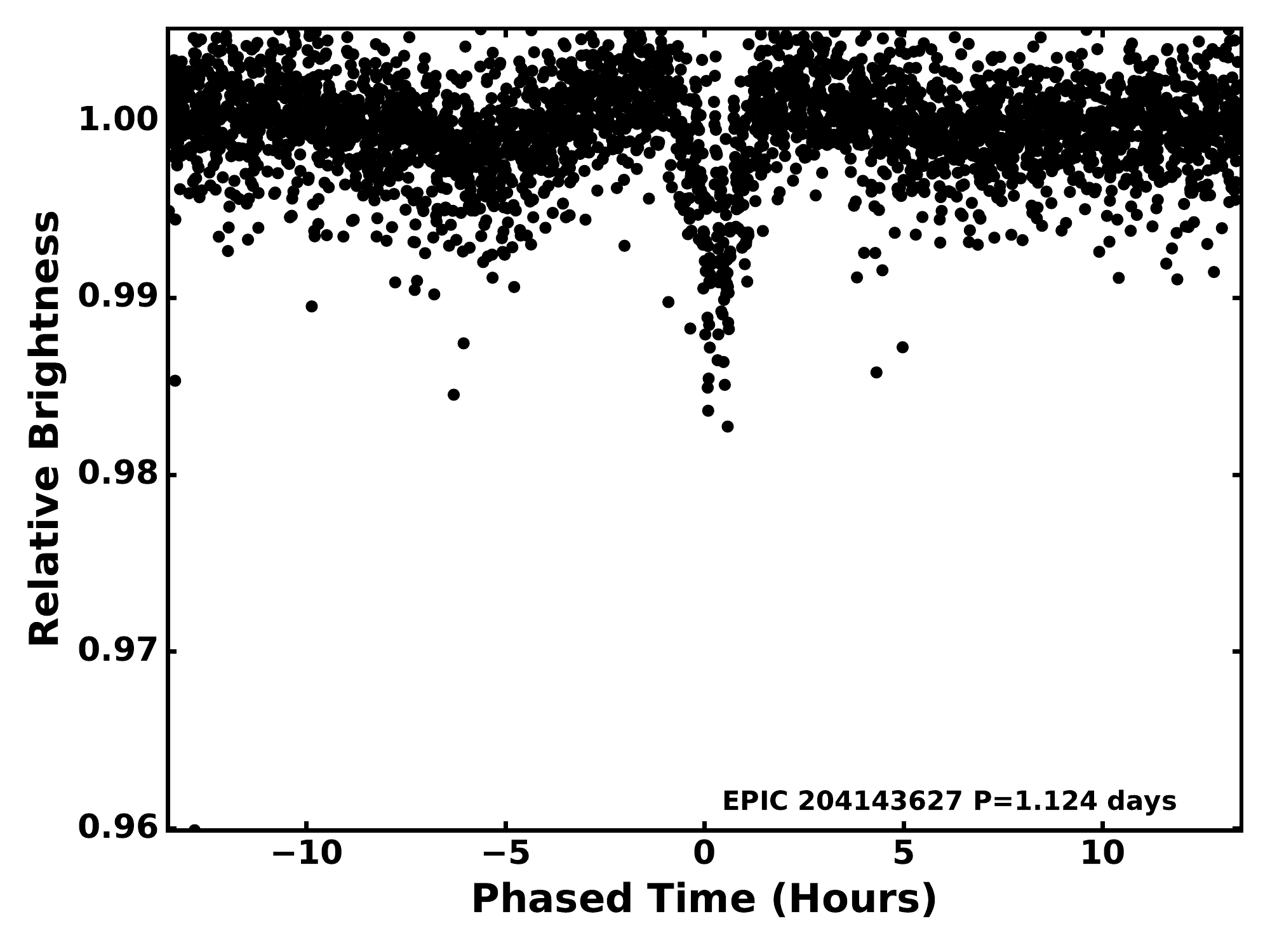} 
\includegraphics[width= 0.35\textwidth]{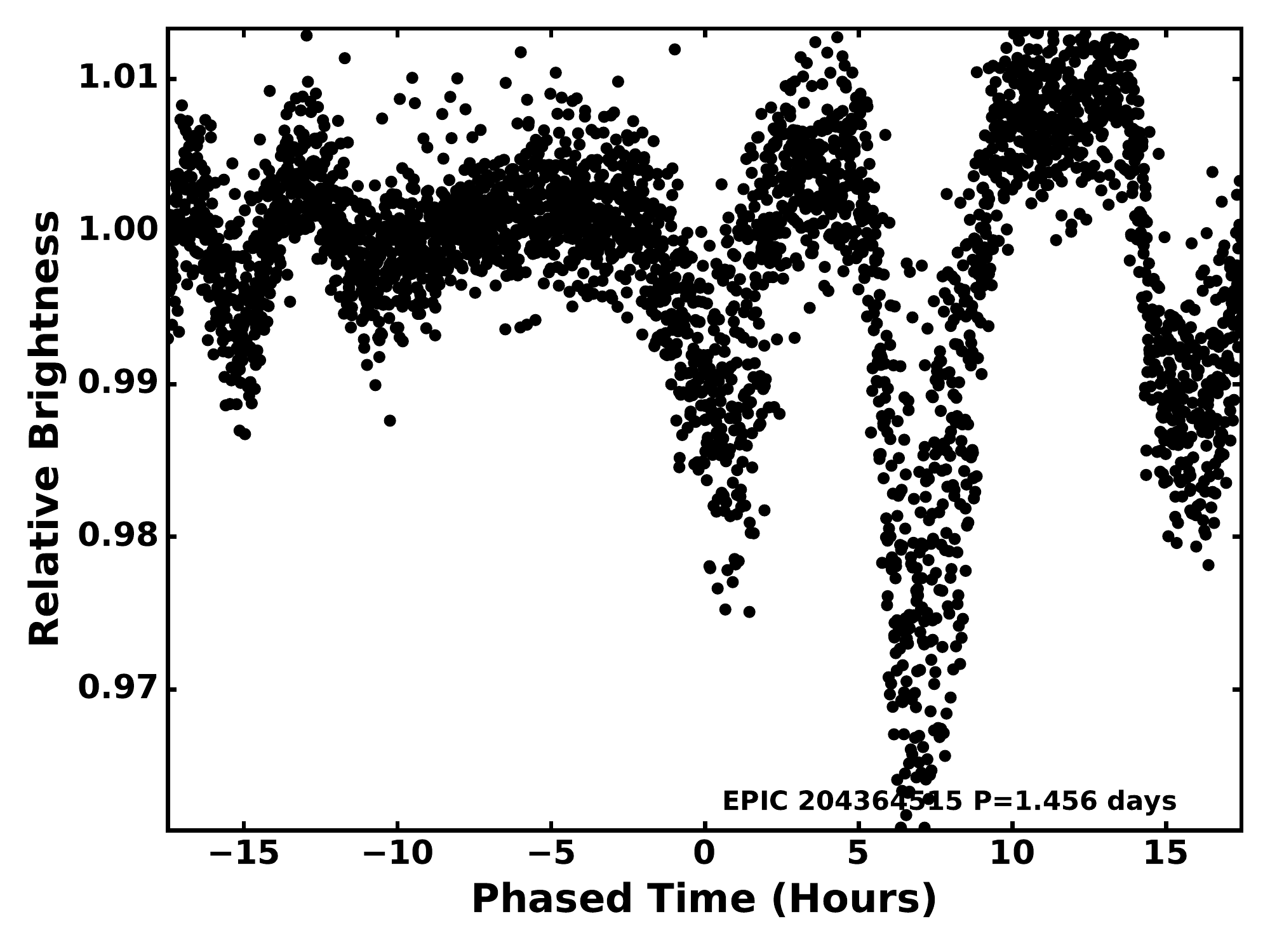} \\  
\includegraphics[width= 0.35\textwidth]{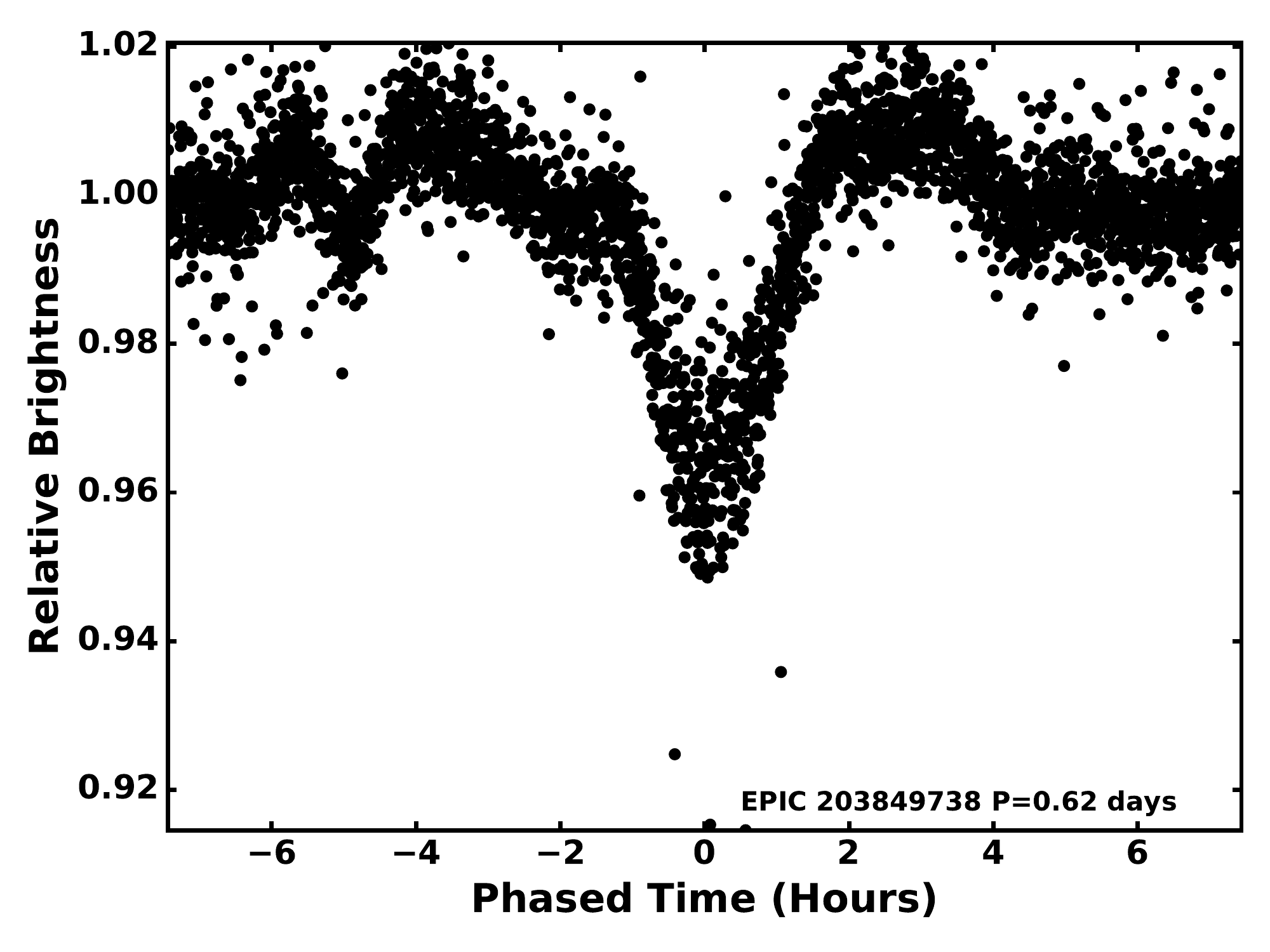} 
\includegraphics[width= 0.35\textwidth]{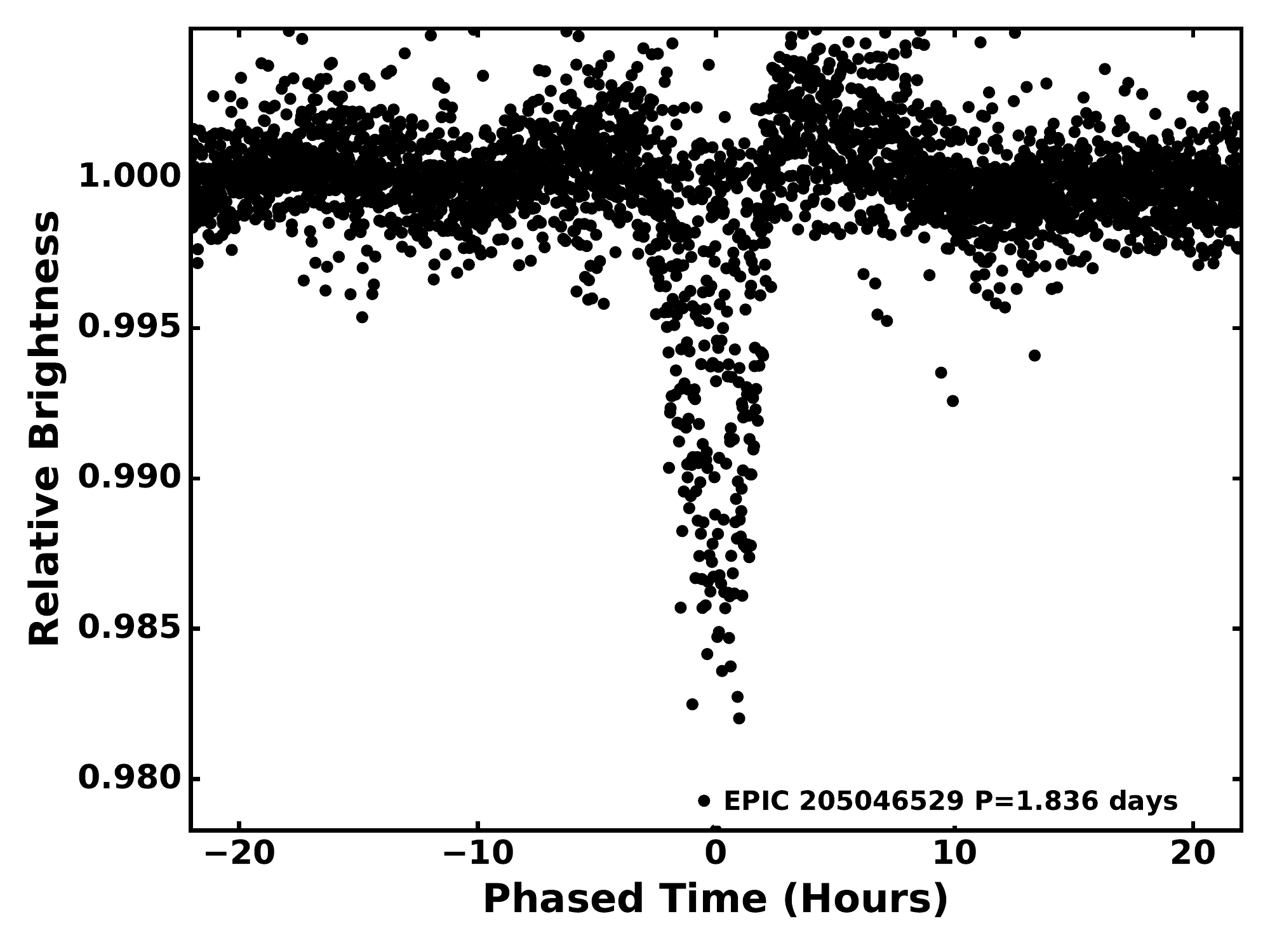} \\  
\includegraphics[width= 0.35\textwidth]{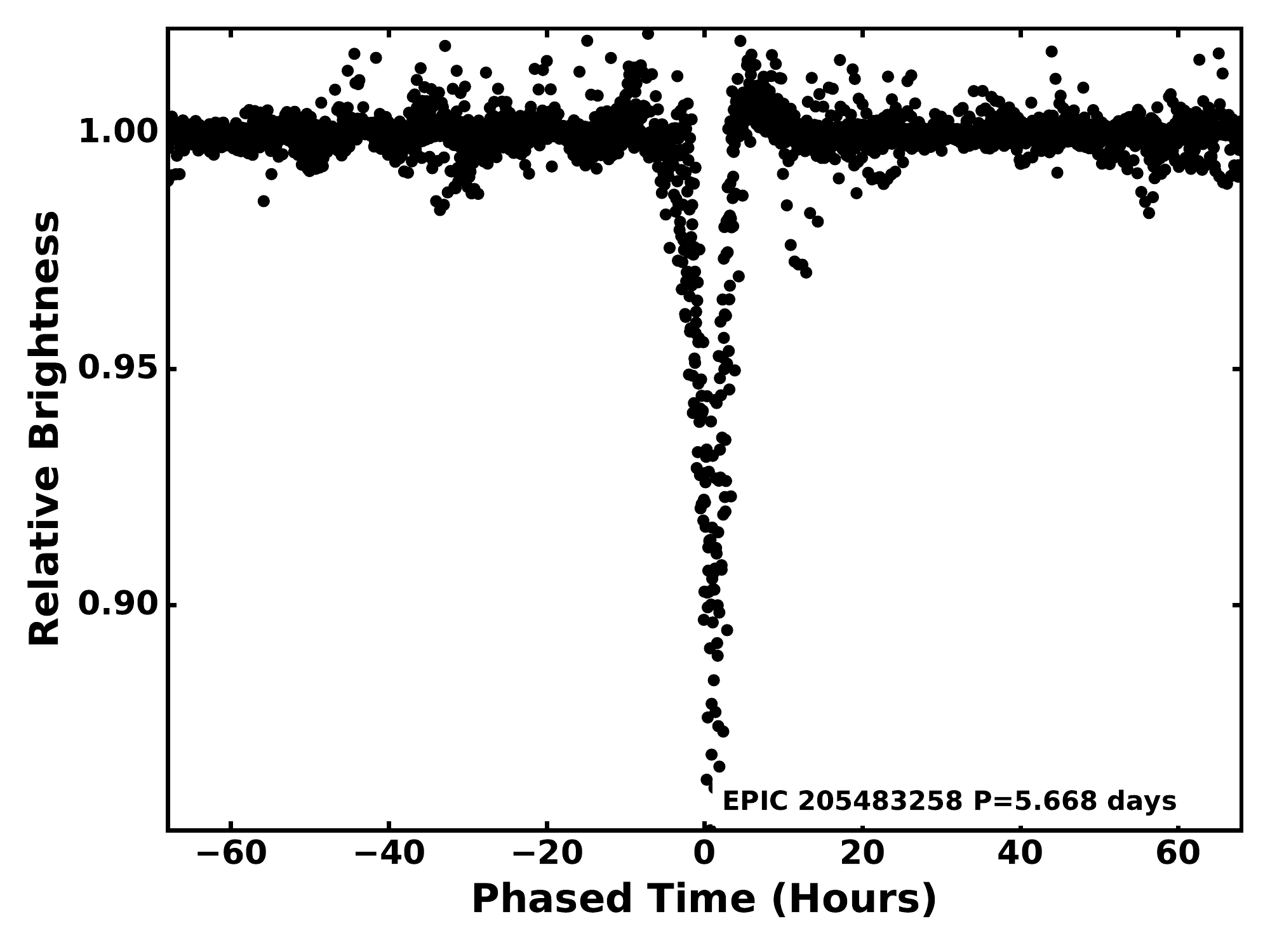} 
\includegraphics[width= 0.35\textwidth]{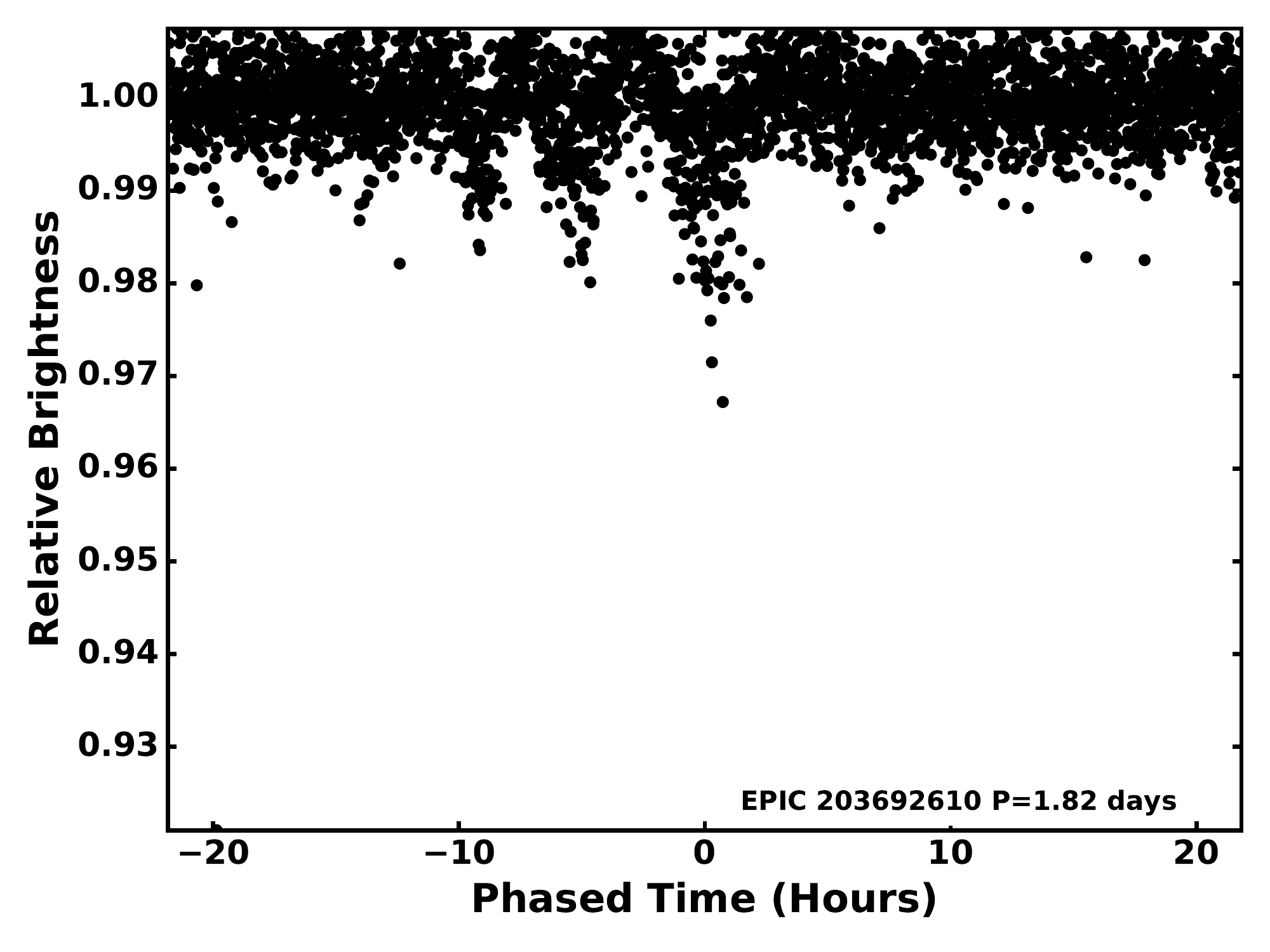} \\ 
\includegraphics[width= 0.35\textwidth]{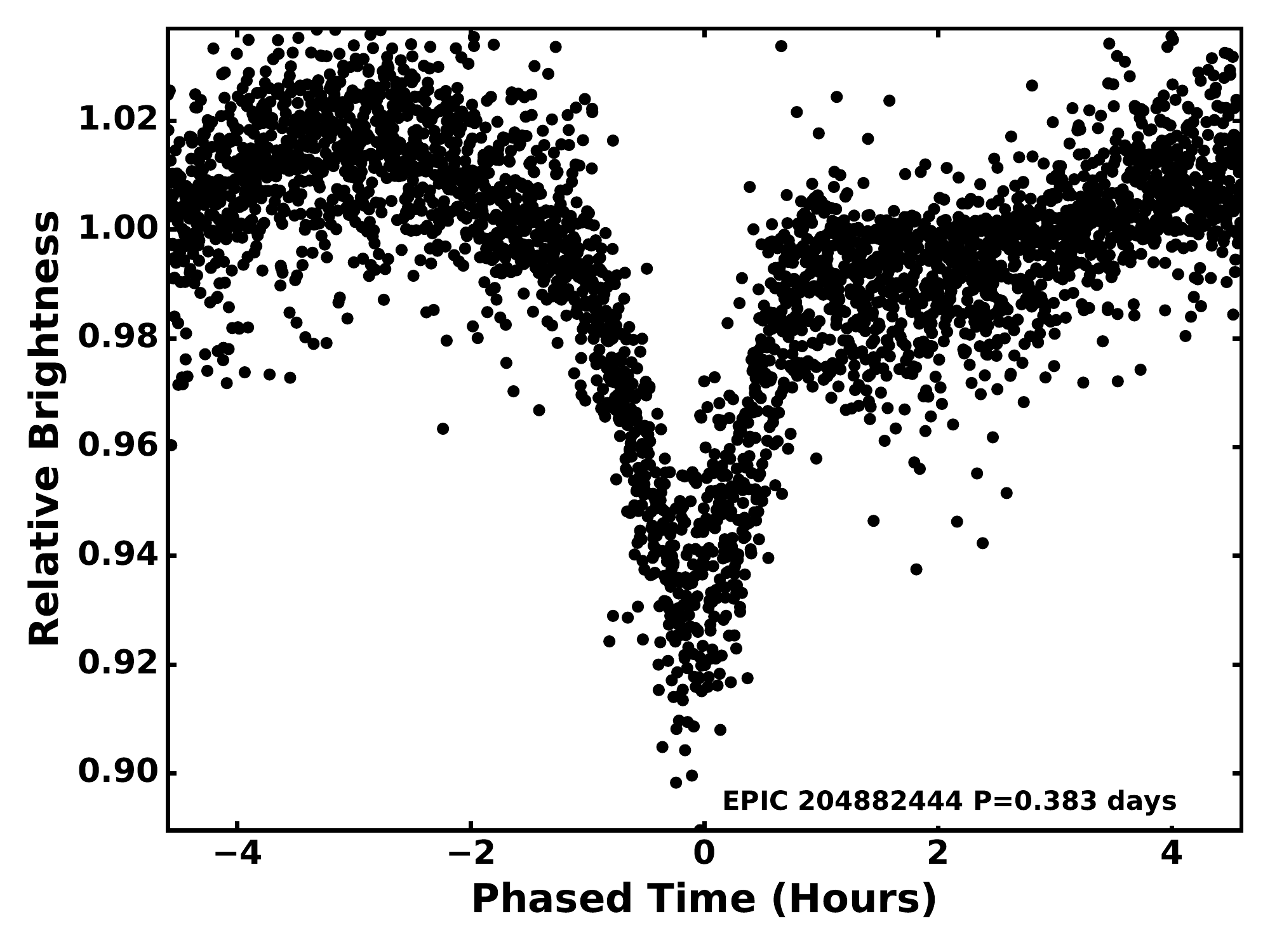} 
\caption{Periodic signals detected with the notch-filter pipeline determined not the be caused by eclipsing binaries or transiting exoplanets.} 
\label{figs:det5} 
\end{figure*}

%% file: detfigurecode_lcr.tex
\begin{figure*} 
\centering 
\includegraphics[width= 0.35\textwidth]{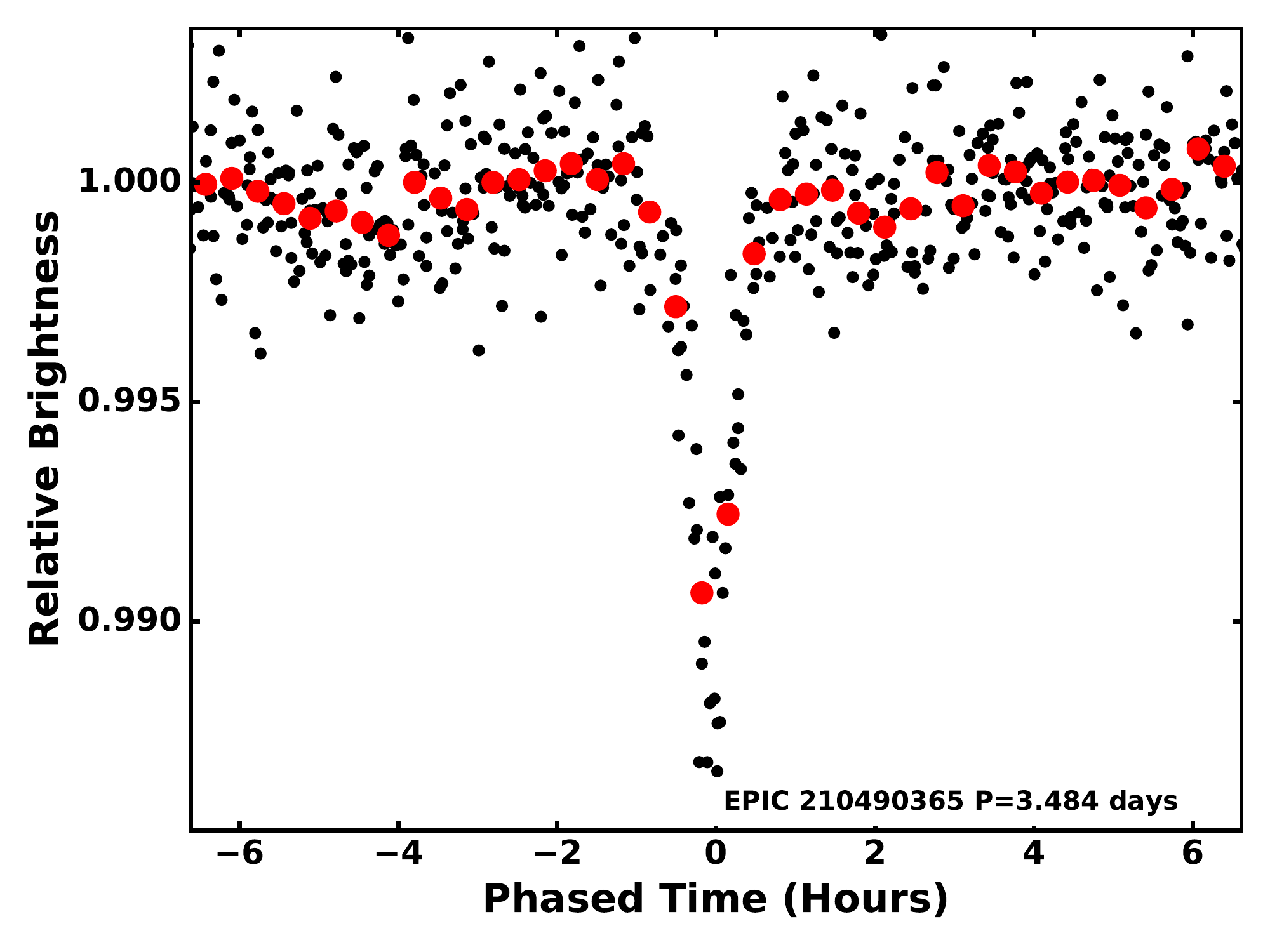} 
\includegraphics[width= 0.35\textwidth]{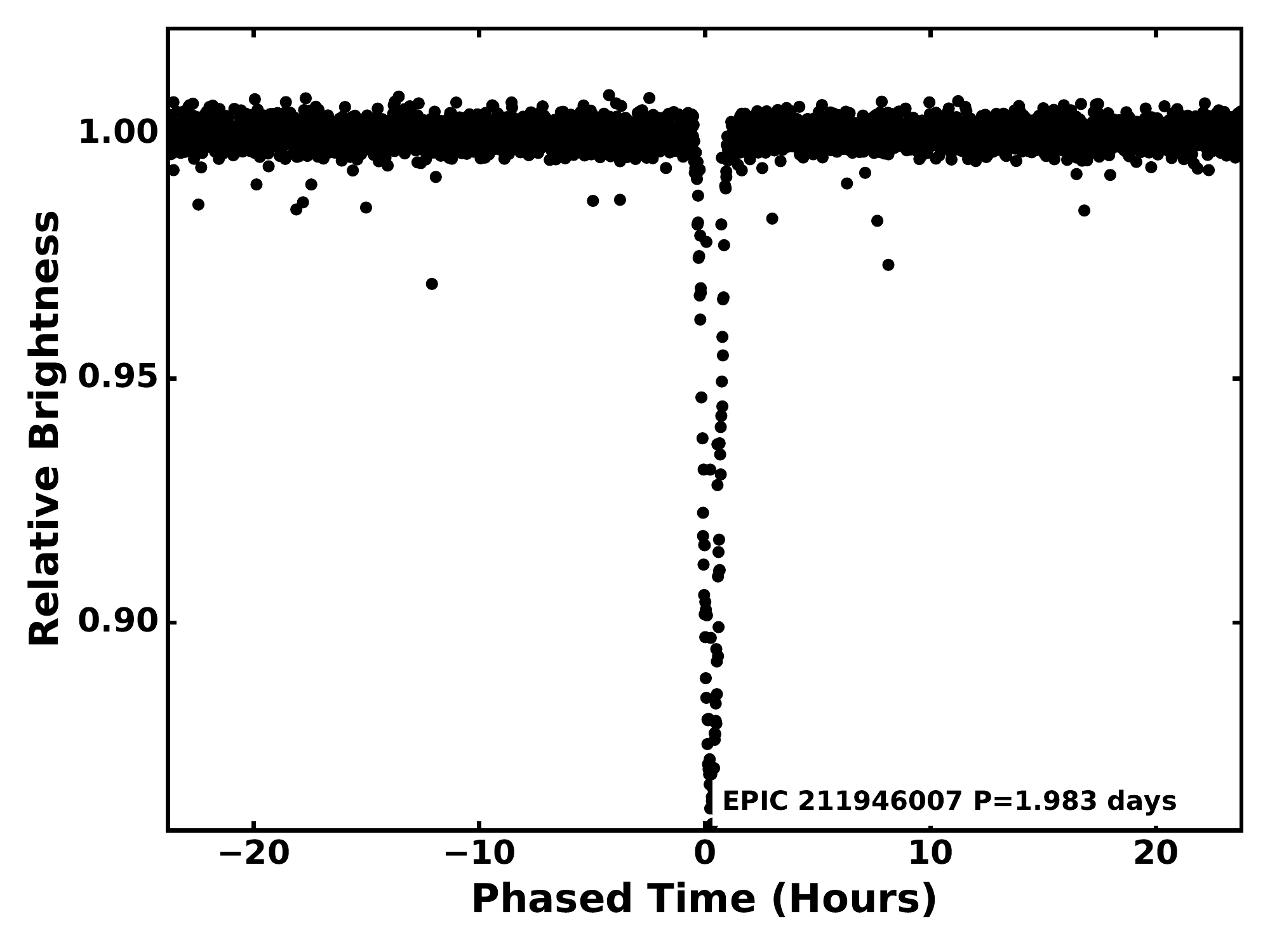} \\  
\includegraphics[width= 0.35\textwidth]{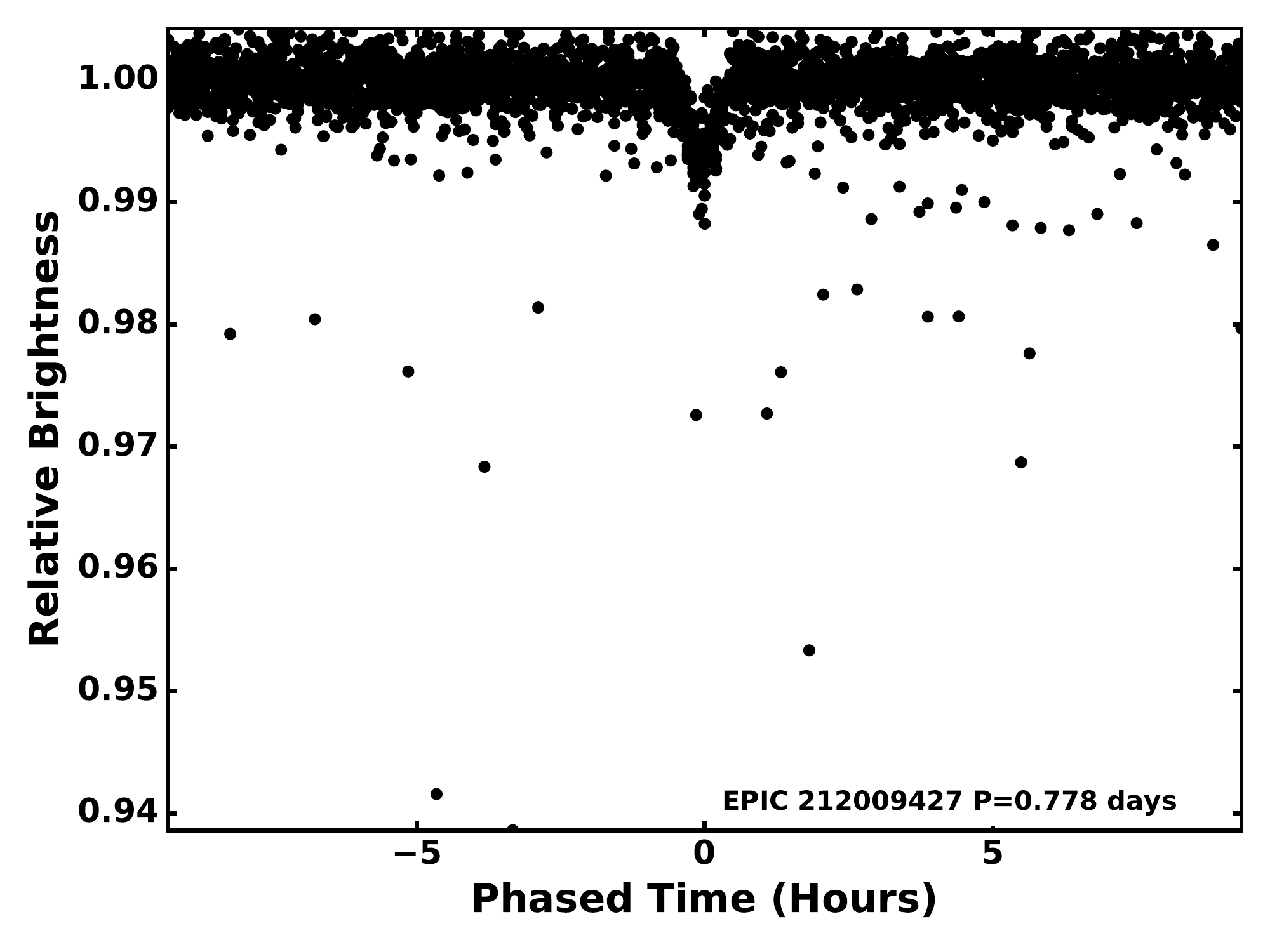} 
\includegraphics[width= 0.35\textwidth]{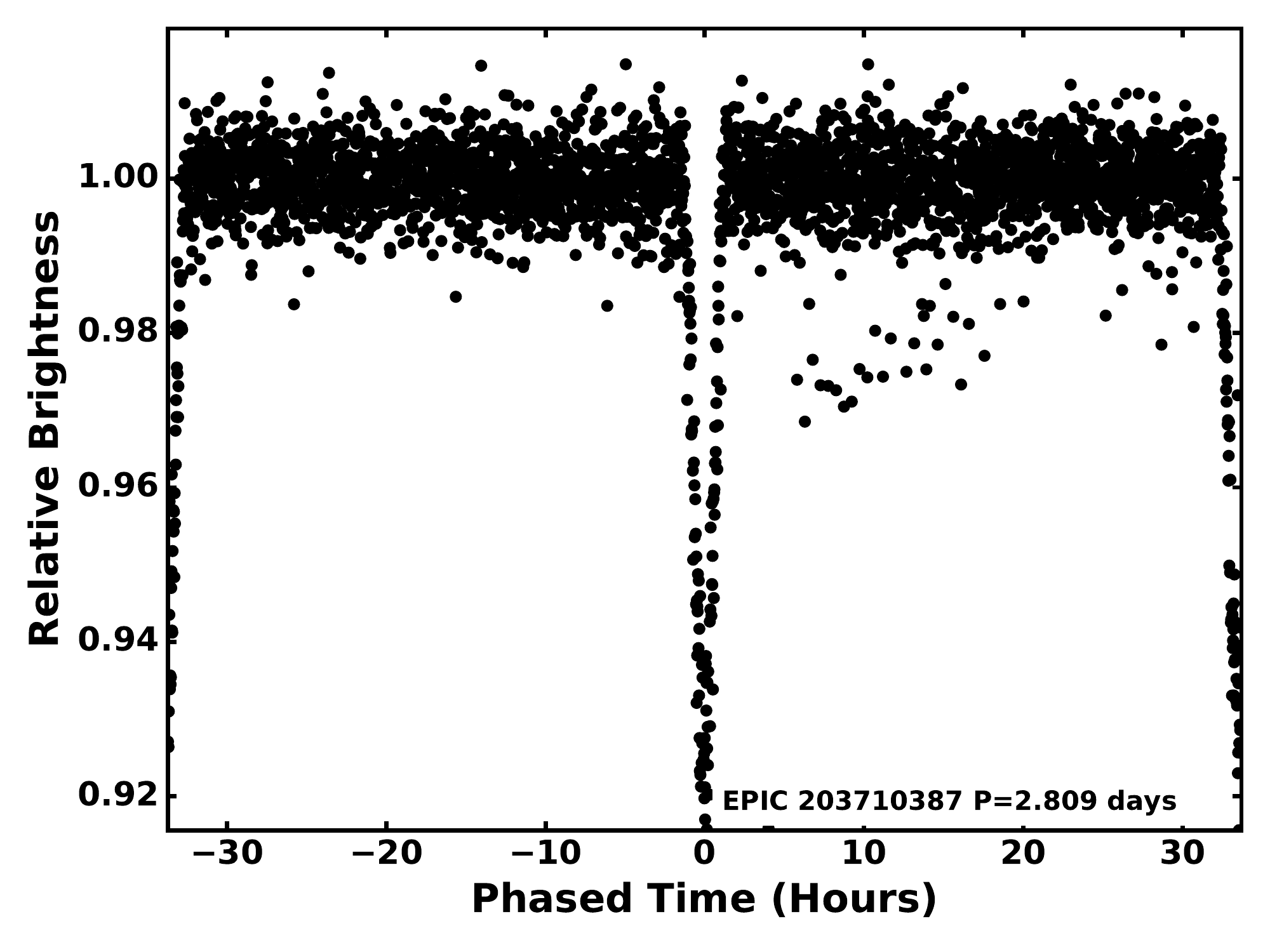}\\
\includegraphics[width= 0.35\textwidth]{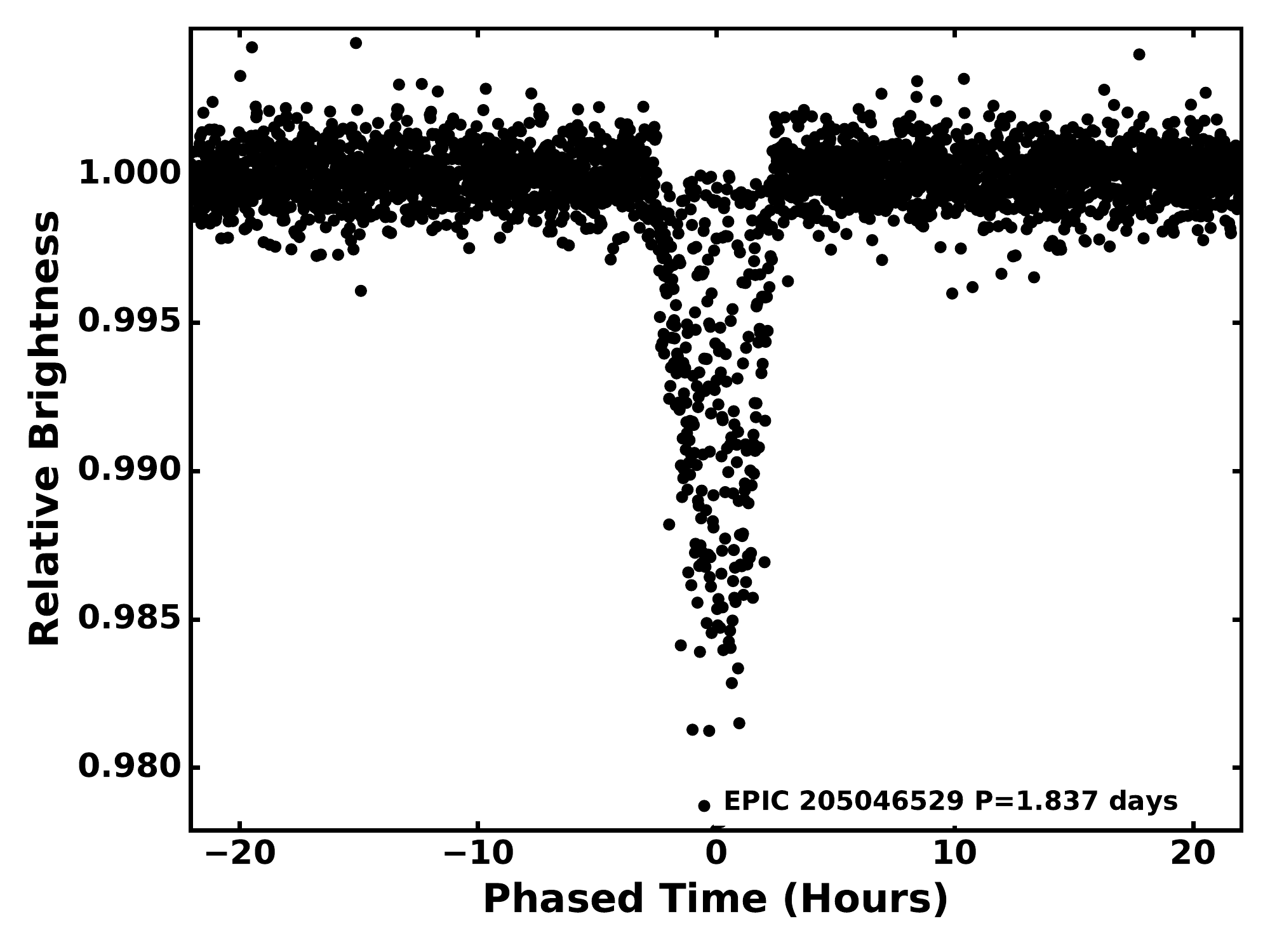} 
\caption{Period signals detected using the LOCoR pipeline, phased to the detected period. In all panels black points are the detrended \emph{K2} observations.} 
\label{figs:det1} 
\end{figure*}